\documentclass[12pt]{article}

\usepackage[titletoc,title]{appendix}
\usepackage[margin=1in]{geometry}
\usepackage{array}
\usepackage{color}
\usepackage{amsmath}
\usepackage{bbm}
\usepackage{amssymb}
\usepackage{amsthm}
\usepackage{amsfonts}
\usepackage{enumerate}
\usepackage[toc]{glossaries}
\usepackage{hyperref}
\usepackage{listings}
\usepackage{natbib}
\usepackage{url}
\usepackage{graphicx}
\usepackage{graphics}
\usepackage{subcaption}
\usepackage{fancyhdr}
\usepackage{rotating}
\usepackage{multirow}
\usepackage[usenames,dvipsnames]{xcolor}
\usepackage{sectsty}
\usepackage{titlesec}
\usepackage{booktabs}
\usepackage{datetime}
\usepackage{tikz}
\usetikzlibrary{arrows,automata,positioning,chains,fit,shapes,calc,patterns}
\usepackage{type1cm}
\usepackage{lettrine}
\usepackage{comment}
\usepackage{morefloats}
\usepackage{longtable}
\usepackage{psfrag}
\usepackage{changepage}
\usepackage{xstring}
\usepackage[ampersand]{}
\usepackage{mathtools}
\usepackage{multicol}
\usepackage{setspace}
\usepackage{censor}
\usepackage{tabularx}
\usepackage[space]{grffile}
\usepackage{blkarray, bigstrut} %
\usepackage[normalem]{ulem}
\useunder{\uline}{\ul}{}

\settimeformat{ampmtime}
\mmddyyyydate

\def\ig{\iota\gamma}
\def\g{\gamma}
\def\i{\iota}
\def\o{\omega}

\DeclareMathOperator*{\argmax}{arg\,max}

\usepackage{multirow,bigdelim}

\IfFileExists{/Users/jfogel/Networks/bibtex/bibtex.bib}{
	\def\bceta{\bibliography{/Users/jfogel/Networks/bibtex/bibtex}}

}{
	\def\bceta{\bibliography{/home/bm/Networks/bibtex/bibtex}}

}

\definecolor{linkcolor}{rgb}{0,0,0.50}
\setcitestyle{authoryear, round, semicolon, aysep={},yysep={;},notesep={,}}
\hypersetup{
    bookmarks=true,                 
    unicode=false,                  
    pdftoolbar=true,                
    pdfmenubar=true,                
    pdffitwindow=true,              
    pdfstartview={FitH},            
    pdftitle={},                    
    pdfauthor={BPS team},           
    pdfnewwindow=true,              
    colorlinks=true,                
    linkcolor=black,                 
    citecolor=black,                
    filecolor=magenta,              
    urlcolor=blue,                  
    breaklinks=true
}

\IfFileExists{/Users/jfogel/Networks/bibtex/bibtex.bib}{
	\def\bceta{\bibliography{/Users/jfogel/Networks/bibtex/bibtex}}}{
	\def\bceta{\bibliography{/home/ber/Networks/bibtex/bibtex}}}

\setcitestyle{authoryear, round, semicolon, aysep={},yysep={;},notesep={,}}

\definecolor{gold}{rgb}{0.85,.66,0}
\definecolor{blue}{rgb}{0,0,1}

\def\bs{\begin{sideways}}
\def\es{\end{sideways}}

\def\Pig{\frac{\exp \left( \frac{\psi_{\ig} 
			w_{\g} + \xi_{\g}}{\nu} \right) }{ \sum\limits_{\g'=0}^{\Gamma} \exp \left( \frac{\psi_{\ig'} w_{\g'} + \xi_{\g'}}{\nu} \right) }}

\usepackage[draft]{fixme}
\fxsetup{layout=footnote, marginclue}

\newcommand{\N}{{\mathcal N}}

\newcommand{\E}{\mathbb {E}}

\renewcommand{\P}{\mathbb {P}}

\usepackage{footmisc}

\theoremstyle{definition}
\newtheorem{theorem}{Theorem}[section]

\theoremstyle{plain}
\newtheorem{assumption}[theorem]{Assumption}

\usepackage{mathrsfs} 

\newcommand\indep{\protect\mathpalette{\protect\independenT}{\perp}}
\def\independenT#1#2{\mathrel{\rlap{$#1#2$}\mkern2mu{#1#2}}}
\def\ve{\varepsilon}

\makeglossaries

\title{What is a Labor Market? \\ Classifying Workers and Jobs Using Network Theory\footnote{Fogel Opportunity Insights, jamiefogel@g.harvard.edu.  Modenesi: University of Michigan, bmodene@umich.edu. This material is based upon work supported by the National Science Foundation Graduate Research Fellowship Program under Grant No. 1256260. Any opinions, findings, and conclusions or recommendations expressed in this material are those of the author(s) and do not necessarily reflect the views of the National Science Foundation. This research is also supported by the Alfred P. Sloan Foundation through the CenHRS project at the University of Michigan. This work is done in partnership with the Brazilian Institute of Applied Economic Research (IPEA). We thank John Bound, Abigail Jacobs, Matthew Shapiro, Mel Stephens, and Sebastian Sotelo for advice and guidance throughout this project. We also thank Charlie Brown, Zach Brown, Raj Chetty, Ying Fan, John Friedman, Florian Gunsilius, Nathan Hendren, Dhiren Patki, Rafael Pereira, Matthew Staiger, Dyanne Vaught, and Jean-Gabriel Young for helpful comments and discussions.  We also received helpful feedback from seminar participants at the University of Michigan, Labo(u)r Day, the Urban Economics Association, Networks 2021, Yale University, Duke University, the Federal Reserve Bank of Boston, Opportunity Insights, and JAM.}}

\author{Jamie Fogel and Bernardo Modenesi}
\date{}

\begin{document}

\maketitle

\begin{abstract}
This paper develops a new data-driven approach to characterizing latent worker skill and job task heterogeneity by applying an empirical tool from network theory to large-scale Brazilian administrative data on worker--job matching. We microfound this tool using a standard equilibrium model of workers matching with jobs according to comparative advantage. Our classifications identify important dimensions of worker and job heterogeneity that standard classifications based on occupations and sectors miss. The equilibrium model based on our classifications more accurately predicts wage changes in response to the 2016 Olympics than a model based on occupations and sectors. Additionally, for a large simulated shock to demand for workers, we show that reduced form estimates of the effects of labor market shock exposure on workers' earnings are nearly 4 times larger when workers and jobs are classified using our classifications as opposed to occupations and sectors. 
\end{abstract}
\clearpage

\clearpage

\onehalfspacing
\section{Introduction}

Many questions in economics require researchers to classify heterogeneous workers and jobs into discrete groups. For example, to study the labor demand shock generated by trade liberalization with China, \citet{AutorDornHanson2013} group workers by commuting zone (CZ) and compare outcomes between workers in CZs with varying levels of exposure to the China shock. Other commonly-used indicators of worker and job similarity include observable variables like age, education, occupation, industry, geography, or skills as measured by the Occupational Information Network (O*NET). Relying on observable indicators of worker and job similarity has well-known limitations: (i) relevant dimensions of worker and job heterogeneity may be unobserved or measured with error, and (ii) it requires researchers to decide which dimensions of heterogeneity are important. Indeed, as Autor and coauthors note, ``according to O*NET, the skill `installation' is equally important to both computer programmers and to plumbers, but, undoubtedly, workers in these occupations are performing very dissimilar tasks.''  \citep{FrankAutorBessenBrynjolfssonCebrianDemingFeldmanGrohLoboMoro2019}

To understand why this matters, consider a simplified version of \citet{AutorDornHanson2013}'s China shock. Suppose there are two CZs, A and B, where we assume A was exposed to the China shock and B was not. To estimate the effect of the China shock, we would compare the difference in pre- to post-shock changes in outcomes between CZs A and B. Suppose further that (i) the truly relevant heterogeneity can be characterized as urban versus rural, (ii) the China shock only affected rural areas, and (iii)  CZ A is 2/3 rural and B is 1/3 rural. In this case, when we compare changes in outcomes between A and B we are comparing outcomes for a ``treatment'' group that is in fact 2/3 treated (rural) and 1/3 control (urban) to outcomes for a ``control'' group that is actually 2/3 control and 1/3 treated. Consequently, our estimated effects may be attenuated. In general, the relevant dimensions of heterogeneity may encompass both skill/task and geographic dimensions, and cut across observable groups. Therefore, instead of relying on observable characteristics, we propose a novel, data-driven approach that classifies workers and jobs into discrete groups based on patterns of similarity revealed by the network of worker--job matches embedded in linked employer-employee data. 

We start with the intuition that workers employed in the same job have similar skills, and jobs employing the same workers require similar tasks. We formalize this intuition using a \citet{Roy1951} model in which workers belong to a discrete set of latent \emph{worker types} and jobs belong to a discrete set of latent \emph{markets}. For ease of exposition, we assume that worker types and markets reflect only skills and tasks, respectively, however they may also reflect factors like geography, credentials, or preferences.  Workers match with jobs according to comparative advantage, which is determined by complementarities between skills and tasks at the worker type--market level. The model implies that all workers in the same worker type have the same vector of match probabilities over jobs, and all jobs in the same market have the same vector of hiring probabilities over workers. We invert this logic and derive a maximum likelihood estimator that assigns workers to worker types and jobs to markets based on the observed network of worker--job matches. The MLE estimator uses realized job matches for each worker and their peers --- coworkers, former coworkers, coworkers' former coworkers, former coworkers' coworkers, etc. --- to approximate each worker's match probability distribution over jobs, and clusters together workers with the most similar distributions into worker types. It simultaneously clusters jobs into markets following a symmetric argument. We show that our MLE is equivalent to a tool from the community detection branch of network theory called the bipartite stochastic block model (BiSBM), and use computational techniques from network theory to solve the model.

By inferring worker and job similarity directly from empirical matching patterns, we bypass the need to (i) directly observe all relevant dimensions of worker and job heterogeneity, and (ii) understand how those dimensions interact in potentially non-linear ways. Moreover, by deriving our clustering algorithm from a Roy model of workers matching with jobs according to comparative advantage, we give our worker types and markets precise economic interpretability. Finally, we embed our Roy model of the labor market within a general equilibrium model with workers, firms, households and exogenous product demand shocks based on \citet{Grigsby2019} in order to microfound the determinants of comparative advantage and allow for counterfactuals representing labor supply or demand shocks.

We estimate our model and conduct empirical analyses using Brazilian administrative records from the Annual Social Information Survey (RAIS) that is managed by the Brazilian labor ministry. The RAIS data contain detailed information about every formal sector employment contract, including worker demographic information, occupation, sector, and earnings. Critically, these data represent a network of worker--job matches in which workers are connected to every job they have ever held, and vice versa.


We identify 446 worker types and 1,371 markets. This level of granularity is similar to defining markets by the intersection of two-digit occupations and meso regions\footnote{Meso regions are administrative divisions that are larger than cities but smaller than states. There are 137 meso regions in Brazil.}, henceforth ``occ2Xmesos'', of which there are 1,480. Therefore, we treat occ2Xmesos as our primary ``status quo'' comparison. We present descriptive facts about our worker types and markets that (i) validate the quality of our worker types and markets, and (ii) identify patterns of worker and job similarity that standard classifications based on occupation, industry or geography would have missed.

Our validation exercises begin with the premise that the ideal worker and job classifications will maximize within-group similarity and minimize across-group similarity. First, we show that workers' labor supply is significantly more concentrated within our markets than within either occ2Xmesos or industries, and symmetrically jobs' hiring is more concentrated within our worker types. Second, we perform an out-of-sample prediction exercise in which our worker types and markets outperform occ2Xmesos in predicting workers' future job-to-job flows. Third, we estimate the worker type--market productivity matrix that governs comparative advantage in our Roy model, which can be interpreted as a representation of worker skills, and show that our worker types do a better job of identifying groups of workers with distinct skills. Finally, we use our general equilibrium model to show that our classifications more accurately predict the wage effects of the labor demand shock created by the 2016 Rio de Janeiro Olympics. 

The clusters we identify are intuitive. For example, we identify one worker type composed primarily of physical education teachers and youth sports coaches and another composed of math and English teachers, despite the fact that these occupations differ at the two-digit level. At the same time, we disaggregate dissimilar workers who are employed in the same occupation. For example, among workers employed in the occupation “Course Instructor,” we infer that some workers are more like coaches and physical education teachers, while others belong with math and English teachers. We find that some markets are defined more by skills and tasks, while others are defined more by geography. Additionally, the distribution of occupations within some markets is highly concentrated, while within others it is dispersed. Traditional market definitions based on occupation and/or geography alone would have missed important details.

Our main empirical application applies our worker types and markets to reduced form Bartik-style regressions and finds that using our classifications significantly increases the magnitude of estimates of the effects of workers' exposure to labor market shocks on their earnings. We estimate the effect of the 2016 Olympics on workers and show that both coefficient estimates and $R^2$ values are significantly larger when workers and jobs are classified using our worker types and markets as opposed to occupations and sectors. We then perform a series of simulations in which we feed shocks through our model to generate data in which we know the true data generating process and estimate the effects of the shocks on workers in the simulated data, first using our network-based classifications, and again using conventional classifications. Across these simulations, the estimated effects of the shocks on workers' earnings are on average 3.7 times larger using our classifications as opposed to conventional classifications. Finally, we perform a detailed case study of a simulated shock to understand why our classifications outperform traditional ones. We show that our worker types more precisely identify groups of workers who experienced similar exposure to labor market shocks than do occupations and our markets more precisely identify groups of jobs that hire similar workers than do sectors. 

While this paper applies our classifications to estimating the effects of local labor market shocks, they are useful for a variety of applications. For example, our classifications may be used to improve labor market definitions when measuring labor market power.\footnote{\citet{BergerHerkenhoffMongey2022,Felix2021,AzarMarinescuSteinbaumTaska2018,BenmelechBergmanKim2018,Rinz2018,AzarMarinescuSteinbaum2019,SchubertStansburyTaska2020,Arnold2020,Lipsius2018,JaroschNimczikSorkin2019}} Similarly,  to characterize two-sided (worker--job) multidimensional heterogeneity, researchers identify groups of workers with similar skills and study how they match with groups of jobs requiring similar tasks. Our classifications may be used in place of conventional classifications --- based on occupations, educational attainment, or low-dimensional measures of skills and tasks --- in this class of models.\footnote{\citet{AutorLevyMurnane2003,AcemogluAutor2011,Autor2013,Tan2018,Lindenlaub2017,Kantenga2018}} 


\textbf{Literature:} We contribute to the large literature measuring the effects of labor market shocks on workers using either reduced form methods \citep{AutorDornHanson2013,Card1990,AutorDornHansonSong2014,Yagan2017,BoundHolzer2000,BlanchardKatz1992,Bartik1991}, or a structural approach \citep{BursteinMoralesVogel2019,CaliendoDvorkinParro2019,GalleRodriguezclareYi2017,KimVogel2021}. Relative to both of these literatures, our contribution is a new approach to classifying workers and jobs based on latent heterogeneity.

Conditional on assigning workers to latent worker types and jobs to latent markets, our model of labor supply is similar to \citet{Grigsby2019} and \citet{BonhommeLamadonManresa2019_distributional}, however our key innovation is identifying worker types in a data-driven way and with considerable greater granularity. Our method for clustering workers and jobs builds upon the bipartite stochastic block model from the community detection branch of the network theory literature \citep{LarremoreClausetJacobs2014,Peixoto2019}. A major contribution of our paper is creating a theoretical link between a labor supply model and the BiSBM, thereby providing microfoundations for using tools from network theory to solve problems in economics and giving these tools clear economic interpretability.

Like \citet{Sorkin2018}, \citet{Nimczik2018}, and \citet{JaroschNimczikSorkin2019}, we use tools from network theory to extract previously unobserved information from LEED. We use the panel of worker--job matches to identify worker and job \emph{similarities}; by contrast, Sorkin exploits the direction of worker flows between firms to identify \emph{differences} between firms. \citet{Nimczik2018}, and \citet{JaroschNimczikSorkin2019} also use network data to identify similarities, however they cluster together only firms, abstracting from worker heterogeneity and within-firm job heterogeneity, while we cluster workers \emph{and} jobs simultaneously. \citet{Schmutte2014} uses a different tool from network theory to cluster workers and firms using survey data, however our microfoundations and detailed data allow us to identify more fine-grained heterogeneity and provide model-based interpretability of our classifications. 

Our approach to modeling multidimensional worker--job heterogeneity is related to the literature on worker--job matching in a skills-tasks framework \citep{AutorLevyMurnane2003,AcemogluAutor2011,Autor2013,Lindenlaub2017,Tan2018,Kantenga2018}. Relative to this literature, we provide a theoretically principled and data-driven way of identifying groups of workers with similar skills and groups of jobs with similar tasks. \citet{Mansfield2019} also studies two-sided matching and integrates skill--task dimensions with geographic dimensions.  Our contribution is to improve identification of clusters of workers and jobs who are similar in terms of high-dimensional latent skills and tasks, respectively.

\textbf{Roadmap:} The paper proceeds as follows.  Section \ref{sec:model} lays out our economic model. Section \ref{sec:bisbm} builds upon the model to derive a maximum likelihood procedure for clustering workers into worker types and jobs into markets.  Section \ref{sec:data} discusses our data and sample restrictions and presents basic summary statistics Section \ref{sec:descriptive_results} demonstrates that our worker types and markets outperform traditional worker and job classifications in a variety of contexts. Section \ref{sec:reduced_form} applies our classifications to Bartik-style regressions and shows that standard methods may be understating the effects of shocks on workers. Section \ref{sec:conclusion} concludes.

\section{Model}
\label{sec:model}

In this section we develop a model based on \citet{Grigsby2019} of workers supplying labor to jobs according to comparative advantage, where comparative advantage is determined by the interaction of potentially high-dimensional worker skills and job tasks. From this model of labor supply we derive our maximum likelihood estimator that clusters workers into worker types and jobs into markets. We embed the labor market model in a general equilibrium model with workers, jobs, firms, and a representative household comprised of workers that consumes firms' output. The general equilibrium model facilitates interpretation of the worker and job types and allows for counterfactuals in which we simulate labor demand shocks.

\subsection{Labor supply}

The labor market consists of workers, $i$, who supply labor to jobs, $j$. There is a unit mass of workers, each of whom belongs to one of $I$ distinct worker types indexed by $i$. All workers in the same worker type are identical form the perspective of jobs. The exogenously determined mass of type $\i$ workers is denoted $m_{\i}$. Jobs are nested within firms and each job represents a set of tasks. A single job may employ multiple workers simultaneously. Each job belongs to one of $\Gamma$ distinct markets indexed by $\g$ and all jobs in the same market are identical from the perspective of workers.

For simplicity of exposition, we assume that worker types and markets are defined entirely by skills and tasks, respectively: all workers in the same worker type have the same set of skills and all jobs in the same market consist of the same set of tasks.\footnote{As we discuss in Section \ref{sec:bisbm_discussion}, it is straightforward to generalize worker types and markets to represent the intersection of skills/tasks, geography, preferences, credentials, and more.} Type $\i$ workers supply $\psi_{\ig}$ efficiency units of labor to jobs in market $\g$, where $\psi_{\ig}$ is a reduced form representation of the skill level of a type $\i$ worker in the various tasks required by a job in market $\g$. See \citet{Grigsby2019} for details.

Units of human capital are perfectly substitutable, meaning that if type 1 workers are twice as productive as type 2 workers in a particular market $\g$ (i.e. $\psi_{1\g} = 2\psi_{2\g}$), firms are indifferent between hiring one type 1 worker and two type 2 workers at a given wage per efficiency unit of labor, $w_{\g}$. Therefore, the law of one price holds for each market, and a type $\i$ worker employed in a job in market $\g$ is paid $\psi_{\ig}w_{\g}$. Because workers' time is indivisible, each worker may supply labor to only one market in each period and we do not consider the hours margin.

Workers' only decisions are their market choices. Workers are indifferent between individual jobs in the same market, meaning that individual jobs face perfectly elastic labor supply at the wage for their market, $w_{\g}$, which is determined in general equilibrium.\footnote{If workers do not view all jobs of the same type as identical, then individual jobs would face an upward-sloping labor supply curve, and would thus have some degree of market power. We explore this in concurrent work \citep{ModenesiFogel2021}.} In addition to earnings, each market $\g$ has a fixed amenity value to workers, $\xi_{\g}$; $\Xi = \begin{bmatrix}	\xi_1 & \xi_2 & \cdots &  \xi_{\Gamma}	\end{bmatrix}$.   Workers may also choose to be non-employed, denoted by $\g=0$, in which case they receive no wages but receive a non-employment benefit, which is normalized to 0 without loss of generality. Finally, each worker $i$ has an idiosyncratic preference for market $\g$ jobs at time $t$, $\ve_{i\g t}$. Therefore, worker $i$ chooses a market by solving
\begin{align}
\g_{it} = \argmax_{\g \in \{0,1,\dots,\Gamma\}} \psi_{\ig} w_{\g t} + \xi_{\g} + \ve_{i\g t} \label{eq:worker_max}
\end{align}
where $\g_{it}$ denotes the market worker $i$ chooses to supply labor to at time $t$. We assume that $\ve_{i\g t}$ is iid type 1 extreme value with scale parameter $\nu$:

\begin{assumption}[Distribution of preference shocks]
	\label{ass:taste_shocks}
	Idiosyncratic preference shocks $\ve_{i\g t}$ are drawn from a type-I extreme value distribution with dispersion parameter $\nu$ and are serially uncorrelated and independent of all other variables in the model.
\end{assumption}

 This gives us a functional form for the probability that a type $\i$ worker chooses a job in market $\g$:
\begin{align}
\P_{\i}[\g_{it}|\Psi,\vec{w_t},\Xi,\nu]= \frac{\exp \left( \frac{\psi_{\ig} w_{\g t} + \xi_{\g}}{\nu} \right) }{ \sum\limits_{\g'=0}^{\Gamma} \exp \left( \frac{\psi_{\ig'} w_{\g' t} + \xi_{\g'}}{\nu} \right) }. \label{eq:emp_probs}
\end{align} 

We derive labor supply to market $\g$ by aggregating equation (\ref{eq:emp_probs}) over all worker types, weighting by the mass of workers in worker type $\i$, $m_{\i}$, and the efficiency units of labor supplied by iota workers to gamma jobs, $\psi_{ig}$:
\begin{align}
LS_{\g}(\vec{w_t}) = \sum_{\i} m_{\i} \P_{\i}[\g_{it}|\Psi,\vec{w_t},\Xi,\nu] \psi_{\ig} = \sum_{\i} m_{\i} \left( \frac{\exp \left( \frac{\psi_{\ig} w_{\g t} + \xi_{\g}}{\nu} \right) }{ \sum\limits_{\g'=0}^{\Gamma} \exp \left( \frac{\psi_{\ig'} w_{\g' t} + \xi_{\g'}}{\nu} \right) } \right) \psi_{\ig} \label{eq:labor_supply}
\end{align}

\subsection{Timing}

We observe the economy for $T$ periods.  In each period a worker may draw a Poisson-distributed exogenous separation shock, denoted $c_{it} = \mathbbm{1}_{j(i,t)\neq j(i,t-1)}$ where $j(i,t)$ is the job employing worker $i$ at time $t$ (Assumption \ref{ass:mobility}). Workers who draw a separation shock receive a new set of idiosyncratic preference shocks $\ve_{i\g t}$ and search again following the same optimization problem defined in equation (\ref{eq:worker_max}). We assume that the labor market parameters, $\{\Psi, \Xi, \nu\}$, and the demand shifters $\vec{a}$, are fixed across all $T$ time periods we use for estimation (Assumption \ref{ass:constant_parameters}). These restrictions make the model a reasonable approximation for relatively short periods of time, but it would be inappropriate for studying long-run changes when labor supply parameters may be changing.

\begin{assumption}[Exogenous separations]
	\label{ass:mobility}
	Job separations for worker $i$, $c_{it}$, arrive at a worker-specific Poisson rate $d_i$, and are serially uncorrelated and independent of all other variables in the model. 
\end{assumption}

\begin{assumption}[Constant parameters]
	\label{ass:constant_parameters}
	The labor supply parameters, $\{\Psi, \Xi, \nu\}$, are constant over the periods in which we estimate the model and perform counterfactuals. The product demand shifters, $\vec{a}$, are constant over the periods in which we estimate the model.
\end{assumption}

The timing of the model is as follows. In each period $t$:
\begin{enumerate}
	\item Each employed worker draws an exogenous separation shock with probability $d_i$; workers who do not receive a separation shock remain in their current job
	\item Separated workers receive new preference shocks $\ve_{i\g t}$
	\item Separated workers choose a market $\g_{it}$ according to $\P_{\i}[\g_{it} |  \vec{w}]$
	\item Separated workers randomly match with a job within their chosen market $\g$
\end{enumerate}

Assumptions \ref{ass:mobility} and \ref{ass:constant_parameters} allow workers to move between jobs over time, generating the network of worker--job matches that is key to identifying worker types and markets. They also imply that worker movement between jobs is idiosyncratic, meaning that each of a worker's jobs represent i.i.d. draws from the same match probability distribution. We discuss this further in Section \ref{sec:bisbm_discussion}.

\subsection{Firms}
\label{sec:model_firms}

There are $S$ sectors indexed by $s$. Each sector $s$ consists of a continuum of firms producing the same sector-specific good in a competitive sector-level product market. Each firm, indexed by $f$, has a Cobb-Douglas production function which aggregates tasks from different labor markets, $\g$.  The quantity of the sector $s$ good produced by firm $f$, $y_{sf}$, is therefore given by
\begin{align}
	y_{sf} = \prod_{\g} \ell_{\g f}^{\beta_{\g s}}  \label{eq:int_prod}
\end{align}
where $\ell_{\g f}$ is the number of efficiency units of labor firm $f$ employs in jobs in market $\g$, and $\beta_{\g s}$ is the elasticity of sector $s$ output with respect to labor employed in market $\g$ in sector $s$. We define a job, indexed by $j$, as a firm-market pair. Therefore, we can replace the $\g f$ indices with $j$ in the equations above: $\ell_{\g f} \equiv \ell_j$. We denote the market to which job $j$ belongs as $\g(j)$. Therefore, the production function becomes 
\begin{align*}
	y_{sf} = \prod_{\{j\}_{j \in f}} \ell_{j}^{\beta_{\g s}}.
\end{align*}

The firm chooses labor inputs in order to maximize profits, taking as given the price of output $p_s$, a vector of wages per efficiency unit of labor $w_{\g}$, and a production function, equation (\ref{eq:int_prod}). Therefore, the firm solves
\begin{align}
	\pi_f = \max_{ \{ \ell_{\g f} \}_{\g=1}^{\Gamma} } \quad  p_s \cdot   \prod_{\g} \ell_{\g f}^{\beta_{\g s}}  - \sum_{\g} w_{\g} \ell_{\g f} . \label{eq:int_profit_max}
\end{align}
Production exhibits decreasing returns to scale because
\[ \sum_{\g} \beta_{\g s} =\alpha < 1 \quad \forall s \]
where $\alpha$ denotes the labor share. 

Total profits in the economy are the sum of all firms' profits: $\Pi = \sum_{s=1}^S \sum_{f \in s} \pi_f$.

\subsection{Household}

A representative household consumes output from each sector as inputs to a constant elasticity of substitution (CES) utility function. Utility is given by
\begin{align}
	U =\left ( \sum_{s=1}^S a_s^\frac{1}{\eta} y_s^{\frac{\eta-1}{\eta}} \right )^{\frac{\eta}{\eta-1}} \label{eq:utility}
\end{align}
where $C$ is a numeraire aggregate consumption good, $y_s$ is the household's consumption of sector $s$'s output, $\eta$ is the elasticity of substitution between sectors' output, and $a_s$ is a demand shifter for the sector $s$ good. In our counterfactual analyses we generate labor demand shocks by changing the vector of sector demand shifters $\vec{a}$.  It follows that the demand curve for sector $s$'s output is given by 
\begin{align}
	y_s^D =  \frac{ a_s }{ \sum_{s'} \left(\frac{p_s}{p_s'}\right)^{\eta} \left(a_{s'} p_{s'}\right)}Y \label{eq:consumer_demand} 
\end{align}
where $Y$ is total income.

The household consumes its entire income each period, meaning that $Y= \sum_{s} p_s y_s^D$. Because all workers belong to the household and the household owns all firms, total income is the sum of all labor income and profits in the economy: $Y = \bar W + \Pi$.

\subsection{Definition of equilibrium}

The model solution consists of vectors of goods prices $\vec{p} := \{p_{s}\}_{s=1}^S$ and wages per efficiency unit of labor $\vec{w} := \{w_{\g}\}_{\g=1}^\Gamma$ that satisfy all equilibrium conditions in each period. Since our model can be solved one period at a time with no cross-time dependence and the fundamentals of the economy are assumed to be constant over our estimation window, the equilibrium conditions below are the same in every period. Our equilibrium has the following components:
\begin{enumerate} 
	\item The labor demand functions $\ell_{\g f}$ solve the firms' problem (\ref{eq:int_profit_max})
	\item Labor supply is consistent with workers' expected utility maximization (\ref{eq:emp_probs})
	\item Goods markets clear. Specifically, demand from the representative household $y_s^D$ equals supply created by evaluating the production function at the optimal level of labor inputs and aggregating over all firms in the sector: $y_s = \sum_{f \in s} \prod_{j \in f} \ell_{j}^{\beta_{\g s}} $ (\ref{eq:int_prod}).
	\item The labor market clears for each market $\g$: $LS_\g = LD_\g := \sum_s \sum_{f \in s} \sum_{j \in f | \g(j)=\g} \ell_{j}$
	\item Aggregate consumption is equal to income: $Y = \sum_{s} p_s y_s^D = \bar W + \Pi$.
\end{enumerate}
We solve the model numerically, as described in Appendix \ref{sec:model_solution}.

\section{Classifying workers and jobs}

In this section, we start with the model of labor supply developed in the previous section and derive our maximum likelihood procedure for assigning workers to worker types, $\i$, and jobs to markets, $\g$. The only data used by the procedure is the set of realized worker--job matches. The procedure formalizes the intuition that two workers belong to the same worker type $\i$ if they have the same vectors of match probabilities over markets, and two jobs belong to the same market $\g$ if they have the same vectors of match probabilities over worker types.

\label{sec:bisbm}

\subsection{Assigning workers to worker types and jobs to markets}

As stated in equation (\ref{eq:emp_probs}), when any worker $i$ belonging to type $\i$ searches for a job, the probability that they choose a job in market $\g$ is
\begin{align*}
\P_{\i}[\g_{it}|\Psi,\vec{w_t},\Xi,\nu]= \frac{\exp \left( \frac{\psi_{\ig} w_{\g t} + \xi_{\g}}{\nu} \right) }{ \sum\limits_{\g'=0}^{\Gamma} \exp \left( \frac{\psi_{\ig'} w_{\g' t} + \xi_{\g'}}{\nu} \right) }
\end{align*}
This quantity corresponds to a discrete choice at a specific time, $t$. Our assumption that the labor supply parameters ($\Psi$, $\Xi$, and $\nu$) and demand shifters ($\vec{a}$) are unchanging during our estimation period, combined with the fact that $\vec{w_t}$ is determined in equilibrium by the labor supply parameters and demand shifters, means that this choice probability does not depend on the time period. Therefore, we drop the time subscript $t$ in what follows. All workers make this choice in period 1, and workers subsequently make another choice following this distribution any time they experience an exogenous separation. 

The quantity in equation (\ref{eq:emp_probs}), $\P_{\i}[\g_{it}|\Psi,\vec{w_t},\Xi,\nu]$, refers to the probability of an individual worker $i$ matching with \emph{any} job in market $\g$, not a particular job $j$. To obtain the probability that worker $i$ matches with a \emph{specific} job $j$ in market $\g$, we multiply the choice probability in equation (\ref{eq:emp_probs}) by the probability that worker $i$ matches with job $j$, conditional on choosing a job in market $\g$. Because we have assumed that all jobs in the same market are identical from the perspective of workers, this probability is equal to job $j$'s share of market $\g$ employment. Let $d_j$ denote the number of workers employed by job $j$ during our estimation period. Then job $j$'s share of all market $\g$ employment can be written 
\begin{align}
	\P[j|\g] = \frac{d_j}{\sum_{j' \in \g} d_{j'}^{J}}. \label{eq:P_j_g}
\end{align}  
Therefore, when worker $i$ of type $\i$ searches, the probability that the search results in worker $i$ matched with job $j$ is the product of the probabilities in equation (\ref{eq:emp_probs}) and equation (\ref{eq:P_j_g}): 
\begin{align}
\P_{ij}  &= 
\overset{\P_i[\g | \Psi,\vec{w},\Xi,\nu] }{\overbrace{\Pig }} 
\times \overset{\P[j|\g]}{\overbrace{\underset{\substack{1/ \text{ type } \g \\ \text{employment}  }}{ 	
			\underbrace{ \vphantom{\Pig} \frac{1}{\sum_{j' \in \g} d_{j'}^{J} } }} 
		\times \underset{\substack{\text{Job }j \\  \text{employment}}}{\underbrace{ \vphantom{\Pig} d_j}}    }} .
\end{align}
The first term represents the probability that worker $i$ chooses market $\gamma$, while the second represents the probability that worker $i$ chooses job $j$ conditional on choosing market $\gamma$. We can rewrite this expression as the product of a term that depends only on the worker's type and job's market, which we denote $\mathcal{P}_{\ig}$, and a job-specific term $d_j$:
\begin{align}
\P_{ij} &= \overset{ := \mathcal{P}_{\ig}}{\overbrace{
		\Pig 
		\times  \underset{\substack{1/ \text{ type } \g \\ \text{employment}  }}{ 	
			\underbrace{{ \vphantom{\Pig} \frac{1}{\sum_{j' \in \g} d_{j'}^{J} } }}}    }} \times \underset{\substack{\text{Job }j \\  \text{employment}}}{\underbrace{ \vphantom{\Pig} d_j}} \\
		&=  \mathcal{P}_{\ig} d_j . \nonumber \label{eq:P_ij}
\end{align}
Define $A_{ij}$ as the number of times worker $i$ matches with job $j$ across \emph{all} of $i$'s searches. Since the number of times worker $i$ searches depends on the number of separation shocks they draw from a $Poisson(d_i)$ distribution, it follows that $A_{ij}$ also follows a Poisson distribution:
\begin{align}
 A_{ij} \sim  Poisson \left( d_i d_j \mathcal{P}_{\ig}  \right).
\end{align}
For a complete proof, see Appendix \ref{app:poisson_proof}. Finally, define $\mathbf{A}$ as the matrix with typical element $A_{ij}$. $\mathbf{A}$ represents the full set of observed worker--job matches and is known as the adjacency matrix in network theory parlance. Since the elements of $\mathbf{A}$ are independent, we can write the density of $\mathbf{A}$ as 
\begin{align} 
P \bigg(\mathbf{A} \bigg|\vec{\i}, \vec{\g}, \vec{d_i}, \vec{d_j} , \mathbf{\mathcal{P}} \bigg)  
&= \prod_{ i,j } \frac{\left(d_i d_j \mathcal{P}_{\i(i)\g(j)}\right)^{A_{ij}}}{A_{ij}!} \exp \left(d_i d_i^J \mathcal{P}_{\i(i)\g(j)} \right)  \label{eq:BiSBM}
\end{align} 
where $\vec{\i} = \{\i(i)\}_{i=1}^N$ is the vector assigning each worker to a worker type, $\vec{\g} = \{\g(j)\}_{j=1}^J$ is the vector assigning each job to a market, $\vec{d_i} = \{d_i\}_{i=1}^N$, $\vec{d_j} = \{d_j\}_{j=1}^J$,  and $\mathbf{\mathcal{P}}$ is the matrix with typical element $\mathcal{P}_{\i\g}$. Using this, we estimate the worker type and market assignments for all workers and jobs, $\vec{\i}$ and $\vec{\g}$ respectively, using maximum likelihood.
\begin{align}
\vec{\i},\vec{\g} = \argmax_{\begin{matrix} \{\vec{\i}=\i(i)\}_{i=1}^N, \\ \{\vec{\g}=\g(j)\}_{j=1}^J \end{matrix}}  \prod_{ i,j } \frac{\left(d_i d_j  \mathcal{P}_{\i(i)\g(j)}\right)^{A_{ij}}}{A_{ij}!} \exp \left(d_i d_i^J  \mathcal{P}_{\i(i)\g(j)}\right) \label{eq:BiSBM_max_likelihood}
\end{align}
This problem actually has five sets of parameters: the worker type and market assignments $\vec{\i}$ and $\vec{\g}$, the worker and job match frequencies $\vec{d_i}$ and $\vec{d_j}$, and the type-specific match probabilities $\mathcal{P}_{\ig}$. The worker and job match frequencies, $\vec{d_i}$ and $\vec{d_j}$, are directly observable in the data so we use their actual values. The worker and market assignments, $\vec{\i}$ and $\vec{\g}$, are the parameters we choose in order to maximize the likelihood. Conditional on group assignments, the number of matches between each worker type--market pair is observable. Therefore, for a given set of $\vec{\i}$ and $\vec{\g}$, we use these observed worker type-market match counts to compute observed match probabilities, which we use as our estimate of the true probabilities, $\mathcal{P}_{\ig}$. We iteratively update these parameters until we reach an optimum, as described below.

Equation (\ref{eq:BiSBM_max_likelihood}) corresponds to the degree-corrected bipartite stochastic block model (BiSBM), a workhorse model in the community detection branch of network theory (see appendix \ref{app:network_theory} for details). It defines a combinatorial optimization problem, whereby if we had infinite computing resources, we would test all possible  assignments of workers to worker types and jobs to markets and choose the one that maximizes the likelihood in equation (\ref{eq:BiSBM_max_likelihood}). This is not computationally feasible for large networks like ours. Therefore, we use a Markov chain Monte Carlo (MCMC) approach in which we modify the assignment of each worker to a worker type and each job to a market in a random fashion and accept or reject each modification with a probability given as a function of the change in the likelihood. We repeat the procedure for multiple different starting values to reduce the chances of finding local maxima. We implement the procedure using a Python package called graph-tool. (\url{https://graph-tool.skewed.de/}. See \citet{Peixoto2014_efficient} for details.)

Equation (\ref{eq:BiSBM_max_likelihood}) assumes that we know the number of worker types and markets \emph{a priori}, however this is rarely the case in real world applications. Therefore we must choose the number of worker types and markets, $I$ and $\Gamma$ respectively. We do so using the principle of minimum description length (MDL), an information theoretic approach that is commonly used in the network theory literature. MDL chooses the number of worker types and markets to minimize the total amount of information necessary to describe the data, where the total includes both the complexity of the model conditional on the parameters \emph{and} the complexity of the parameter space itself. MDL penalizes a model that overfits the data by using a large number of parameters (corresponding to a large number of worker types and markets) by effectively adding a penalty term in our objective function, such that our algorithm finds a parsimonious model. This method has been found to work well in a number of real world networks \citep{Peixoto2013,Peixoto2014,RosvallBergstrom2007}. See appendix \ref{sec:MDL_details} for greater detail.

\subsection{Visual intuition of the BiSBM}

\begin{figure}[!htbp]
	\caption{Network representation of the labor market}
	\label{fig:sample_network}
	\centering
	\begin{subfigure}{\textwidth}
		\caption{}
		\begin{tikzpicture}
		\draw (0, 0) node[inner sep=0] {\includegraphics[width=.8\textwidth]{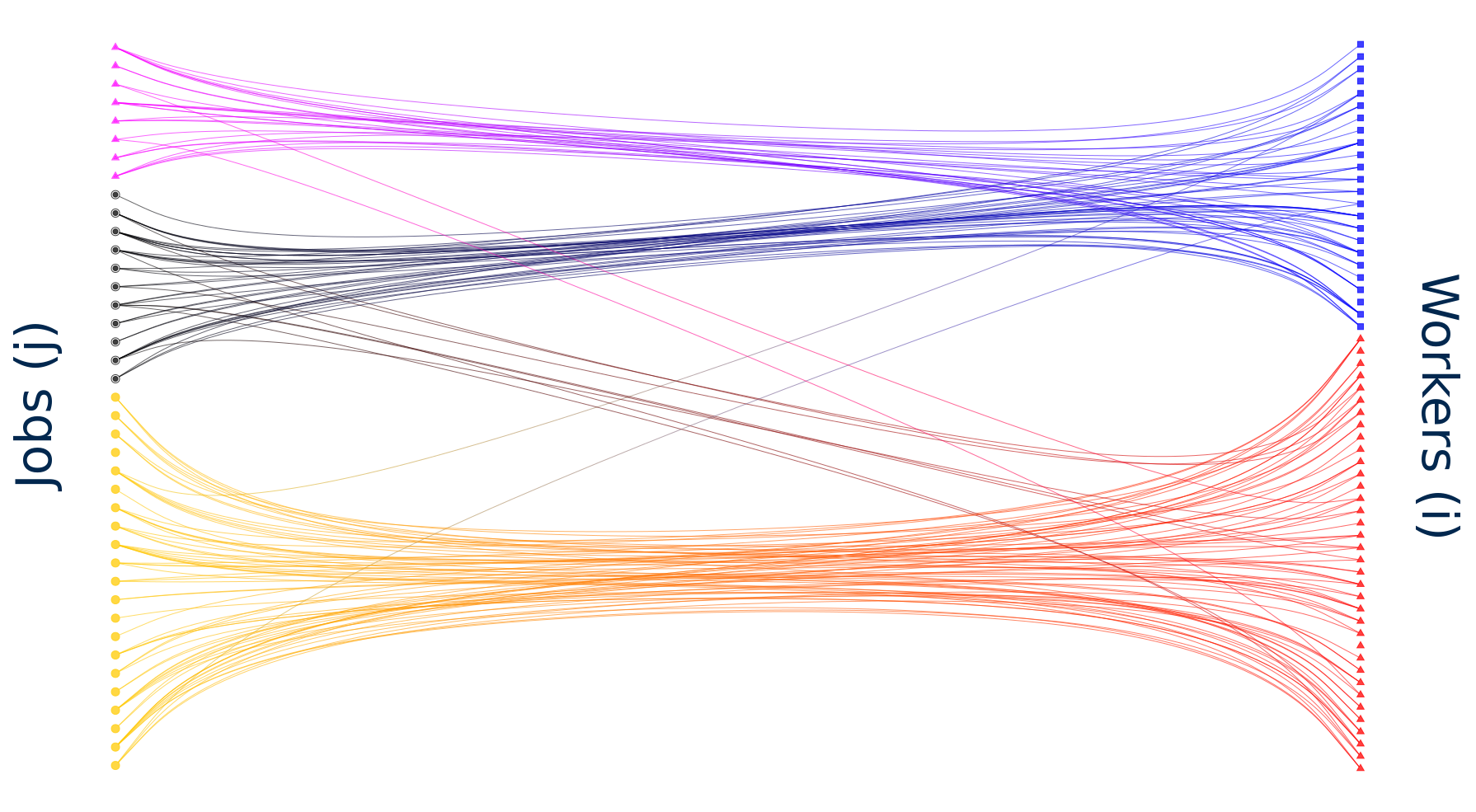}};
		\draw (-7.5, 2.3) node[inner sep=0] { $\g=1$};
		\draw (-7.5, 0) node { $\g=2$};
		\draw (-7.5, -2.3) node {  $\g=3$};
		\draw (7, 1.6) node { $\i=1$};
		\draw (7, -1.7) node { $\i=2$};
		\end{tikzpicture}\\
	\end{subfigure}
	\begin{subfigure}{\textwidth}
		\caption{}
		\centering
		\begin{tikzpicture}
		\draw (0, 0) node[inner sep=0] {\includegraphics[width=.8\textwidth]{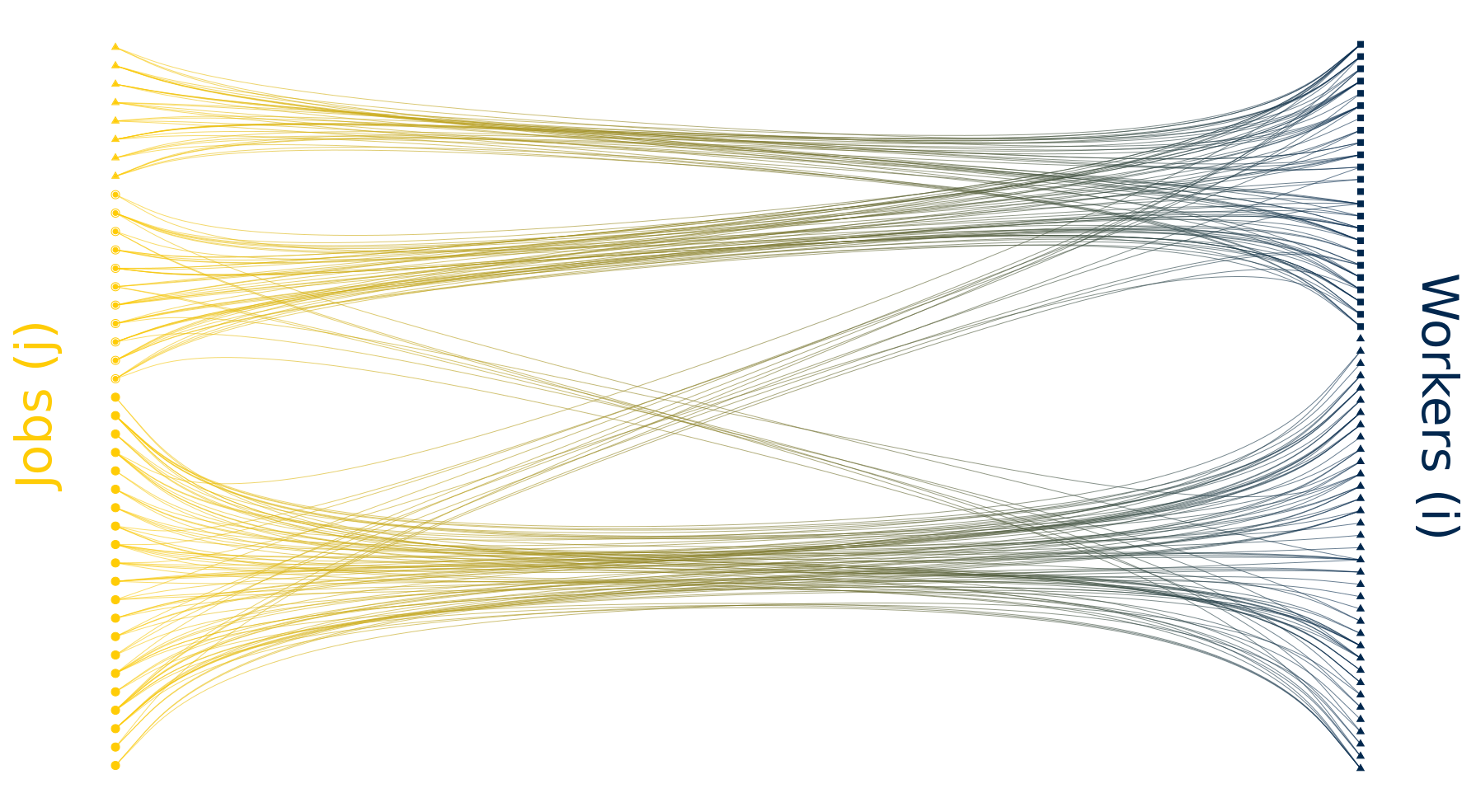}};
		\end{tikzpicture}\\
	\end{subfigure}
	\flushleft \footnotesize Dots represent individual workers/jobs; lines represent employment contracts. Network drawn according to 
	\begin{align*} 
	P \bigg(\mathbf{A} \bigg|\vec{\i}, \vec{\g}, \vec{d_i}, \vec{d_j} , \mathbf{\mathcal{P}} \bigg)  
	&= \prod_{ i,j } \frac{\left(d_i d_j \mathcal{P}_{\i(i)\g(j)}\right)^{A_{ij}}}{A_{ij}!} \exp \left(d_i d_i^J \mathcal{P}_{\i(i)\g(j)} \right) 
	\end{align*} 
	where
	\[P_{\i}[\g_{it}|\vec{w}] = 
	\begin{blockarray}{lccc}
	\g=1	&\g=2	& \g=3	& 	\\
	\begin{block}{(ccc)r}
	0.3		& 0.5	& 0.2	& \i=1	\\
	0.15	& 0.05	& 0.8	&  \i=2	\\ 
	\end{block}
	\end{blockarray}\]
\end{figure}
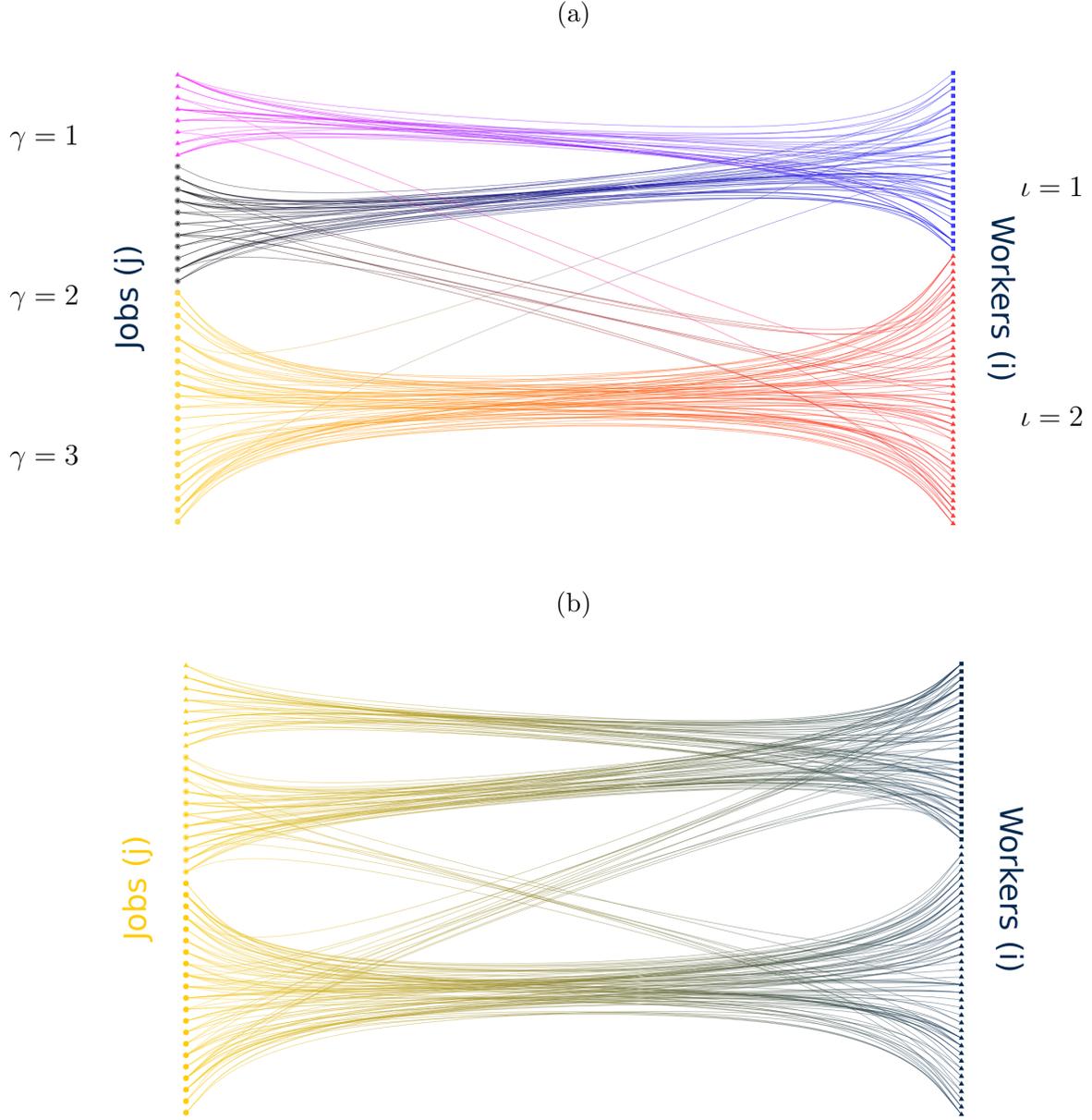

Figure \ref{fig:sample_network} panel (a) provides a simplified visual representation of how our model generates a network of worker--job matches. We assume that there are 2 worker types, 3 markets, and matches are drawn from a sample match probability distribution
\[\mathcal{P}_{\ig} = 
\begin{blockarray}{lccc}
\g=1	&\g=2	& \g=3	& 	\\
\begin{block}{(ccc)r}
0.3		& 0.5	& 0.2	& \i=1	\\
0.15	& 0.05	& 0.8	&  \i=2	\\ 
\end{block}
\end{blockarray}\]
Dots on the left axis represent individual jobs $j$ and dots on the right axis represent individual workers $i$. Workers belong to one of two worker types ($\i\in\{1,2\}$) and jobs belong to one of three markets ($\g\in\{1,2,3\}$).  Lines represent employment contracts between individual workers and jobs. A line connects worker $i$ and job $j$ if $A_{ij}>0$, while i and j are not connected if $A_{ij}=0$.  Consistent with $\mathcal{P}_{\ig}$, we see that type $\i=1$ workers match with all 3 markets with somewhat similar probabilities, while type $\i=2$ workers overwhelmingly match with type $\g=3$ jobs. In our actual data, we observe neither worker types and markets, nor worker type-market match probabilities. We only observe matches between individual workers and jobs, as represented by $A_{ij}$, and visualized here in panel (b) of Figure \ref{fig:sample_network}. Therefore, our task, formalized in the maximum likelihood procedure defined in equation (\ref{eq:BiSBM_max_likelihood}), is to take the data represented by panel (b) and label it as we do in panel (a). Intuitively, two workers belong to the same worker type if they have approximately the same vectors of match probabilities over all markets, and two jobs belong to the same market if they have approximately the same vector of match probabilities over all worker types.

\subsection{Discussion}

\label{sec:bisbm_discussion}

Our approach rests on the insight that workers with similar propensities to match with particular jobs have similar skills, while jobs with similar propensities to hire particular workers require similar tasks. We formalize this by making three major assumptions. First, our model implicitly assumes that workers match with jobs according to comparative advantage, where comparative advantage is governed by the productivity of the worker's skills when employed in the job's tasks (equation \ref{eq:emp_probs}). Second, Assumption \ref{ass:constant_parameters} states that the fundamentals of the economy --- the labor supply parameters $\Psi$, $\Xi$, and $\nu$, and the demand shifters $\vec{a}$ --- are fixed throughout our estimation window.  Third, combining the assumptions of i.i.d. T1EV preference shocks (Assumption \ref{ass:taste_shocks}) and exogenous separations (Assumption \ref{ass:mobility}), we assume that movement of workers between jobs represents idiosyncratic lateral moves. This allows us to treat a worker's multiple spells of employment as repeated draws from the same distribution, however, as we discuss below, this comes at the cost of ignoring the possibility that workers are climbing the career ladder or that worker flows represent structural shifts in the economy. These assumptions allow us to write the data generating process of the linked employer-employee data in equation (\ref{eq:BiSBM}), which in turn implies a maximum likelihood estimation strategy. Now, we address the ramifications of these assumptions in turn.

The first major assumption is that workers and jobs match according to a Roy model in which match probabilities are driven by skill-task match productivity. Since workers and jobs are clustered according to match probabilities, to the extent that match probabilities are determined by factors other than skills and tasks, we are clustering on the basis of these other factors. For example, if two groups of workers have very similar skills but rarely end up in the same jobs because they have different credentials, they would be assigned to different worker types, reflecting heterogeneity in credentials rather than skills. Similarly, we may identify groups of workers with similar skills but different preferences. For example, liberal and conservative political consultants may have very similar skills, but consider entirely disjoint sets of jobs due to their preferences. If this is true, our model would assign them to different worker types. If there is discrimination, for example on the basis of race or gender, this would be reflected in our productivity measure: our model would assume that certain workers are not being hired because they have low productivity, when in reality they are being discriminated against. Finally, our ``skills'' and ``tasks'' may also reflect geographic location and associated commuting costs. Therefore, what we call ``skills'' should be interpreted more generally as worker characteristics valued by jobs in the labor market, and similarly for ``tasks.'' This is an appealing feature of our method because our agnostic approach to defining labor market relevant worker characteristics allows us to identify clusters of workers who are viewed by the market as approximately perfect substitutes, and these clusters are the relevant units of analysis when considering the effects of shocks on workers. Our method would, however, be inappropriate for studying changes in how worker characteristics are viewed by the market, for example changes in occupational licensing laws or discrimination.  A similar logic applies to jobs and tasks.

The second assumption is that the fundamentals of the economy --- the assignments of individual workers and jobs to worker types and markets, the labor supply parameters $\Psi$, $\Xi$, and $\nu$, and the demand shifters $\vec{a}$ --- are fixed throughout our estimation window. This assumption is key to identification because it implies that the set of worker--job matches is drawn independently from an unchanging probability matrix $\mathcal{P}$, meaning that if two workers have the same vector of match probabilities it must be because they have the same vector of skills, and similarly for jobs. The static fundamentals assumption implies that we must estimate the model during a period of time in which the labor market experiences no large shocks.\footnote{Endogenously determined wages also drive observed matching patterns, but this is not a problem for our identification strategy. As long as the fundamentals of the economy are fixed, workers of the same type will still display similar matching probabilities and will be clustered together according to our method. In other words, even though the wage distribution shapes the matching patterns in the labor market, similar workers will still behave similarly if fundamentals are fixed.}


Finally, we assume exogenous separation shocks in order to rationalize the fact that while worker--job matches are somewhat persistent, we still observe job-to-job transitions even when the fundamentals of the economy are unchanging. We could have alternatively rationalized persistent matches by allowing for endogenous separations alongside persistent idiosyncratic preferences $\ve_{it}$, however exogenous separations are more tractable.\footnote{See \citet[ Appendix D]{Grigsby2019} for details on this alternative approach.} An implication of the exogenous separations assumption is that a worker's match probabilities are independent of their job history, conditional on their type.\footnote{This rules out job ladders in which the identity of a worker's next job depends on the identity of their current job. We view this as a reasonable approximation for two reasons. First, our model is intended to analyze relatively short periods of time, over which workers skills are fixed and promotions up the career ladder are less frequent. Second, our aim is to identify groups of workers and jobs which are similar in the sense of being substitutable for each other. If one job lies directly above another on the career ladder, meaning that the higher job routinely hires workers from the lower job, then these jobs hire workers with similar skills, and therefore likely require similar tasks. If there was a large increase in employment at jobs on the higher level of the ladder, many of these workers would presumably be hired from jobs at the lower level of the ladder, implying that these workers can reasonably be assigned to the same type. This is effectively a question of whether or not to merge two similar worker types, and we answer it using MDL.  However, it would be possible to extend our model to allow for job ladders by modeling the temporal relationship between a worker's multiple job matches.} 

\section{Data} 
\label{sec:data}

We use the Brazilian linked employer-employee data set RAIS, which contains detailed data on all employment contracts in the Brazilian formal sector. Each observation in the data set represents a unique employment contract and includes a unique worker ID variable, an establishment ID, an occupation code, and earnings. Our sample includes all workers between the ages of 25 and 55 employed in the formal sector in the states of Sao Paulo, Minas Gerais, or Rio de Janeiro at least once between 2009 and 2012. These states are the 3 most populous and highest-GDP states in Brazil, are contiguous, and have a combined population of approximately 80 million; restricting to this set makes estimation computationally tractable.  We exclude public sector and military employment because institutional barriers make flows between the Brazilian public and private sectors rare. We also exclude the small number of jobs that do not pay workers on a monthly basis.

We define a job as an occupation--establishment pair and generate a unique ``Job ID'' for each job by concatenating the establishment ID code and the 4-digit occupation code.  Although we use occupation to define jobs, we do not use occupation as an input to our algorithm for classifying workers and jobs. This gives us a set of worker--job pairs that we use to cluster workers into worker types and jobs into markets. We restrict to jobs employing at least 5 unique workers during our estimation window, though the 5 workers need not be employed by the job simultaneously. This restriction eliminates jobs that are not sufficiently connected to the rest of the network of worker--job matches to infer their match probabilities and assign them to markets.

Once we have assigned workers to worker types and jobs to markets, we create a balanced panel of workers with one observation per worker per year. Our earnings variable is the real hourly log wage in December, defined as total December earnings divided by hours worked. We deflate earnings using the CPI. We exclude workers who were not employed for the entire month of December because we do not have accurate hours worked information for such workers. If a worker is employed in more than one job in December, we keep the job with greater hours. If the worker worked the same number of hours in both jobs, we pick the job with the greatest earnings. If tied on both, we choose randomly. Workers who are not matched with a job are defined as matching with the outside option, denoted $\g=0$, which includes non-employment and employment in the informal sector.

The RAIS data  cover only the formal sector of the Brazilian economy. We cannot distinguish between non-employment and informal employment. Therefore, our outside option includes both non-employment and informal sector employment.  In 2019, 32.1\% of employment in the Rio de Janeiro metropolitan area was in the informal sector.\footnote{IBGE - INSTITUTO BRASILEIRO DE GEOGRAFIA E ESTATÍSTICA. Indicadores de subutilização da força de trabalho e de informalidade no mercado de trabalho brasileiro. Rio de Janeiro: IBGE, 2019.} However, transitions between the formal and informal sectors are relatively rare: during our sample period, in a given year, fewer than 2\% of formal sector workers moved to the informal sector, and approximately 10\% of informal sector workers moved to the formal sector.\footnote{See \citet[ Figure 21]{EngbomGonzagaMoserOlivieri2021}.}

We calibrate demand shocks using annual data on real output per sector for the state of Rio de Janeiro from the Brazilian Institute of Geography and Statistics  (IBGE). These data are available for 15 sectors, the most disaggregated sector definitions for which annual state-level data are available. The 15 sectors are listed in Table \ref{table:IBGE_sectors}. 


\begin{table}
	\centering
	\caption{IBGE Sectors}
	\begin{tabular}{cl}
		\toprule
		& Sector name \\
		\midrule
		1	& Agriculture, livestock, forestry, fisheries and aquaculture  \\
		2	& Extractive industries  \\
		3	& Manufacturing industries  \\
		4	& Electricity and gas, water, sewage, waste mgmt and decontamination  \\
		5	& Construction  \\
		6	& Retail, Wholesale and Vehicle Repair  \\
		7	& Transport, storage and mail  \\
		8	& Accommodation and food  \\
		9	& Information and communication  \\
		10	& Financial, insurance and related services  \\
		11	& Real estate activities  \\
		12	& Professional, scientific and technical, admin and complementary svcs  \\
		13	& Public admin, defense, educ and health and soc security  \\
		14	& Private health and education  \\
		15	& Arts, culture, sports and recreation and other svcs \\
		\bottomrule
	\end{tabular}
	\label{table:IBGE_sectors}
\end{table}


\subsection{Summary statistics}

\begin{figure}
	\centering
	\caption{Distributions of Number of Matches Per Worker and Job}
	\begin{subfigure}{\textwidth}
		\centering
		\caption{Workers}
		\includegraphics[width=0.7\linewidth]{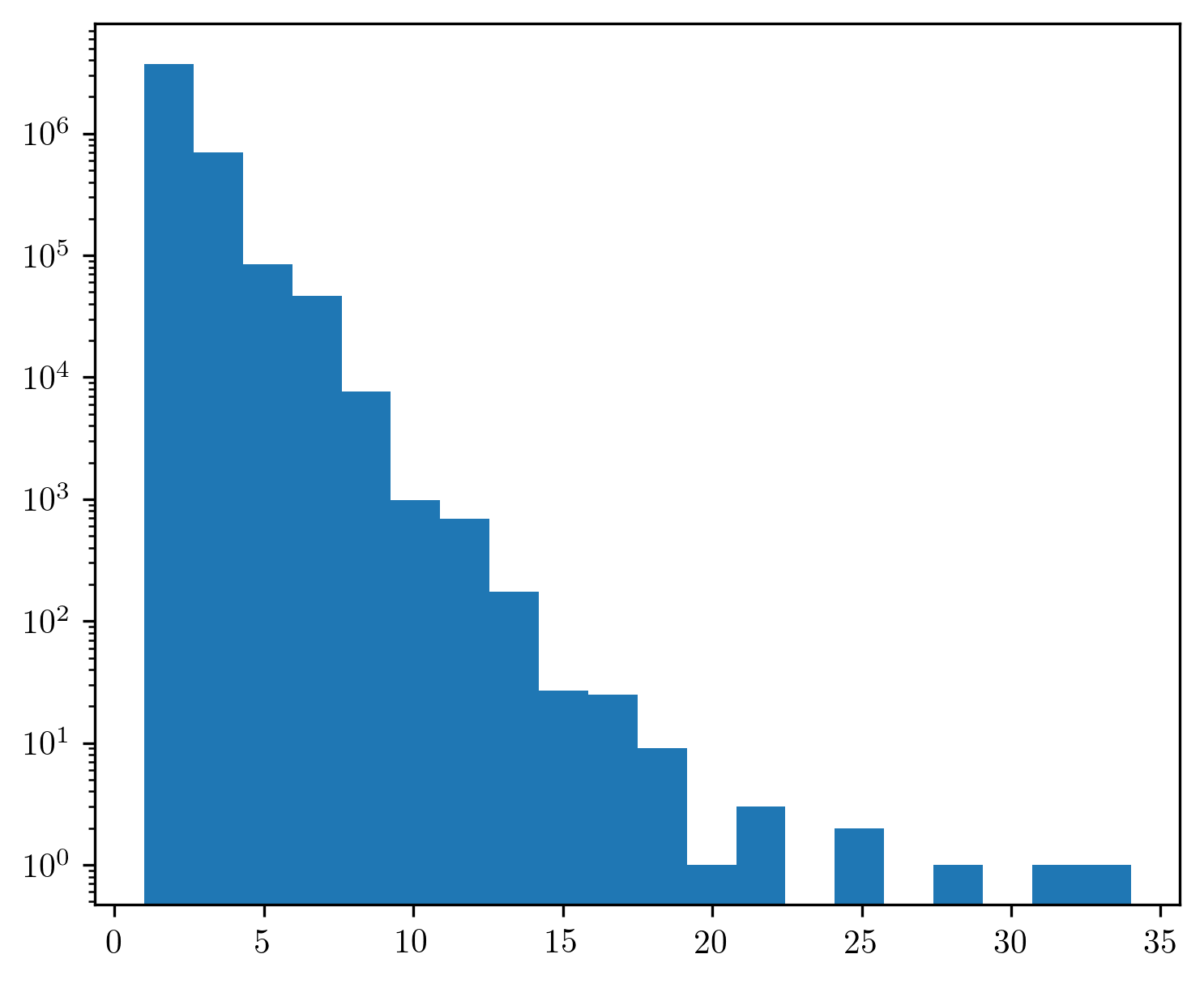}
		\label{fig:worker_degree_distribution_hist}
	\end{subfigure}
	\begin{subfigure}{\textwidth}
		\centering
		\caption{Jobs}
		\includegraphics[width=0.7\linewidth]{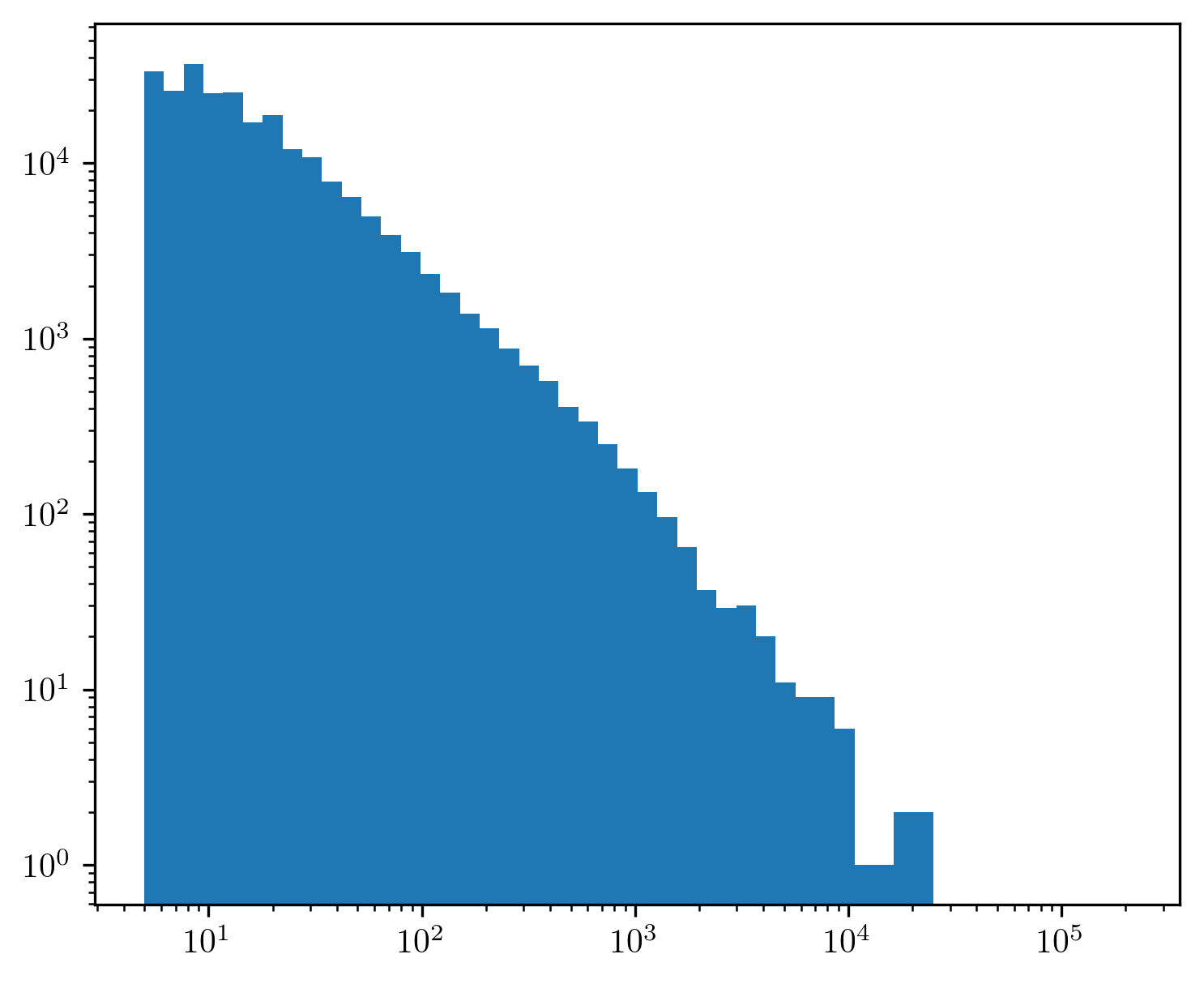}
		\label{fig:job_degree_distribution_hist}
	\end{subfigure}
	\label{fig:degree_distribution_hist}
	\footnotesize\flushleft \emph{Notes:} Figure presents histograms of the number of matches for workers and jobs, respectively. In network theory parlance, these are known as degree distributions. Vertical axes presented in log scale. Horizontal axis of bottom panel also presented in log scale. Number of matches per worker and job computed from the network of worker--job matches described in Section \ref{sec:data}.
\end{figure}

Our data contain 13,694,535 unique workers, 552,109 unique jobs, and 20,815,674 unique worker--job matches. The average worker matches with 1.73 jobs and the average job matches with 27.4 workers. 42\% of workers match with more than one job during our sample. Figure \ref{fig:degree_distribution_hist} presents histograms of the number of matches for workers and jobs, respectively. In network theory parlance, these are known as degree distributions.

\begin{figure}
	\centering
	\caption{Worker Type ($\i$) and Market ($\g$) Size Distributions}
	\begin{subfigure}{\textwidth}
		\centering
		\caption{Number of Workers Per Worker Type ($\i$)}
		\includegraphics[width=0.7\linewidth]{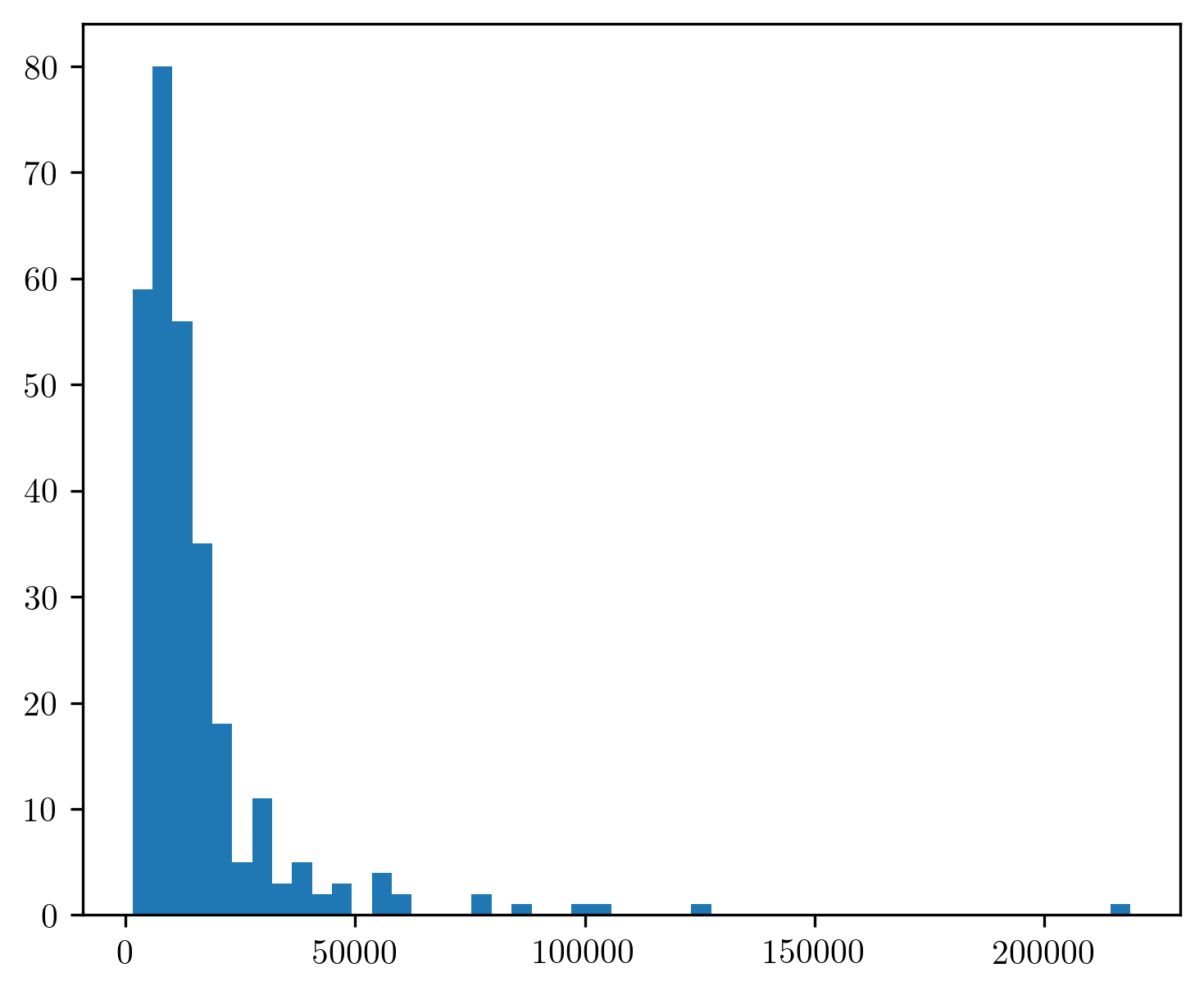}
		\label{fig:iota_size_distribution}
	\end{subfigure}
	\begin{subfigure}{\textwidth}
		\centering
		\caption{Number of Jobs Per Market ($\g$)}
		\includegraphics[width=0.7\linewidth]{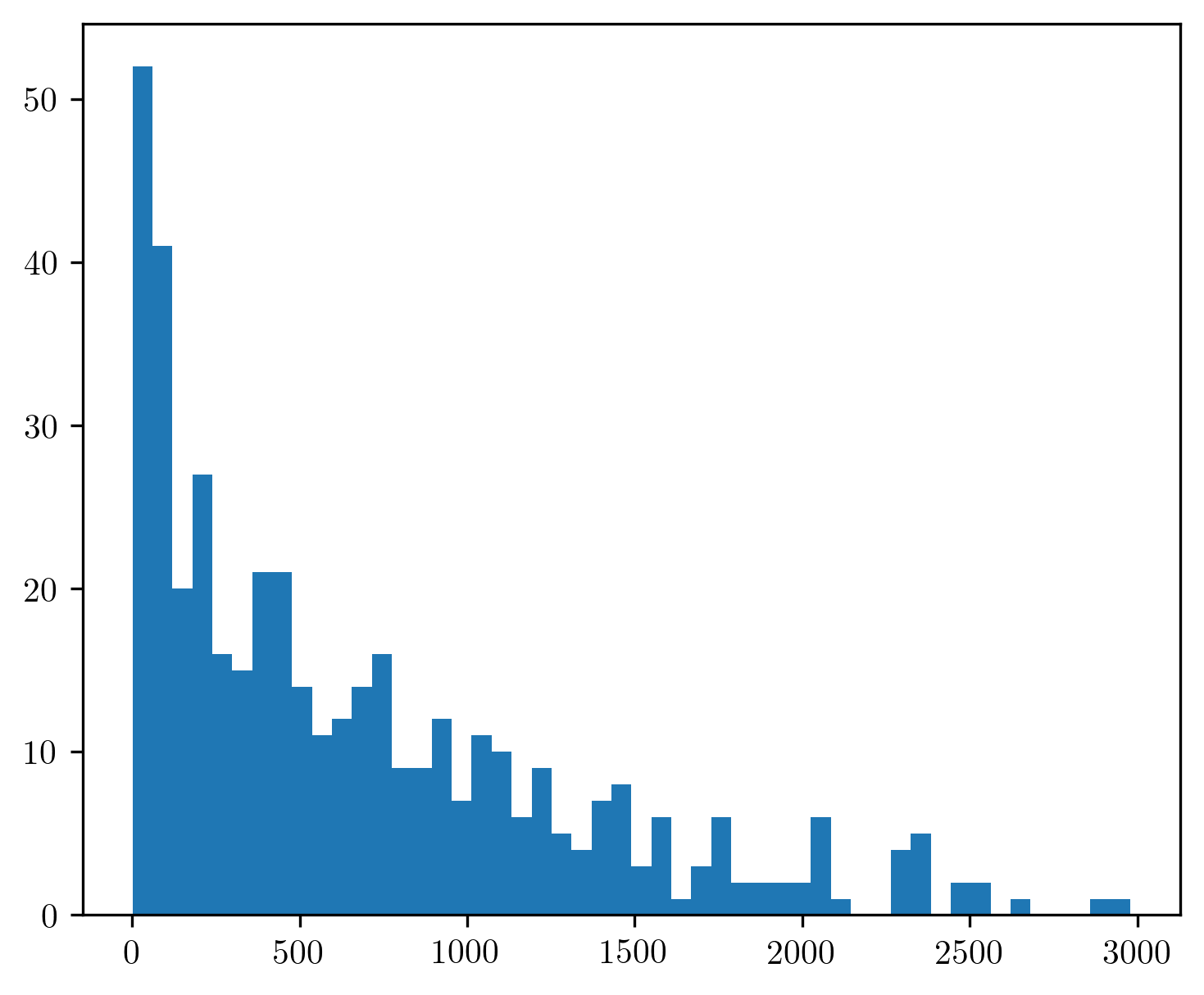}
		\label{fig:gamma_size_distribution}
	\end{subfigure}
	\label{fig:iota_gamma_size_distribution}
	\footnotesize\flushleft \emph{Notes:} Figure presents histograms of the number of workers per worker type $\i$ and jobs per market ($\g$). The units of analysis are worker types in the upper panel and markets in the lower panel. Computed using assignments of workers to worker types and jobs to markets as described in Section \ref{sec:bisbm}.
\end{figure}

Our network-based classification algorithm identifies 446 worker types ($\i$) and 1,371 markets ($\g$). Figure \ref{fig:iota_gamma_size_distribution} presents histograms of the number of workers per worker type and jobs per market. The average worker belongs to a worker type with 40,978 unique workers and the median worker belongs to a worker type with 20,413 unique workers. The average job belongs to a market with 1,273 unique jobs and the median job belongs to a market with 1,188 unique jobs. 

Table \ref{table:transitions_table} presents the fraction of job changes that also involve a change in occupation, sector, market, firm, or establishment. The column ``All Job Changes'' computes the probability that a worker changes occupation, industry, sector, market, firm, or establishment conditional on changing jobs. The column ``Firm Change Only'' presents the same quantities restricting to the set of job changes that also involve a change in firm. The column ``No Firm Change'' restricts to job changes that do \emph{not} involve a change in firm. Recall that we define a job as a 4-digit occupation--establishment pair. Table \ref{table:transitions_table} shows that 65\% of job changes also involve a change in establishment and 54\% change firm. This tells us that job changes are not dominated by workers ``climbing the job ladder'' by changing occupations within a firm. 

Table \ref{table:transitions_table} also shows that job changes are frequently associated with occupation, industry, and sector changes. 41\% of job changes involve a change in 1-digit occupation (most aggregated) and 73\% involve a change in 6-digit occupation (most disaggregated). Since occupation, industry, and sector changes are so frequent, it is unlikely that any of these variables precisely measure workers' skills, since workers' skills are unlikely to evolve so quickly. Similarly, the fact that job transitions frequently (59\% of the time) involve moving to a job in a different market ($\g$) demonstrates the value of allowing workers to costlessly change the market to which they supply labor, a feature that our model incorporates.

\begin{table}[h!] \centering
	\caption{Occupation/Sector/Market Transition Frequencies}
	\begin{tabular}{lrrr}
\toprule
           Variable &  All Job Changes &  Firm Change Only &  No Firm Change \\
\midrule
 1-digit Occupation &     0.410 &                   0.345 &                      0.484 \\
 2-digit Occupation &     0.496 &                   0.422 &                      0.580 \\
 4-digit Occupation &     0.676 &                   0.563 &                      0.807 \\
 6-digit Occupation &     0.725 &                   0.648 &                      0.814 \\
   5-digit Industry &     0.418 &                   0.708 &                      0.083 \\
      Sector (IBGE) &     0.262 &                   0.456 &                      0.039 \\
      Market ($\g$) &     0.591 &                   0.727 &                      0.434 \\
               Firm &     0.536 &                   1.000 &                      0.000 \\
      Establishment &     0.645 &                   0.996 &                      0.240 \\
\bottomrule
\end{tabular}  
	\footnotesize\flushleft \emph{Notes:} This table presents the fraction of job changes that also involve a change in occupation, sector, market, firm, or establishment. The column ``All Job Changes'' computes the probability that a worker changes occupation, industry, sector, market, firm, or establishment conditional on changing jobs. The column ``Firm Change Only'' presents the same quantities restricting to the set of job changes that also involve a change in firm. The column ``No Firm Change'' restricts to job changes that do \emph{not} involve a change in firm. Since the fraction of job changes that involve a firm change is 0.536, values in the column ``All Job Changes'' equal 0.536 $\times$ ``Firm Change Only'' + (1-0.536) $\times$ ``No Firm Change.'' 5-digit sectors refer to narrow industry codes, while there are 15 IBGE sectors, defined in Table \ref{table:IBGE_sectors}, taken from the Brazilian Institute of Geography and Statistics (IBGE). Values computed using the worker earnings panel described in Section \ref{sec:data} using RAIS data from 2009--2012. 
	\label{table:transitions_table}
\end{table}

\section{Validation of our worker types and markets}
\label{sec:descriptive_results}

An ideal labor market definition is one that maximizes the similarity of workers and jobs within the same groups, respectively, and minimizes the similarity of workers and jobs in different groups. More formally, the ideal definition maximizes the within-market substitution elasticity and minimizes the cross-market substitution elasticity.\footnote{This is most easily understood if we extended our model in Section \ref{sec:model} to have a nested logit structure. Then, in addition to the parameter $\nu$ which governs the variance of idiosyncratic preference shocks \emph{across} markets, we would have an additional parameter $\eta$ that governs the correlation of idiosyncratic preference shocks \emph{within} markets. To the extent that we incorrectly classify jobs incorrectly, $\nu$ and $\eta$ will be biased towards each other. To understand this, consider the extreme case in which jobs are classified at random. Then the market classifications carry no economic information and, in expectation, two jobs in the same market will be equally substitutable from the worker's perspective as two jobs in different markets, meaning that $\nu=\eta$. Alternatively, if we have perfectly classified the most similar jobs into the same markets, then jobs in the same market will be viewed as close substitutes (small $\eta$) and jobs in different markets will be more distant substitutes (large $\nu$). We expand this argument and apply our market definitions to the question of measuring labor market power in ongoing work.} Directly estimating these elasticities is conceptually and computationally difficult, and is beyond the scope of this paper. Therefore, in this section we provide indirect evidence that our network-based classifications outperform standard classifications in identifying similar workers and similar jobs. 

We provide five such pieces of evidence. First, we provide descriptive examples of the occupation and geographic distributions of jobs within our markets and demonstrate that they capture reasonable but potentially difficult to observe, forms of structure in the labor market that traditional market definitions miss. Second, we show that workers' labor supply is more concentrated  within our markets than traditional definitions and jobs' hiring is more concentrated within our worker types. Third, we perform an out-of-sample prediction exercise in which our worker types and markets outperform traditional definitions in predicting workers' job-to-job flows. Fourth, we show that our worker types do a better job of identifying groups of workers with distinct skills as measured by the worker productivity parameter, $\psi$. Finally, we use the full general equilibrium extension of our model and show that our classifications more accurately predict the wage effects of the labor demand shock created by the 2016 Rio de Janeiro Olympics.


\subsection{Occupation count tables}
\label{sec:occ_count_tables}

Our method simultaneously clusters together workers in different occupations who are revealed by the network structure of the labor market to have similar skills, \emph{and} disaggregates workers employed in the same occupation who are revealed to have different skills. As a concrete example, consider the occupation identified by the code 3331-10 in the Brazilian occupation classification system. This occupation is called\footnote{Occupation names and descriptions are translated from Portuguese using Google Translate.} ``course instructor'' and is described as 
\begin{quote}
	\textbf{Summary description}\\
	The professionals in this occupational family must be able to create and plan courses, develop programs for companies and clients, define teaching materials, teach classes, evaluate students and suggest structural changes in courses.
\end{quote}
Despite this being the most disaggregated level of the occupation classification system (6-digit), there may be considerable heterogeneity within this occupation. This occupation may include, for example, both math tutors and personal fitness trainers --- two sets of workers with very different skills. At the same time, it is not obvious what distinguishes a  course instructor from a personal trainer (occupation code 2241-20) or an elementary school teacher (occupation code 2312-10). However, if we can identify a cluster of course instructors who at other times in their career work as personal trainers and another cluster who have also worked as elementary school teachers, then we can simultaneously \emph{disaggregate} course instructors with distinct skills, and  \emph{aggregate} different subsets of course instructors with other workers in different occupations who have similar skills.  We pursue these examples in Tables \ref{occ_counts_iota360} and \ref{occ_counts_iota557}. 

To illustrate this point more clearly, We focus on a specific worker type, $\i=17$, in which many workers are employed as course instructors. Table \ref{occ_counts_iota360} presents the 10 occupations in which workers belonging to worker type $\i=17$ are most frequently employed.\footnote{To interpret this table, recall that we have assigned each individual worker to a worker type, $\i$. Each worker may be employed by one or more jobs in our sample, and each job is assigned an occupation code by the Brazilian statistical agency. A worker who has multiple jobs during the sample may have a different occupation associated with each job.}  The most frequently occurring- occupation is course instructr, however most of these occupations are related to physical fitness, education, or both. The fact that these course instructors tend to match with similar jobs as other workers whom we observe employed as personal trainers, physical education teachers, and sports coaches allows us to infer that these course instructors have skills more closely related to physical education than math. 

Table \ref{occ_counts_iota557} presents the distribution of occupations with a different worker type that contains many course instructors, $\i=52$. Unlike the previous example, where the workers appear to have physical education skills, the other frequently-occurring occupations in this worker type are teachers of more traditional academic subjects. If we had relied upon occupation codes alone, we would have assumed that all course instructors have the same skills, whereas our clustering approach tells us that there are at least two different types of course instructors: physical education and academic education.

In addition to disaggregating workers in the same occupation with different skills, these tables display our  in aggregating workers in different occupations with similar skills. For most of the occupations in these tables, it makes intuitive sense that they should be clustered together. For example, it is not surprising that physical education teachers, sports coaches, and personal trainers would have similar skills.  Relying on occupation codes --- even the highly-aggregated two-digit occupation codes --- would not have grouped these workers together. More generally, we view the fact that our worker types imperfectly align with occupation codes as suggestive evidence of our success in identifying groups of workers with similar skills. Workers with similar skills are likely to be employed in similar occupations, so it would be concerning if our worker types did not overlap with occupations. However, the fact that they only partially overlap with occupations suggests that they capture important dimensions of worker heterogeneity that occupations miss. We develop this argument further in the rest of the paper.

\begin{table}[tbp] \centering
\newcolumntype{C}{>{\centering\arraybackslash}X}

\caption{Top Ten Occupations for Worker Type $\i=17$}
\label{occ_counts_iota360}
\begin{tabularx}{\textwidth}{llr}

\toprule
{Occ-6}&{Occupation Name}&{Share} \tabularnewline
\midrule\addlinespace[1.5ex]
333110&Course Instructor&.15 \tabularnewline
224120&Personal trainer&.11 \tabularnewline
231315&Physical Education Teacher in Primary School&.08 \tabularnewline
224125&Coach (except for soccer)&.06 \tabularnewline
234410&Physical Education Teacher in Higher Education \hspace{2.12cm}&.05 \tabularnewline
224105&Fitness monitor&.05 \tabularnewline
333115&Teacher (with High School degree)&.05 \tabularnewline
234520&Education Teacher (with College degree) & .03 \tabularnewline
371410&Recreational Activities Coordinator&.03 \tabularnewline
377105&Professional Athlete (various modalities)&.02 \tabularnewline
\bottomrule \addlinespace[1.5ex]
\end{tabularx}
\footnotesize\flushleft \emph{Notes:} Table reports the 6-digit occupations in which workers assigned to worker type $\i=17$ are most frequently observed, showing only the 10 most frequent. Values computed using the worker earnings panel described in Section \ref{sec:data} using RAIS data from 2009--2012. Occupation classification codes defined according to the Brazilian occupation classification system, \emph{CBO 2002: Classificacao Brasileira de Ocupacoes} and translated from Portuguese to English using Google Translate. 
\end{table}

\begin{table}[tbp] \centering
\newcolumntype{C}{>{\centering\arraybackslash}X}

\caption{Top Ten Occupations for Worker Type $\i=52$}
\label{occ_counts_iota557}
\begin{tabularx}{\textwidth}{llr}

\toprule
{Occ-6}&{Occupation Name}&{Share} \tabularnewline
\midrule\addlinespace[1.5ex]
331205&Elementary School Teacher &.07 \tabularnewline
333110&Course Instructor&.07 \tabularnewline
231210&Elementary School Teacher (1st to 4th grade)&.06 \tabularnewline
231205&Young and Adult Teacher teaching elementary school content \hspace{.3cm} &.06 \tabularnewline
232115&High School Teacher&.05 \tabularnewline
234616&English Teacher&.04 \tabularnewline
333115&Teacher of Free Courses&.03 \tabularnewline
231305&Elementary School Science and Math Teacher&.03 \tabularnewline
331105&Kindergarten Teacher &.02 \tabularnewline
231310&Art Teacher in Elementary School&.02 \tabularnewline
\bottomrule \addlinespace[1.5ex]
\end{tabularx}
\footnotesize\flushleft \emph{Notes:} Table reports the 6-digit occupations in which workers assigned to worker type $\i=52$ are most frequently observed, showing only the 10 most frequent. Values computed using the worker earnings panel described in Section \ref{sec:data} using RAIS data from 2009--2012. Occupation classification codes defined according to the Brazilian occupation classification system, \emph{CBO 2002: Classificacao Brasileira de Ocupacoes} and translated from Portuguese to English using Google Translate.  
\end{table}


\subsection{Worker types' labor market concentration}

\label{sec:hhi}

If our worker types and markets are successful in identifying groups of workers and jobs that are viewed as similar from the perspective of the labor market, then each worker type's labor supply will be concentrated within specific markets and each job’s hiring will be concentrated among specific worker types. While there will be considerable variation across worker types --- worker types with more specific skills will be more concentrated in a small set of markets than those with more general skills --- if we compare two classification schemes applied to the same underlying set of worker--job matches, the one that does a better job of identifying workers with similar skills and jobs requiring similar tasks will yield more concentrated labor supply and hiring distributions.

We compute each worker type's employment concentration across markets and occ2Xmesos using the Herfindahl-Hirschman index (HHI): 
\[ HHI_{\i}^{Occ2Xmeso} = \sum_s \pi_{\i s}^2 \quad \text{and} \quad  HHI_{\i}^{Market} = \sum_\g \pi_{\i \g}^2 \]
where $s$ indexes Occ2Xmesos, $\g$ indexes markets, and $\pi_{\i s}$ and $\pi_{\i \g}$ are the share of type $\i$ workers employed in occ2Xmeso $s$ and market $\g$, respectively. An HHI close to 0 indicates that type $\i$ employment is spread approximately evenly across sectors/markets, while an HHI close to 1 indicates that type $\i$ employment is very concentrated in a single occ2Xmeso/market. Suppose we classified jobs randomly. Then worker types would not have a comparative advantage in specific markets and therefore would not be concentrated in specific markets. In this case, the HHI for each worker type would converge to $1/\Gamma$, where $\Gamma$ is the total number of markets, indicating a uniform distribution of employment across job classifications. At the other extreme, if each worker type had perfectly specific skills and supplied all of its labor to exactly one market, the HHI would be 1. While we would not expect perfectly specific skills,  larger HHIs are evidence that we have done a better job of classifying similar jobs, whereas smaller HHIs imply that we are closer to simply classifying jobs randomly.

\begin{figure}
	\centering
	\caption{Employment/Hiring Concentration According to Different Worker and Job Classifications}
	\begin{subfigure}{\textwidth}
		\centering
		\caption{Concentration of Worker Types' ($\i$) Employment Within Markets/Occ2Xmesos}
		\includegraphics[width=0.7\linewidth]{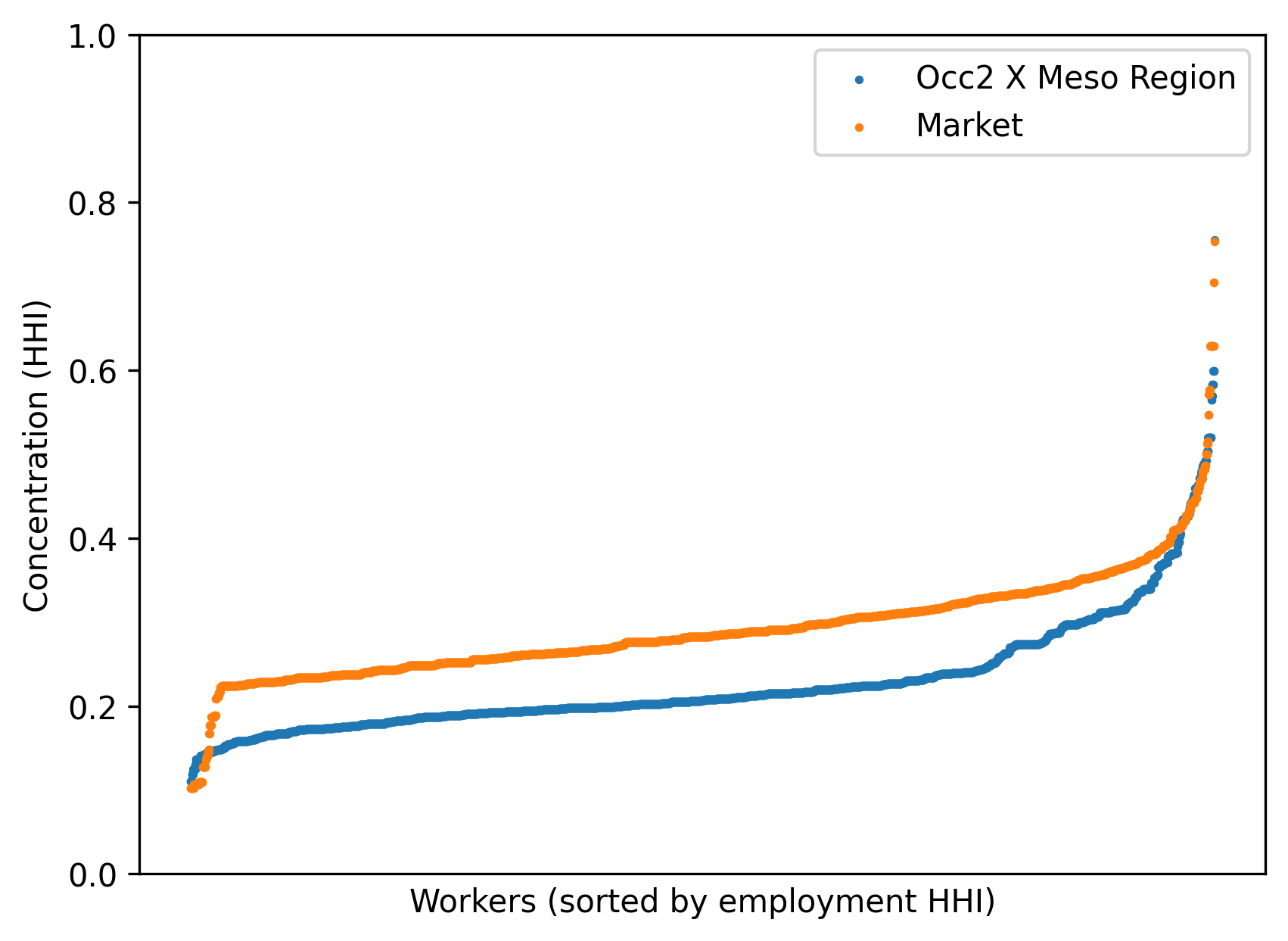}
		\label{fig:concentration_figures__iota__occ2Xmeso__gamma}
	\end{subfigure}
	\begin{subfigure}{\textwidth}
		\centering
		\caption{Concentration of Markets' ($\g$) Hiring Within Worker Types' ($\i$)/Occ2Xmesos}
		\includegraphics[width=0.7\linewidth]{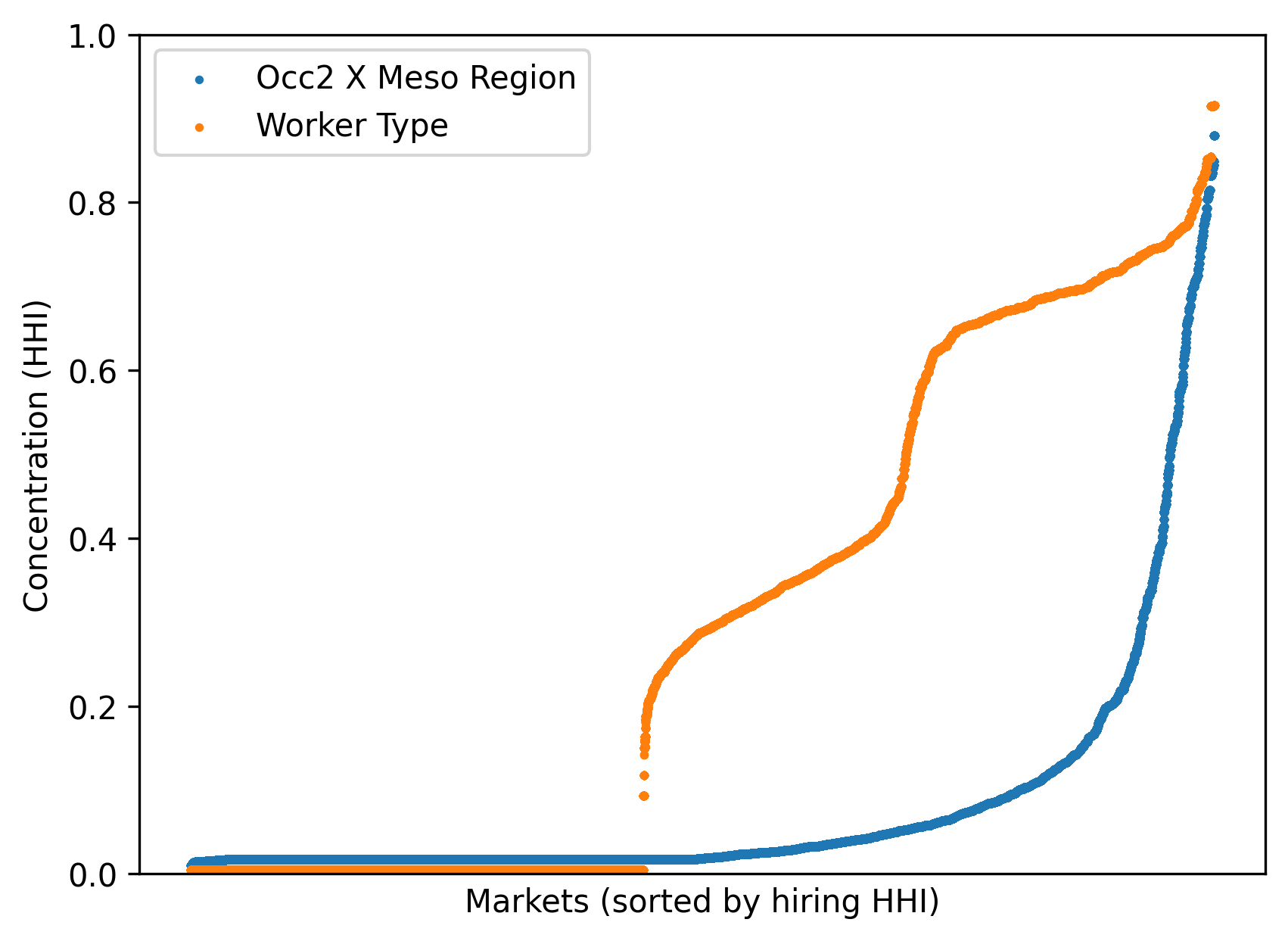}
		\label{fig:concentration_figures__gamma__occ2Xmeso_first__iota}
	\end{subfigure}
	\label{fig:concentration_figures_iota_gamma_hhi}
	\footnotesize\flushleft \emph{Notes:} Figure presents concentration, defined as a Herfindahl-Hirschman Index (HHI), of worker types' employment within individual markets (orange lines) and sectors (blue line). The figure is weighted by the number of workers in each worker type. Workers are sorted from lowest to highest HHI along the horizontal axis. HHIs computed from the 2009-2012 RAIS worker earnings panel described in Section \ref{sec:data}.
\end{figure}

Figure \ref{fig:concentration_figures__iota__occ2Xmeso__gamma} presents $ HHI_{\i}^{Occ2Xmeso}$ and $ HHI_{\i}^{Market}$ for each worker type, sorted from least concentrated to most concentrated. Most worker types' labor supply is more concentrated among markets than among Occ2Xmeso, which according to the argument above, indicates that markets identify groups of jobs that have more homogenous tasks than do sectors. Figure \ref{fig:concentration_figures__gamma__occ2Xmeso_first__iota} repeats this analysis, instead focusing the hiring distribution of markets, classifying workers alternately by their worker type and by the first Occ2Xmeso in which they are ever ever observed. The story is analogous, with markets' hiring significantly more concentrated within worker types than within workers' first Occ2Xmeso. 

\subsection{Predicting out-of-sample job-to-job flows}

Another test of a market definition's success at identifying similar jobs is its ability to successfully predict job-to-job flows. As noted in Section \ref{sec:model}, we assume that every time a worker changes jobs, they draw a new job from the same distribution. This implies that workers are more likely to transition to new jobs that have similar task distributions to their old jobs.  Therefore, a market definition that does a better job of predicting job-to-job flows will be one that better captures latent job similarity. In this subsection we predict out-of-sample job-to-job flows using both our network-based market definitions and traditional definitions and show that our definitions outperform traditional ones.

Consider a market definition $\mathbf{M}$ and the matrix of out-of-sample job-to-job empirical transition probabilities $\mathbb{P}_{oos}[j'|j]$. The ability to predict future job-to-job transitions from the structure of $\mathbf{M}$ can be assessed as follows:

\begin{enumerate}
	\item Compute a matrix of empirical market-to-market transition probabilities denoted by $\mathbb{P}_{\mathbf{M}}[m'|m]$
	\item For each job $j$ in market $m$, compute the probability of transitioning to $j'$ in market $m'$, $Prob(j' \in m'|j \in m)$, as follows:
	\[Prob(j' \in m'|j \in m) := \mathbb{P}_{\mathbf{M}}[m'|m] \frac{d_{j'}}{\sum_{k\in m'} d_{k}} \]
	Intuitively, the probability that a worker in job $j$ transitions to job $j'$ is the product of the market-level transition probabilities $Prob(j' \in m'|j \in m)$ and the probability that job $j'$ is chosen conditional on market $m'$ being chosen. The latter probability is simply job $j'$'s share of employment in market $m'$, $ \frac{d_{j'}}{\sum_{k\in m'} d_{k}}$.
	\item Stack the probabilities computed in the previous step for each job, resulting in a predicted transitions matrix $\mathbb{\hat P}_{\mathbf{M}}[j'|j]$ with identical dimensions to the matrix $\mathbb{P}_{oos}[j'|j]$.
	\item Compute a measure of fit, using a matrix norm of preference:
	\[ \Omega_{\mathbf{M}} := \begin{Vmatrix}
		\mathbb{\hat P}_{\mathbf{M}}[j'|j] - \mathbb{P}_{oos}[j'|j]
	\end{Vmatrix} \]
\end{enumerate}

We perform this exercise using empirical transitions from our estimation period of 2009--2011 and compute out-of-sample transitions using 2012 and 2013. We compare our network-based market definitions to a traditional alternative defined as the intersection of two-digit occupations and meso regions. We obtain lower prediction error using our network-based market definitions according to both the $L_1$ and $L_2$ norms. Specifically, the average $L_1$ prediction errors using markets and occ2Xmesos are 15.5 and 15.6, respectively. For the  $L_2$ norm the corresponding values are 237.9 and 317.7. The greater discrepancy for the  $L_2$ norm reflects the fact that predictions based on occ2Xmesos are much more likely to generate very large errors.

\subsection{Worker type skill correlations}

\label{sec:correlograms}

While Section \ref{sec:occ_count_tables} provided a qualitative example of our method's success in identifying clusters of workers with similar skills, we now provide quantitative evidence of our success in this regard. An ideal worker skills classification scheme will maximize the variance in skills across different worker classifications and minimize the variance of skills within a worker classification. While we do not directly observe individual-level skills and therefore cannot measure within-classification skills variance, we do have a measure of across-classification skills variation. Each element of $\Psi$ represents the productivity of a type $\i$ worker employed in market $\g$. Therefore, $\psi_{\ig}$ is a summary measure of a type $\i$ worker's skill at jobs in market $\g$, and a full row vector of $\Psi$, $\psi_{\i\cdot}$, summarizes a type $\i$ worker's skills in \emph{all} markets. This yields a natural metric for skill similarity across worker types: two worker types, $\i$ and $\i'$, have similar skills if their associated productivity vectors $\psi_{\i\cdot}$ and $\psi_{\i'\cdot}$ are highly correlated.

We estimate $\Psi$ using maximum likelihood. Identification comes from two sources: earnings for all employed workers, and market choices for all workers in period $t=1$ and workers who receive exogenous separation shocks in periods $t>1$. Intuitively, ($\i,\g$) matches that pay more and occur more frequently are revealed to be more productive. For details, see Appendix \ref{sec:MLE}.

If we have done a good job of clustering workers with similar skills into the same type, then the correlations of skills across different worker types will be low. To understand this, consider an extreme example in which workers were clustered randomly. In this case, all clusters would, in expectation, have exactly the same skills --- because the skills of each cluster would just be the average skills of the entire population --- and all pairs of productivity vectors would be approximately perfectly correlated. That is, $corr(\psi_{\i\cdot},\psi_{\i'\cdot}) \approx 1$ for all $\i,\i'$.  Alternatively, we might have two clusters of worker types --- for example those intensive in manual skills and those intensive in cognitive skills --- such that worker types in the same cluster have highly-correlated skills and those in different clusters have  negatively correlated skills. At the other extreme, if skills were perfectly specific (meaning that $\Psi$ was close to a diagonal matrix), skill correlations would be close to zero. For this reason, we argue that if the pairwise correlations between different worker types’ skills vectors tend to be close to zero, this is an indication that we have successfully identified groups of workers with distinct skills. 


We summarize the correlations between different worker types' productivity vectors in Figure \ref{fig:correlograms_sorted}. We do this in two ways. In the left column we present correlation coefficients between all pairs of the $I=446$ worker types in a lower triangular $446\times 446$ matrix (the upper triangular portion is redundant and therefore omitted). Dark red points represent large positive correlations, dark blue points represent large negative correlations, and lighter colors represent smaller correlations. Worker types are sorted by mean earnings, from smallest to largest. In the right column, we present histograms of the correlation coefficients in the left column, along with the standard deviation of the correlation coefficients. The first row presents correlations in which workers are classified by worker type and jobs by market. We provide context for these figures by repeating this exercise using versions of $\hat \Psi$ in which workers and jobs are classified using the standard labels in the data: occupation and sector. To do this, we estimate a different version of $\Psi$ using the same maximum likelihood estimation described above and detailed in Appendix \ref{sec:MLE}, except we classify workers and jobs by occupation and sector, rather than worker type and market. Row 2 of Figure \ref{fig:correlograms_sorted} shows workers classified by 4-digit occupation and jobs by sector. Row 3 shows workers classified by four-digit occupation and jobs by market ($\g$). We choose 4-digit occupations as our primary ``status quo'' benchmark to compare our method to because occupations are a frequently-used measure of granular worker heterogeneity and because the number of 4-digit occupations in our data (306) is similar to the number of worker types (446), allowing for comparisons at a similar level of granularity.

Figure \ref{fig:correlograms_sorted} shows that correlations between different worker types' productivity vectors are smaller in magnitude when we use our model's ($\i,\g$) classifications rather than classifications based on labels available in the data, occupation and sector.  This is because the network-based clusters of workers are more successful at segregating workers with distinct skills than are standard occupations. Connecting this to the example in the previous section, if high school and middle school math teachers have similar skills but are classified as distinct worker types, we would observe large correlations (dark red) between their productivity vectors. By contrast, our worker types disentangle teachers into physical education teachers --- including coaches and personal trainers --- and teachers in traditional academic subjects. Physical education and academic teachers have less correlated skills than do elementary and middle school teachers.  Because we have done a better job of segregating workers with disparate skills, and aggregating workers with similar skills, we observe fewer clusters of highly-correlated worker types.

\def\w{.45}
\begin{figure}[!htbp]
	\centering
	\caption{Skill Correlation Across Worker Types and Occupations}
	\begin{subfigure}[b]{\w\textwidth}
		\centering
		\caption{($\i,\g$) correlogram}
		\includegraphics[height=.25\textheight]{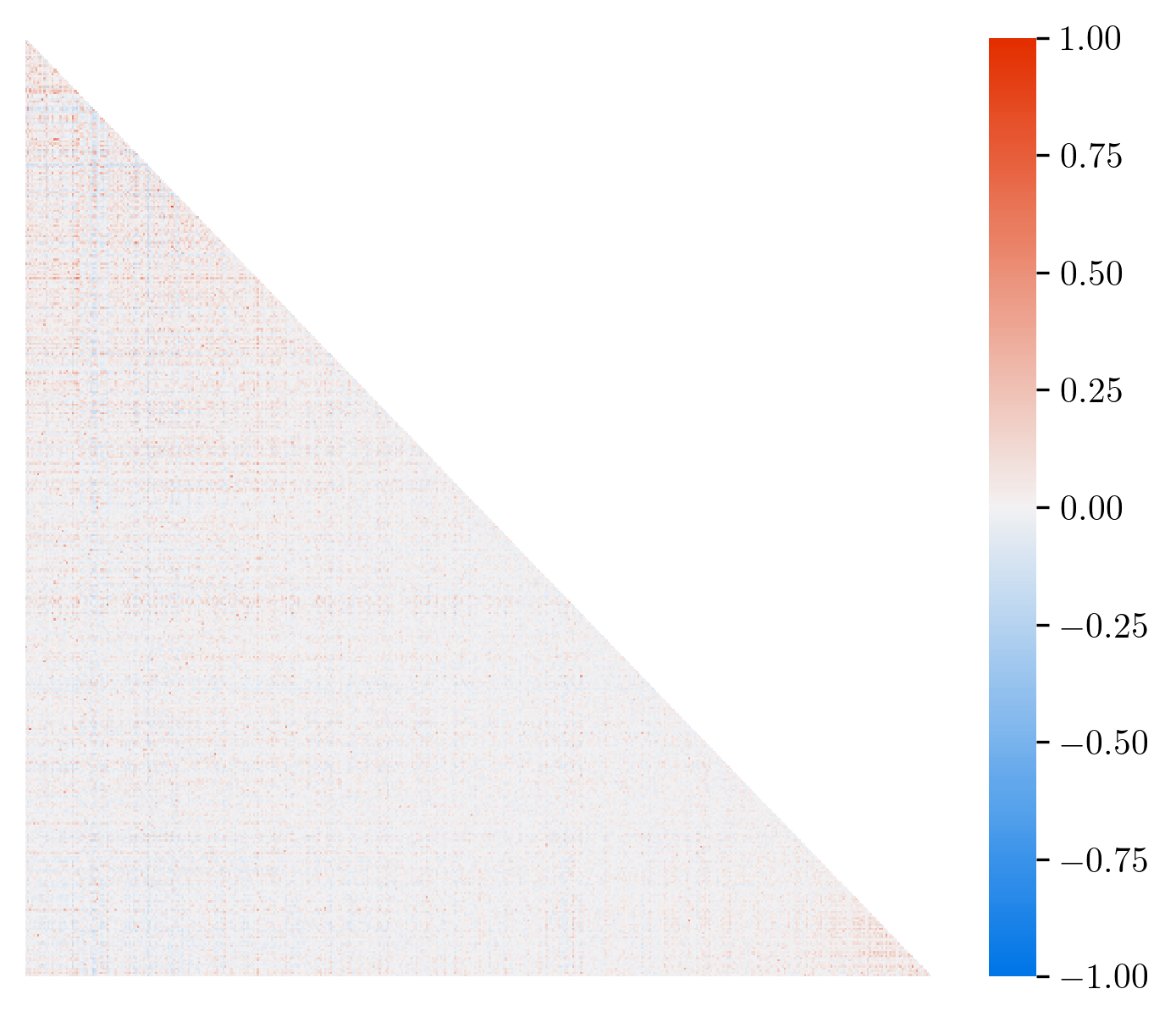}
		\label{fig:correlograms_iota_gamma_sorted}
	\end{subfigure}\hfill
	\begin{subfigure}[b]{\w\textwidth}
		\centering
		\caption{($\i,\g$) histogram}
		\includegraphics[height=.25\textheight]{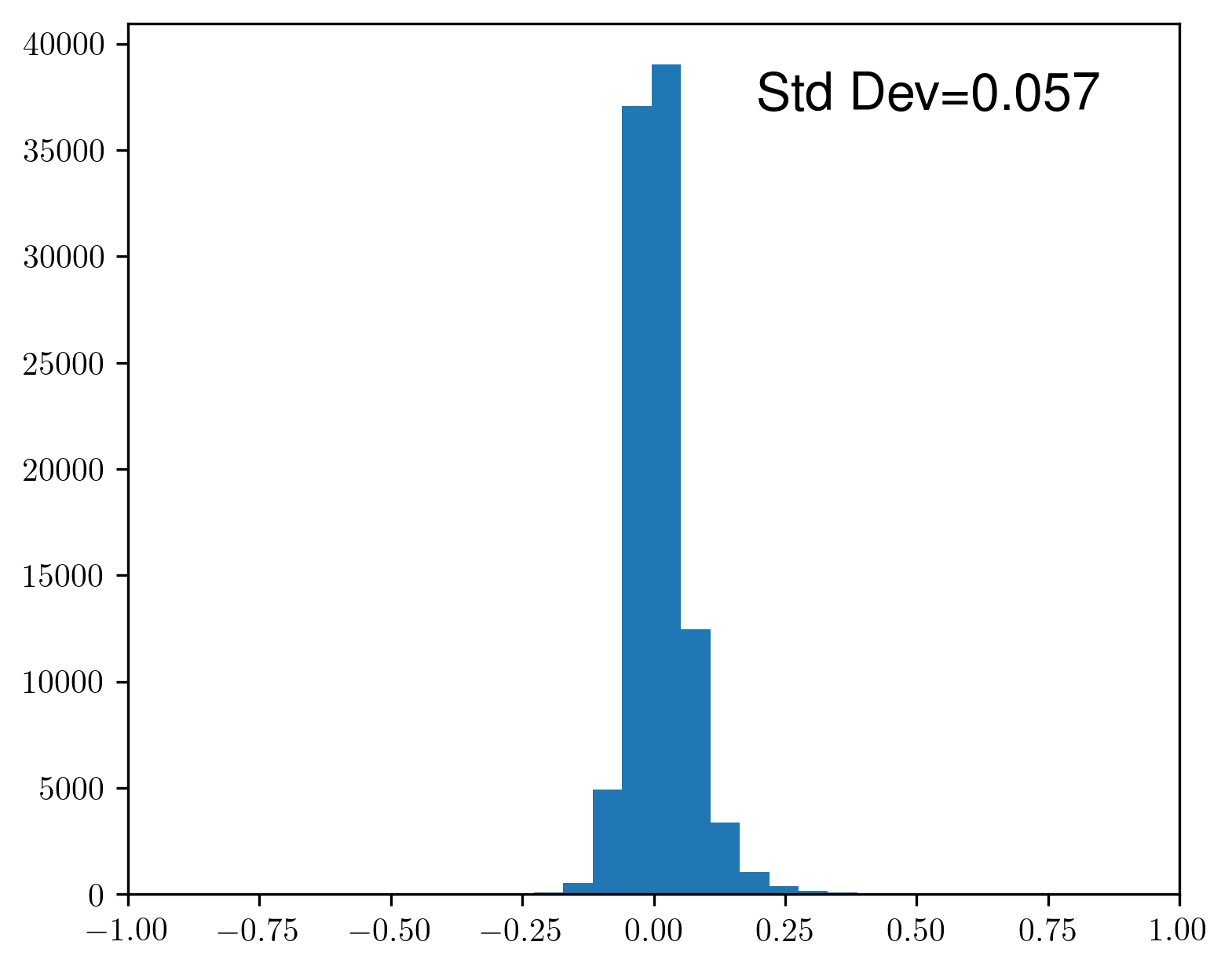}
		\label{fig:correlograms_hist_iota_gamma_sorted}
	\end{subfigure}\vfill
	\begin{subfigure}[b]{\w\textwidth}
		\centering
		\caption{(Occ4, Sector) correlogram}
		\includegraphics[height=.25\textheight]{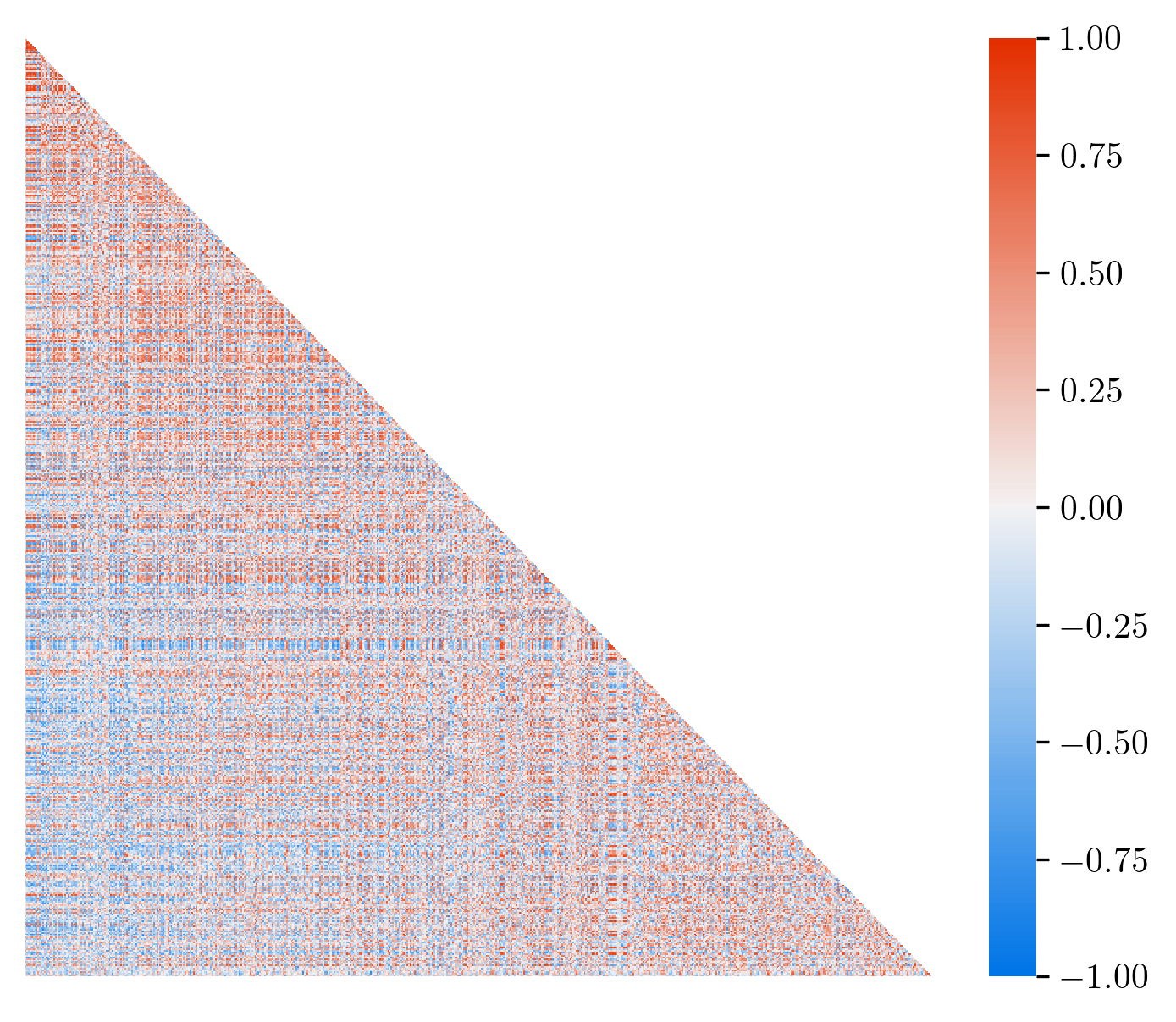}
		\label{fig:correlograms_occ4_first_recode_sector_IBGE_sorted}
	\end{subfigure}\hfill
	\begin{subfigure}[b]{\w\textwidth}
		\centering
		\caption{(Occ4, Sector) histogram}
		\includegraphics[height=.25\textheight]{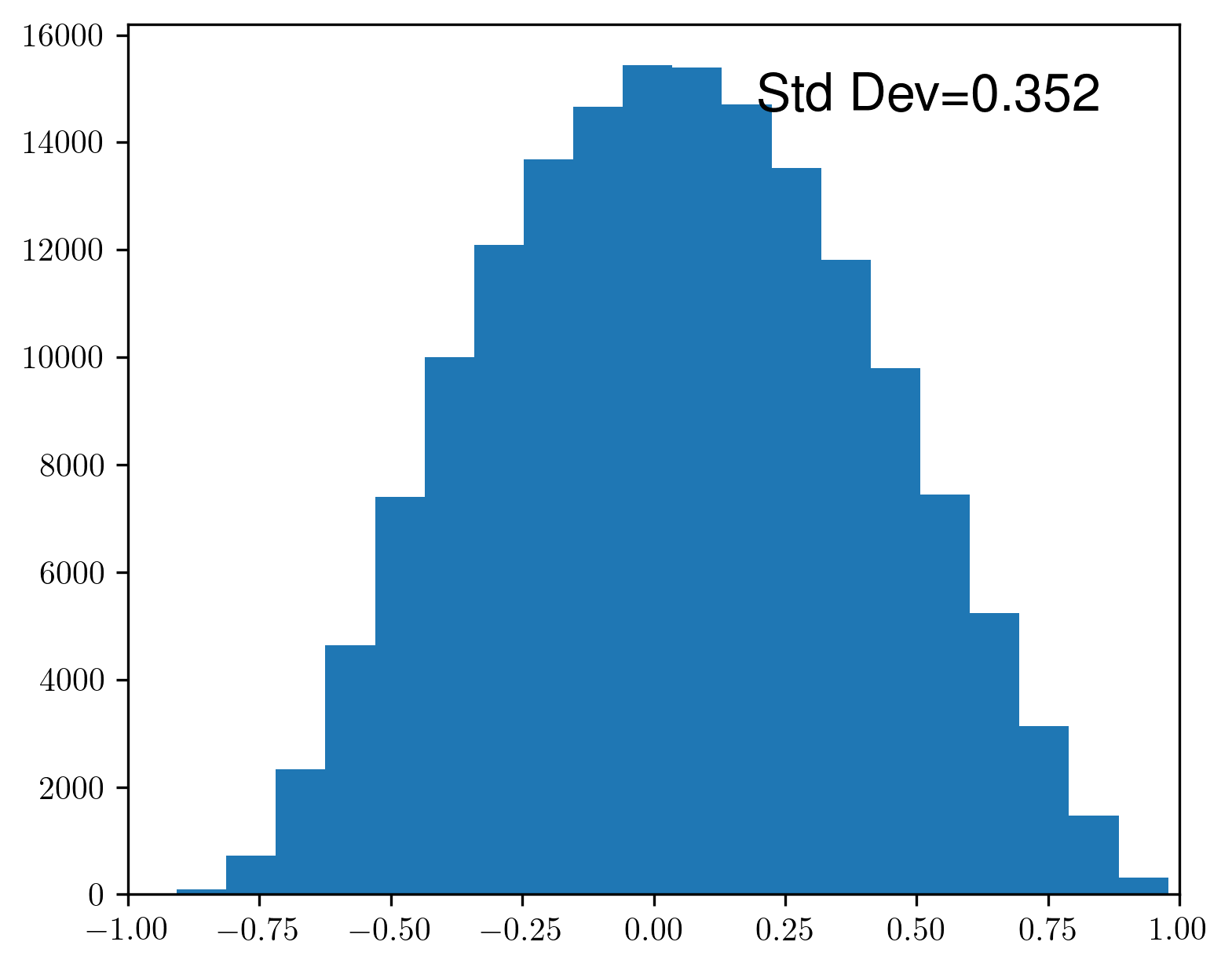}
		\label{fig:correlograms_hist_occ4_first_recode_sector_IBGE_sorted}
	\end{subfigure}\vfill
	\begin{subfigure}[b]{\w\textwidth}
		\centering
		\caption{(Occ4, $\g$) correlogram}
		\includegraphics[height=.25\textheight]{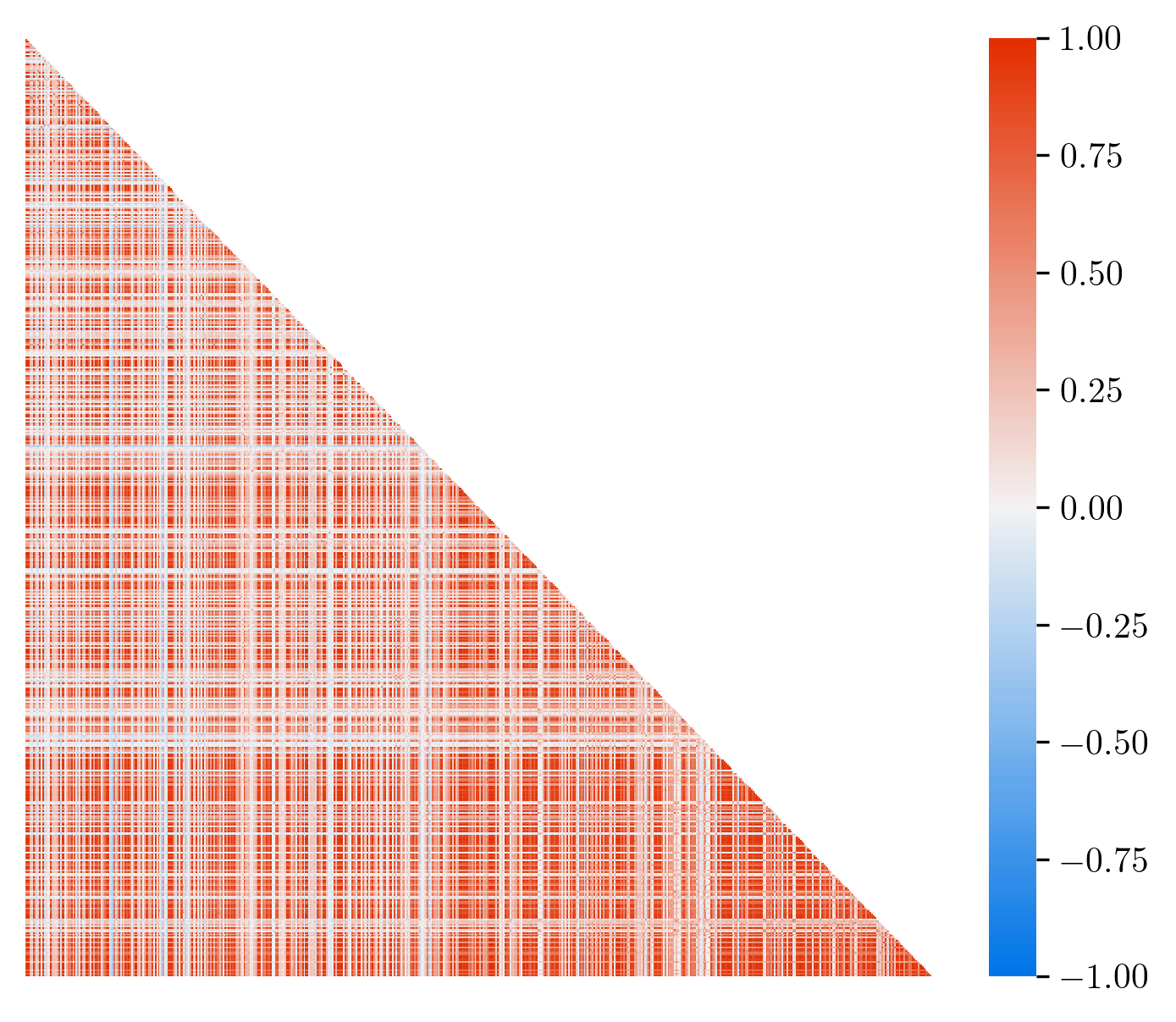}
		\label{fig:correlograms_occ4_first_recode_gamma_sorted}
	\end{subfigure}\hfill
	\begin{subfigure}[b]{\w\textwidth}
		\centering
		\caption{(Occ4, $\g$) histogram}
		\includegraphics[height=.25\textheight]{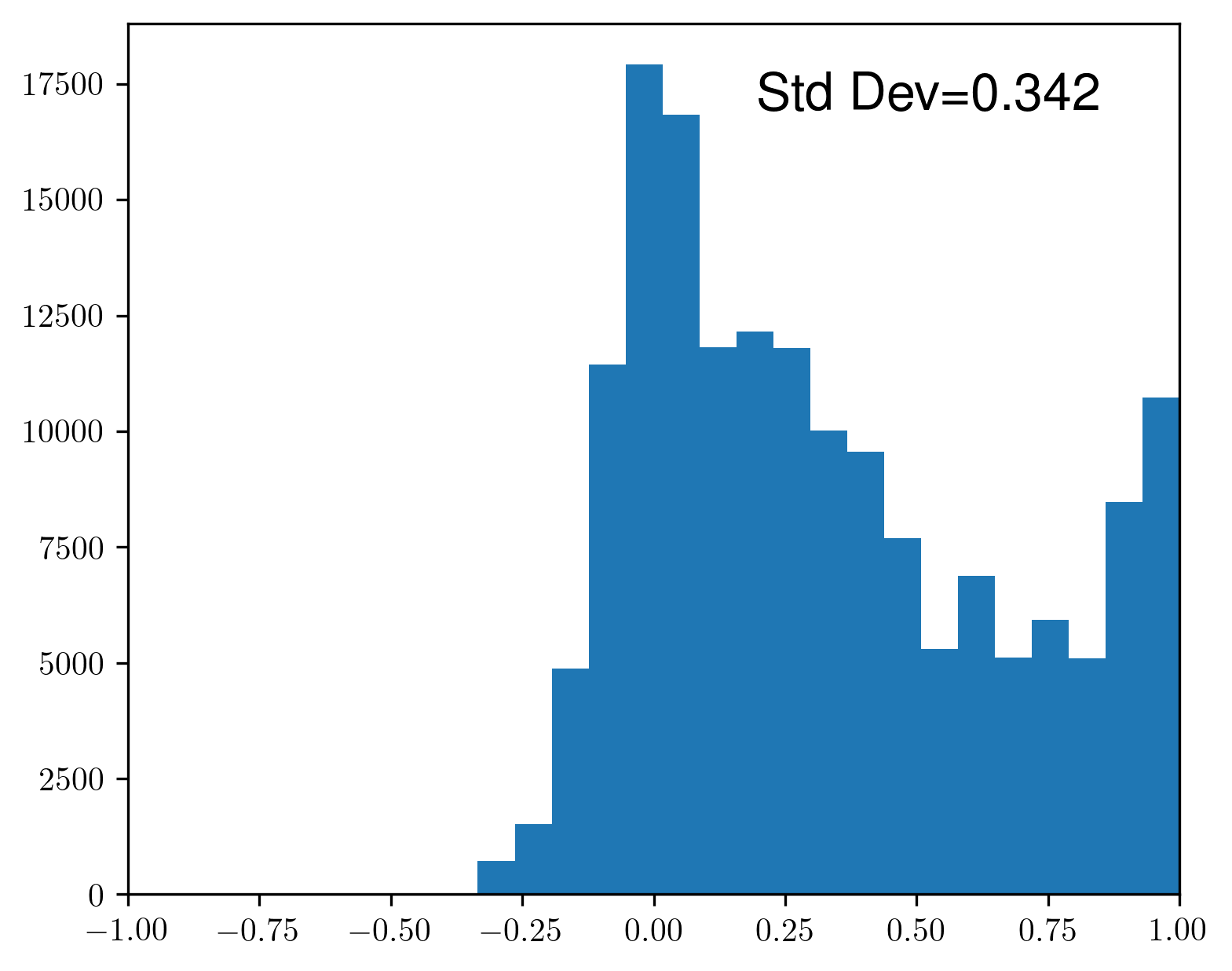}
		\label{fig:correlograms_hist_occ4_first_recode_gamma_sorted}
	\end{subfigure}
	\label{fig:correlograms_sorted}
	\flushleft\footnotesize \emph{Notes:} Figure presents pairwise skills vector correlations (left column) and histograms of these skill correlations (right column) for all pairs of worker types $\i$ (row 1) and 4-digit occupations (rows 2 and 3). In the left column, dark red squares indicate large positive correlations, while dark blue squares represent large negative correlations. ``Skills'' defined as row vectors of the matrix $\Psi$, $\psi_{\i\cdot}$, where $\Psi$ is estimated as described in Section \ref{sec:estimating_psi} using the 2009-2012 RAIS worker earnings panel described in Section \ref{sec:data}. Workers classified by worker types $\i$ in row 1 and by 4-digit occupation in rows 2 and 3. Jobs classified by market $\g$ in rows 1 and 3, and by sector in row 2. Figures in the left column are sorted by worker type mean earnings (smallest to largest).
\end{figure}

\subsection{Predicting general equilibrium effects of Rio de Janeiro Olympics}

\label{sec:model_fit}

We test our model's ability to predict the effects of shocks in the context of the infrastructure investment and other preparations for the 2016 Rio de Janeiro Olympics. The Olympics were announced in late 2009 and construction of new venues and infrastructure were in full effect by 2014. Therefore, we define 2009 as our pre-shock period and 2014 as our ``shock'' period. We calibrate demand shifters $\vec{a}^{2009}$ and $\vec{a}^{2014}$ to fit sector-level product output in those years, feed these demand shifters through our model and solve for the equilibrium to compute model-implied earnings for each worker type for each year, $\hat y_\i^{2009}$ and $\hat y_\i^{2014}$, and then take the difference $\Delta\hat y = \hat y_\i^{2014}-\hat y_\i^{2009}$. We also compute the \emph{actual} mean earnings changes for each worker type, $\Delta y = y_\i^{2014}- y_\i^{2009}$. Finally, we regress actual changes in mean earnings on model-predicted changes in mean earnings for each worker type. 
\begin{align}
	\Delta y  = \beta_0 + \beta_1 \Delta\hat y + \ve \label{eq:model_fit_reg}
\end{align}
If our model is able to perfectly predict the actual effects of the Rio Olympics shock, the slope would be 1 and the intercept 0. As shown in the first column of Table \ref{table:model_fit} the slope of the best fit line is 0.982 and the intercept is -0.003, very close to our goals of 1 and 0, respectively.\footnote{The standard errors in this regression are large, but this is not surprising. There is significant variation that we are unable to predict because a number of important margins of adjustment are outside of our model. However, the fact that we estimate a slope close to 1 and an intercept close to 0 is consistent with these other factors being approximately orthogonal to our classifications. These other factors may include job amenities and non-monetary compensation, migration into or out of the Rio de Janeiro metro area, worker retraining, and changes in the tasks required by each job. Moreover, our model excludes linkages between sectors in the product market, which could affect demand for different types of labor, although our model could be expanded to include product market linkages by adding sector-level intermediate goods as inputs to firms' production functions (equation \ref{eq:int_prod}).}

\begin{table}[h!] \centering
	\caption{Predicted Effect of Olympics on Wages: Network-Based vs. Standard Classifications}
	\begin{tabular}{lccccc}
\toprule 
Worker classification 	 & $\iota$ 	 & Occ4 	 & Occ4 	 & k-means  	 & k-means  \\ 
Market classification 	 & $\gamma$ 	 & Sector 	 & $\gamma$ 	 & Sector  	 & $\gamma$  \\ 
\midrule 
Intercept & -0.003 & -0.001 & -0.002 & -0.000 & -0.002 \\ 
 & ( 0.009 ) & ( 0.009 ) & ( 0.009 ) & ( 0.011 ) & ( 0.010 ) \\ 
Model-implied $\Delta$ log earnings & 0.982 & 0.148 & 0.428 & 0.234 & 0.560 \\ 
 & (0.551) & (0.434) & (0.185) & (0.575) & (0.259) \\ 
\midrule 
MSE & 0.021 & 0.025 & 0.025 & 0.023 & 0.023 \\ 
Observations & 290 & 306 & 306 & 214 & 214 \\ 
\bottomrule 
\end{tabular}
	\label{table:model_fit}
	\flushleft\footnotesize \emph{Notes:} Table presents results from estimating equation (\ref{eq:model_fit_reg}) for various worker and job classifications. Workers classified by worker type ($\i$) in column 1, 4-digit occupation in columns 2 and 3, and by k-means clusters of 6-digit occupations in columns 4 and 5. K-means clustering done on the basis of occupation specific skills defined by the U.S. O*NET, which is applied to Brazilian occupations using a crosswalk creaged by Aguinaldo Maciente \citep{Maciente2013}. Jobs are classified by market ($\g$) in columns 1, 3, and 5, and by IBGE sector in columns 2 and 4. Standard errors reported in parentheses. Independent and dependent variables defined at the worker classification level as described in Section \ref{sec:model_fit}. Dependent variables based on data from the 2009-2012 RAIS worker earnings panel described in Section \ref{sec:data}. Independent computed by solving the model described in Section \ref{sec:model} using parameters estimated in Section \ref{sec:estimating_psi} and calibrated in Section \ref{sec:calibration}. Regressions are weighted by the number of workers per classification.
\end{table}

We further assess our model's predictive ability by comparing it to a series of standard approaches, which use our model but classify worker and job heterogeneity using commonly-used observable variables. Our first two standard approaches classify workers using 4-digit occupation codes instead of our network-based worker types. After dropping occupations with fewer than 5,000 employees for computational reasons,\footnote{This is necessary because we can only use occupations that are observed both pre-shock and post-shock. \iffalse XX provide details here and review exactly why we did this and how sensitive results are.\fi}  we are left with 306 4-digit occupations, yielding a level of disaggregation similar to the 446 worker types. The second two benchmarks characterize worker heterogeneity using k-means clusters of 6-digit occupations based on 225 O*NET skills, where the number of clusters is chosen to match the number of worker types $\i$, however we have to drop some of the resulting clusters because they are very small and are not observed in both the pre-shock and post-shock periods.\footnote{O*NET is defined for the U.S., but we use a crosswalk from the U.S. O*NET to the Brazilian occupation classification system created by Aguinaldo Maciente \citep{Maciente2013}. The clustering method yields a highly skewed cluster size distribution and we must drop some of the smallest clusters because they are not observed in both the pre-shock and post-shock periods. Therefore the actual number of clusters is somewhat smaller than the number of $\i$'s.} We classify job heterogeneity using sector in the first and third benchmark and using our network-based markets in the second and fourth. We present the results of these standard approaches in columns 2--5 of Table \ref{table:model_fit}.

While our network-based classifications yield an approximately unbiased prediction of the actual shock-induced changes in earnings, the standard classifications do not. The coefficients on model-implied earnings changes are far below 1 in all four of the standard classifications. Moreover, the mean squared error (MSE) of our network-based classifications is below all four standard classifications. We interpret this as evidence in favor of our network-based classifications since they do a better job of predicting actual changes in the data than reasonable standard classifications.

\section{Reduced form estimation of labor market shocks} 

\label{sec:reduced_form}


A standard way of estimating the effects of labor demand shocks on workers is through the use of a Bartik instrument. A typical Bartik instrument measures the exposure of different groups of workers to labor demand shocks within groups of jobs. It can be written as 
\begin{align}
	Bartik_{g} = \sum_s \pi_{g s} Shock_s \label{eq:bartik}
\end{align}
where $g$ defines a group of workers, $s$ defines a group of jobs, $\pi_{gs}$ is the fraction of group $g$ workers employed in group $s$ jobs before the shock, and $Shock_s$ is the size of the shock to group $s$ jobs. For example, in \citeauthor{AutorDornHanson2013}'s ``China shock,'' $g$ represents commuting zones, $s$ indexes sectors, $\pi_{gs}$ is commuting zone $g$'s share of sector $s$ employment, and $Shock_s$ is the growth in Chinese imports in sector $s$. $Shock_s$ is a proxy for the size of the labor demand shock in sector $s$ jobs created by Chinese import growth, while $\pi_{gs}$ governs which workers are affected by the shock. Both $Shock_s$ and $\pi_{gs}$ depend upon the researcher's choice of classifications, $g$ and $s$, and therefore estimated effects of shocks are sensitive to these choices. In this section we study how the researcher's choice of worker and job classifications affect results. 

We compare Bartik instruments based on our network-based worker types and markets to Bartik instruments based on occupations and sectors. First, we show that estimated effects of shocks on workers are significantly larger, as are $R^2$ values, when using our network-based classifications. Second, we provide a case study of a simulated shock in which we demonstrate that the reason why our worker types and markets yield larger coefficient estimates and $R^2$ values is that they more precisely identify which jobs experienced a change in demand for labor, and which workers were exposed to those jobs.


\subsection{Analysis of the 2016 Rio de Janeiro Olympics}

\label{sec:reduced_form_real_olympics}


We begin by once again considering the labor demand shock created by the preparations for the Rio de Janeiro Olympics. As in Section \ref{sec:model_fit}, we define 2009 as the pre-shock period and 2014 as the post-shock period. We regress 2009 to 2014 changes in worker group $g$ earnings on the Bartik instrument defined in equation (\ref{eq:bartik}). 
\begin{align}
	\Delta Earnings_g = \beta_0 + \beta_1 Bartik_g	+ \ve_g \label{eq:bartik_reg}
\end{align}
We have four specifications using all four combinations of our two worker classifications $g\in\{\text{worker type, occupation}\}$ and our two job classifications $s\in\{\text{market, sector}\}$. We normalize all of the Bartik instruments to have mean 0 and standard deviation 1 so that coefficients are directly comparable and can be interpreted as the effects of a 1 standard deviation change in the Bartik instrument on log earnings.\footnote{Nonemployment is treated as 0 log earnings, so these regressions capture both movements in and out of employment and changes in earnings conditional on employment.} We measure $\pi_{gs}$ as the fraction of group $g$ workers who are employed in group $s$ jobs. $Shock_s$ is alternatively defined as the change in sector-level product output or changes in the market-level labor input, $\ell_{\g}$. 

The results, presented in  Table \ref{table:real_data_iota_occ4_exposure_regs_ln_wage_N}, show that estimated effects of the shock are highly sensitive to worker and job classifications. In column 1 we present our network-based classifications: workers are classified by worker type and jobs by market. In this specification, the effect of the shock on workers' earnings is positive and statistically significant, and the $R^2$ is large. The coefficient implies that a 1 standard deviation increase in exposure to the Olympics shock leads to an approximately 15.5\% increase in earnings. Columns 2--4 present specifications using standard classifications. These specifications consistently find smaller (and in some cases negative) effects of the shock on workers, and have less explanatory power for variation in worker earnings, as shown by the smaller $R^2$ values. These results are consistent with occupation and sector doing a worse job of characterizing worker skill and job task heterogeneity than worker types and markets, and this misclassification leading to attenuated estimates and worse model fit. 

While our results indicate that classifying worker and job heterogeneity with error yields attenuated estimates of effects \emph{in this case}, it is not necessarily the case that classification errors of this sort yield estimates that are biased towards zero in general. Since we do not have classical measurement error, the intuition of measurement error leading to attenuation bias does not apply. In fact, there is no theoretical prediction about the direction of the bias due to misclasssification of workers and jobs in our context \citep{Mahajan2006,Hu2008}. We confirm this through a series of simulations in which we generate a data set according to the data generating process implied by our model, randomly misclassify varying percentages of workers and jobs, and then estimate the Bartik regression, equation (\ref{eq:bartik_reg}). We find no clear relationship between the amount of misclassification and the slope coefficient $\hat\beta$. However, we do find that the $R^2$ values decline approximately monotonically with the fraction of workers and jobs misclassified. Therefore, we interpret the larger $R^2$ values from estimating equation (\ref{eq:bartik_reg}) using our network-based classifications as evidence that the network-based classifications classify worker and job heterogeneity with less error than the standard classifications. By contrast, the larger coefficient estimate when we use our network-based classifications is an empirical finding about the implications of misclassification in this context. See Appendix \ref{app:misclassification} for details on these simulations.

Although the focus of this paper is classification of workers rather than identification of shocks, it is possible that the Olympics shock we study in this section may have been confounded by labor supply or other shocks. For example, workers may have anticipated the shock and migrated to Rio de Janeiro from other parts of Brazil. Therefore, in the next subsection we replicate the analysis in this subsection using simulated data in which we control the data generating process. 

\begin{table}
	\centering
	\caption{Effects of exposure to Rio Olympics shock}
	\begin{tabular}{@{\extracolsep{5pt}}lcccc}
\\[-1.8ex]\hline
\hline \\[-1.8ex]
\hline \\[-1.8ex]
 Exposure: & Market ($\gamma$) & Sector & Market ($\gamma$) & Sector \\
 Worker classification: & Worker type ($\iota$) & Worker type ($\iota$) & Occ4 & Occ4 \\
 \hline &  &  &  &  \\
 Intercept & -0.297$^{***}$ & -0.297$^{***}$ & -0.313$^{***}$ & -0.313$^{***}$ \\
& (0.006) & (0.006) & (0.009) & (0.009) \\
 iota exposure (market) & 0.058$^{}$ & & & \\
& (0.006) & & & \\
 iota exposure (sector) & & -0.030$^{}$ & & \\
& & (0.006) & & \\
 occ4 exposure (market) & & & 0.038$^{}$ & \\
& & & (0.009) & \\
 occ4 exposure (sector) & & & & -0.006$^{}$ \\
& & & & (0.009) \\
\hline \\[-1.8ex]
 Observations & 446 & 446 & 570 & 570 \\
 $R^2$ & 0.198 & 0.051 & 0.030 & 0.001 \\
\hline
\hline \\[-1.8ex]
\textit{Note:}\end{tabular} 
	\label{table:real_data_iota_occ4_exposure_regs_ln_wage_N} 
	\footnotesize\flushleft \emph{Notes:} Table presents the effect of the 2016 Rio de Janeiro Olympics shock on workers earnings from estimating equation (\ref{eq:bartik_reg}). Independent variables normalized to have mean 0 and standard deviation 1. Workers classified by worker type ($\i$) in columns 1 and 2, and by 4-digit occupation in columns 3 and 4. Standard errors reported in parentheses. Jobs classified by market in columns 1 and 3, and by sector in columns 2 and 4. Estimated using data from the 2009-2012 RAIS worker earnings panel described in Section \ref{sec:data} aggregated to the worker classification level. 
\end{table}

\subsection{Reduced form analysis using simulated data}

\label{sec:reduced_form_fake_olympics}

In this subsection, we demonstrate how estimated effects of shocks are sensitive to worker and job classifications in a setting where we can control the underlying data generating process. We replicate the analysis in the preceding section using simulated data. The simulated data have the same structure as the actual worker earnings panel described in Section \ref{sec:data} that we used to estimate the labor supply parameters and for the empirical exercises in Sections \ref{sec:model_fit} and \ref{sec:reduced_form_real_olympics}, and are drawn from the data generating process defined by our model. Since we control the data generating process, we can be certain that we are observing an exogenous labor demand shock that is unconfounded by, for example, concurrent labor supply changes. 

We generate the simulated data as follows. First, we calibrate demand shifters $\vec{a}^{Pre}$ to match the levels of product demand in each sector in 2009. We then solve the model using the 2009 demand shifters to generate a pre-shock wage vector $\vec{w}^{Pre}$ that clears all markets $\g$. We draw worker types and four-digit occupations from the empirical joint distribution of worker types and four-digit occupations. To generate job matches for each worker recall that, conditional on searching, workers choose a market to supply labor to according to equation (\ref{eq:worker_max}):
\[\g_{it} = \argmax_{\g \in \{0,1,\dots,\Gamma\}} \psi_{\ig} w_{\g t} + \xi_{\g} + \ve_{i\g t} .\] 
This implies that a type $\i$ worker chooses market $\g$ with probability given by equation (\ref{eq:emp_probs}): 
\[\P_{\i}[\g | = \frac{\exp \left( \frac{\hat\psi_{\ig} w_{\g t} + \hat\xi_{\g}}{\hat\nu} \right) }{ \sum\limits_{\g'=0}^{\Gamma} \exp \left( \frac{\hat\psi_{\ig'} w_{\g' t} + \hat\xi_{\g'}}{\hat\nu} \right) },\] 
where we use estimated parameter values $\hat \Psi$, $\hat \Xi$, and $\hat \nu$, estimated as described in Section \ref{sec:MLE}. All workers make this choice in period $t=1$, and in subsequent periods workers search again if they draw a separation shock as described in Assumption \ref{ass:mobility}. In our full model, workers match with individual jobs after choosing markets, however the identity of the worker's individual job $j$ does not affect earnings or employment; it is only useful for classifying workers and jobs according to the BiSBM. Therefore, we do not specify the identity of each worker's specific job when generating our simulated data set. 

Next, we draw sectors for each worker--job match according to the empirical joint distribution of sectors and markets. Finally, we draw earnings according to equation (\ref{eq:observed_wages}):
\[\omega_{it} = \psi_{\i(i)\g_{it}} w_{\g_{it}} e_{it}. \]
where $e_{it}$ is log-normal measurement error. We repeat this exercise using the same labor supply parameters $\hat \Psi$, $\hat \Xi$, and $\hat \nu$ along with a new vector of demand shifters, $\vec{a}^{Post}$, calibrated to match the levels of product demand in each sector in 2014. We stack the two data sets to create a panel data set with both the pre-shock and post-shock periods. 

We repeat the four Bartik-style regressions from the previous section using our simulated data. The results, presented in Table \ref{table:fake_data_rio_iota_occ4_exposure_regs_ln_wage_N}, are qualitatively similar to the results using actual data in the previous section (Table \ref{table:real_data_iota_occ4_exposure_regs_ln_wage_N}), with the exception that the negative coefficients when jobs are classified by sector are now small positive coefficients. We continue to find larger coefficients and $R^2$ values when we define shock exposure according to markets as opposed to sectors, and when we classify workers according to worker type as opposed to 4-digit occupation. These results reiterate our point that misclassifying worker and jobs causes us to significantly understate the effects of shocks on workers in this context. In the next section we demonstrate that this is a more general finding.

\begin{table}
	\centering
	\caption{Effects of exposure to \emph{simulated} Rio Olympics shock}
	\begin{tabular}{@{\extracolsep{5pt}}lcccc}
\\[-1.8ex]\hline
\hline \\[-1.8ex]
\hline \\[-1.8ex]
 Exposure: & Market ($\gamma$) & Sector & Market ($\gamma$) & Sector \\
 Worker classification: & Worker type ($\iota$) & Worker type ($\iota$) & Occ4 & Occ4 \\
 \hline &  &  &  &  \\
 Intercept & -0.239$^{***}$ & -0.239$^{***}$ & -0.264$^{***}$ & -0.264$^{***}$ \\
& (0.003) & (0.003) & (0.007) & (0.007) \\
 iota exposure (market) & 0.028$^{}$ & & & \\
& (0.003) & & & \\
 iota exposure (sector) & & 0.022$^{}$ & & \\
& & (0.003) & & \\
 occ4 exposure (market) & & & 0.004$^{}$ & \\
& & & (0.007) & \\
 occ4 exposure (sector) & & & & 0.035$^{}$ \\
& & & & (0.007) \\
\hline \\[-1.8ex]
 Observations & 446 & 446 & 573 & 573 \\
 $R^2$ & 0.158 & 0.099 & 0.001 & 0.045 \\
\hline
\hline \\[-1.8ex]
\textit{Note:}\end{tabular}
	\label{table:fake_data_rio_iota_occ4_exposure_regs_ln_wage_N}
		\footnotesize\flushleft \emph{Notes:} Table presents the effect of the \emph{simulated} 2016 Rio de Janeiro Olympics shock on workers earnings from estimating equation (\ref{eq:bartik_reg}). Independent variables normalized to have mean 0 and standard deviation 1. Workers classified by worker type ($\i$) in columns 1 and 2, and by 4-digit occupation in columns 3 and 4. Standard errors reported in parentheses. Jobs classified by market in columns 1 and 3, and by sector in columns 2 and 4. Estimated using data generated using our model as the data generating process, as described  in Section \ref{sec:reduced_form_fake_olympics}, and aggregated to the worker classification level. 
\end{table}

\subsection{Simulating many shocks}

\label{sec:many_shocks}

\begin{figure}[htbp]
	\centering
	\caption{Exposure coefficients from all simulated shocks}
	\begin{subfigure}{\textwidth}
		\centering
		\caption{Slope coefficients}
		\includegraphics[width=.7\textwidth]{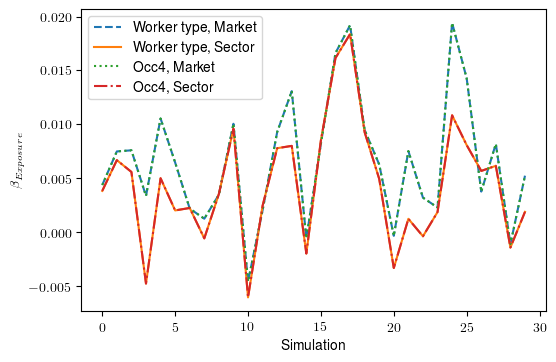} 
		\label{fig:fake_data_all_sector_shocks_coef}
	\end{subfigure}\\
 	\vfill
 	\vspace{1em}
	\begin{subfigure}{\textwidth}
		\centering
		\caption{$R^2$ values}
		\includegraphics[width=.7\textwidth]{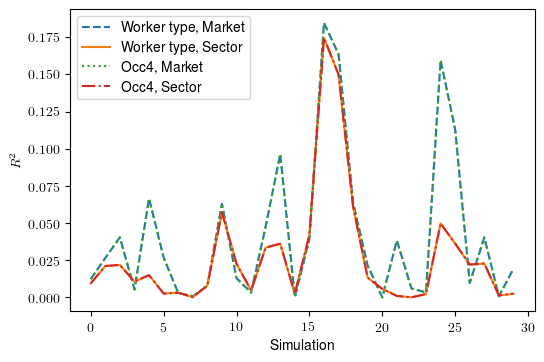} 
		\label{fig:fake_data_all_sector_shocks_r2}
	\end{subfigure}
	\label{fig:fake_data_all_sector_shocks}
	\footnotesize\flushleft \emph{Notes:} Figure presents estimated regression coefficients and $R^2$ values from estimating the Bartik-style regression, equation (\ref{eq:bartik_reg}), for each of the 30 simulated shocks described in Section \ref{sec:many_shocks}.  
\end{figure}

In the previous sections we found that the estimated effects of shocks are larger when using our network-based worker and job classifications than when using standard classifications. To allay any concern that our finding is specific to the Rio Olympics shock, we replicate the analysis in the previous section for a series of different shocks. For each of the 15 sectors, we simulate a positive shock in which the demand shifter for the shocked sector is doubled and the demand shifters for all other sectors are unchanged, and a negative shock in which the demand shifter for the shocked sector is halved and the demand shifters for all other sectors are unchanged. For each shock, we generate a new simulated data set and then use the simulated data to estimate the Bartik-style regression in equation (\ref{eq:bartik_reg}) for each of the four combinations of worker and job classifications: $g\in\{\text{worker type, occupation}\}$ and $s\in\{\text{market, sector}\}$. We present the results in Table \ref{table:all_sector_shocks_means}. We consistently find larger coefficients and $R^2$ values using our network-based classifications. The average coefficient from our network-based classification specification is 3.7 times larger than the average coefficient from the occupation--sector specification, and the average $R^2$ is 11 times larger. Figure \ref{fig:fake_data_all_sector_shocks} presents the slope coefficients and $R^2$ values from each individual regression in these simulations and shows that our network-based classifications yield slope coefficients and $R^2$ values that are uniformly larger than those from standard classifications, not just larger on average.

\begin{table}
\centering
\caption{Means across all simulated shocks}
\label{table:all_sector_shocks_means}
\begin{tabular}{@{\extracolsep{10pt}}llrrrr}
\toprule
Worker Classification & Job Classification & \multicolumn{2}{c}{Coefficient} & \multicolumn{2}{c}{$R^2$} \\ \cline{3-4}\cline{5-6}
               \hfill &             \hfill &        Mean & Std Dev &  Mean & Std Dev \\
\midrule
          Worker type &             Market &       0.007 &   0.006 & 0.043 &   0.052 \\
          Worker type &             Sector &       0.004 &   0.006 & 0.028 &   0.041 \\
                 Occ4 &             Market &       0.007 &   0.006 & 0.043 &   0.052 \\
                 Occ4 &             Sector &       0.004 &   0.006 & 0.028 &   0.041 \\
\bottomrule
\end{tabular}
\end{table}  

\clearpage
\subsection{Case study of shock to the ``Accommodations and Food'' sector}

\label{sec:case_study}

One of the shocks we simulated in the previous section was a 50\% reduction in demand for the output of the Accommodations and Food sector, leaving the demand for all other sectors' output unchanged. This subsection explores that shock in greater detail to elucidate the mechanisms behind the finding that our network-based classifications yield larger estimates of the effects of shocks on workers.  We focus on a shock to a single sector, as opposed to all sectors simultaneously as in the Rio Olympics shock, because this allows us to understand the precise nature of the shock.

%


Table \ref{table:fake_data_AccomFood_iota_occ4_exposure_regs_ln_wage_N} presents the same set of Bartik-style regressions as Tables \ref{table:real_data_iota_occ4_exposure_regs_ln_wage_N} and \ref{table:fake_data_rio_iota_occ4_exposure_regs_ln_wage_N} in the preceding sections. The qualitative story is unchanged: larger coefficients and $R^2$ values when we (i) define job heterogeneity according to markets as opposed to sectors, and (ii) when we define worker heterogeneity according to worker type as opposed to 4-digit occupation.

\begin{table}[h!]
	\centering
	\caption{Effects of exposure to simulated Accommodations and Food sector shock}
	\vspace{-.5cm}
	\begin{tabular}{@{\extracolsep{5pt}}lcccc}
\\[-1.8ex]\hline
\hline \\[-1.8ex]
\hline \\[-1.8ex]
 Exposure: & Market ($\gamma$) & Sector & Market ($\gamma$) & Sector \\
 Worker classification: & Worker type ($\iota$) & Worker type ($\iota$) & Occ4 & Occ4 \\
 \hline &  &  &  &  \\
 Intercept & -0.000$^{}$ & -0.000$^{}$ & -0.000$^{}$ & -0.000$^{}$ \\
  & (0.001) & (0.001) & (0.002) & (0.002) \\
 iota exposure (market) & 0.008$^{}$ & & & \\
  & (0.001) & & & \\
 iota exposure (sector) & & 0.006$^{}$ & & \\
  & & (0.001) & & \\
 occ4 exposure (market) & & & -0.000$^{}$ & \\
  & & & (0.002) & \\
 occ4 exposure (sector) & & & & -0.001$^{}$ \\
  & & & & (0.002) \\
\hline \\[-1.8ex]
 Observations & 290 & 290 & 306 & 306 \\
 $R^2$ & 0.092 & 0.054 & 0.000 & 0.001 \\
\hline
\hline \\[-1.8ex]
\textit{Note:}\end{tabular}
	\label{table:fake_data_AccomFood_iota_occ4_exposure_regs_ln_wage_N}
	\footnotesize\flushleft \emph{Notes:} Table presents the effect of the \emph{simulated} Accommodations and Food sector shock on workers earnings from estimating equation (\ref{eq:bartik_reg}). The shock is a 50\% reduction in demand for the Accommodations and Food sector's output, holding demand for all other sectors' output constant. Independent variables normalized to have mean 0 and standard deviation 1. Workers classified by worker type ($\i$) in columns 1 and 2, and by 4-digit occupation in columns 3 and 4. Standard errors reported in parentheses. Jobs classified by market in columns 1 and 3, and by sector in columns 2 and 4. Estimated using data generated using our model as the data generating process, as described in Section \ref{sec:reduced_form_fake_olympics}, and aggregated to the worker classification level.  
\end{table}

Why does the Bartik instrument have more explanatory power for workers' outcomes when workers are classified by worker types and jobs are classified by markets? On the worker side, it is because, as we argued in Sections \ref{sec:occ_count_tables} and \ref{sec:correlograms}, our worker types do a better job of identifying groups of homogenous workers than do occupations. We see this again by focusing on one of the worker types that was most affected by the shock to the Accommodations and Food sector, worker type $\i=64$. Table \ref{table:shock_case_study_AccomFood_most_shocked_iota_occ_counts_3} tabulates the 10 occupations we most frequently observe type $\i=64$ workers employed in. These occupations tend to be low-pay, low-education service sector occupations. The two most frequent are ``food services assistant'' and ``retail salesperson.'' Our network-based classification method tells us that these retail and food services workers have similar skills despite the fact that they are employed in different occupations. If we had classified workers by occupation and jobs by sector, we would have implicitly assumed that the food services workers were exposed to the Accommodations and Food sector shock, while the retail salespeople were not. In reality, all of these workers were exposed to the shock because they have similar skills;  workers not employed in the shocked sector may still be exposed to and affected by the shock if they are close substitutes for workers in the shocked sector.  As we discussed in Section \ref{sec:reduced_form_real_olympics} and Appendix \ref{app:misclassification}, misclassifying workers such that some workers actually exposed to the shock are assumed not to have been exposed, and vice versa, leads to biased coefficient estimates and attenuated $R^2$ values.

\begin{table}
\centering
\caption{Occupation counts for $\iota=$64}
\label{table:shock_case_study_AccomFood_most_shocked_iota_occ_counts_3}
\begin{tabular}{rlr}
\toprule
Occ Code &          Occ Description & Occ share \\
\midrule
  513505 &  Food services assistant &     0.090 \\
  521110 &       Retail salesperson &     0.072 \\
  411005 &             Office clerk &     0.043 \\
  514320 &                  Janitor &     0.032 \\
  513205 &             General cook &     0.032 \\
  513215 &          Industrial cook &     0.030 \\
  421125 &                  Cashier &     0.028 \\
  411010 & Administrative assistant &     0.026 \\
  763215 &  Sewing machine operator &     0.024 \\
  521125 &              Stock clerk &     0.019 \\
\bottomrule
\end{tabular}
\end{table}

On the jobs side, classifying jobs by market rather than sector more accurately captures the channels through which shocks propagate from jobs to workers. Bartik instruments based on standard classifications assume that workers supply labor directly to sectors; our classifications allow workers to supply labor directly to markets but only indirectly to sectors, by way of markets (see Figure \ref{fig:us_vs_ADH_tikz}). We illustrate why our approach is prefereable by again focusing on type $\i=64$ workers. We have already established that these workers' skills are employable in both retail occupations and food service occupations, but who hires them? Do they supply labor to a retail market and a food services market? Or is there actually a market that includes jobs in both retail and food services? In Table \ref{table:shock_case_study_AccomFood_sector_share_table_3} we present type $\i=64$ workers' labor supply by sector. Type $\i=64$ workers supply labor to a variety of sectors, including Retail, Wholesale and Vehicle Repair (28\%) and Accommodations and Food (14\%). Since these workers supply labor to such a  variety of sectors, no single sector can reasonably approximate the set of jobs to which they supply labor. By contrast, type $\i=64$ workers' labor supply \emph{is} concentrated within specific network-based markets, $\g$.

Table \ref{table:shock_case_study_AccomFood_gamma_share_table_3} presents the percentage of their labor that type $\i=64$ workers supply to each market, restricting to the top 10. Type $\i=64$ workers supply over 60\% of their labor to a single market, market $\g=47$, and there is no other market to which they supply more than 3.5 percent of their labor. In other words, type $\i=64$ workers' labor supply is highly concentrated within a specific market, but not nearly as concentrated in specific sectors, despite the fact that we have vastly more markets (1,371) than sectors (15).  This is a specific example of the more general finding of greater concentration of employment within markets than sectors that we presented in Section \ref{sec:hhi}. Worker types' employment is more concentrated within markets than sectors because our markets are designed to identify groups of jobs that compete for similar workers, whereas sectors are defined by product markets. Therefore, our markets more closely approximate the channels through which shocks propagate through the labor market to workers. By contrast, classifying jobs by sectors introduces error by grouping together jobs with heterogeneous changes in labor demand.  Again, as we discussed in Section \ref{sec:reduced_form_real_olympics} and Appendix \ref{app:misclassification}, misclassifying jobs such that jobs that in fact hire dissimilar workers are assumed to hire similar workers, and vice versa, leads to biased coefficient estimates and attenuated $R^2$ values.

\begin{table}[h!]
	\centering
	\caption{Type $\i=64$ workers' labor supply by sector}
	\begin{tabular}{lr}
\toprule
                                            Sector &  Share (\%) \\
\midrule
Trade and repair of motor vehicles and motorcycles &       28.5 \\
                            Accommodation and food &       12.7 \\
                          Manufacturing industries &       12.2 \\
       Professional, scientific and technical svcs &       11.9 \\
                      Private health and education &        7.0 \\
Arts, culture, sports and recreation and other ... &        6.9 \\
                       Transport, storage and mail &        6.8 \\
                                      Construction &        3.5 \\
                                         Utilities &        3.1 \\
         Financial, insurance and related services &        2.1 \\
                             Extractive industries &        1.8 \\
                     Information and communication &        1.8 \\
Public admin, defense, educ, health and soc sec... &        1.3 \\
                            Real estate activities &        0.2 \\
Agriculture, livestock, forestry, fisheries and... &        0.1 \\
\bottomrule
\end{tabular}
	\label{table:shock_case_study_AccomFood_sector_share_table_3}
	\footnotesize \flushleft \emph{Notes:} Table presents the share of type $\i=64$ workers employed in each sector according to data generated by simulating the Accommodations and Food sector shock. The shock is a 50\% reduction in demand for the Accommodations and Food sector's output, holding demand for all other sectors' output constant.
\end{table}

\begin{table}[h!]
	\centering
	\caption{Type $\i=64$ workers' labor supply by market ($\g$)}
	\begin{tabular}{rr}
\toprule
 Market ($\gamma$) &  Share (\%) \\
\midrule
                47 &        60.1 \\
               189 &         3.5 \\
               116 &         1.7 \\
               242 &         1.5 \\
               418 &         1.3 \\
                83 &         1.3 \\
                36 &         1.2 \\
               138 &         1.1 \\
               125 &         0.9 \\
                45 &         0.8 \\
\bottomrule
\end{tabular}
	\label{table:shock_case_study_AccomFood_gamma_share_table_3}
	\footnotesize \flushleft \emph{Notes:} Table presents the share of type $\i=64$ workers employed in each market ($\g$) according to data generated by simulating the Accommodations and Food sector shock. The shock is a 50\% reduction in demand for the Accommodations and Food sector's output, holding demand for all other sectors' output constant. Only the 10 most frequently occurring markets are shown. 
\end{table}

\begin{figure}[!htbp]
	\centering
	\caption{}
	\begin{subfigure}[t]{0.49\textwidth}
		\centering
		\caption{Standard Classifications}
		\resizebox{3.5cm}{\height}{\definecolor{myblue}{RGB}{80,80,160}
\definecolor{mygreen}{RGB}{80,160,80}

\begin{tikzpicture}[thick,
every node/.style={draw,circle},
isnode/.style={fill=orange},
gsnode/.style={fill=mygreen},
ssnode/.style={fill=myblue},
every fit/.style={rectangle,draw,inner sep=1pt, text height=2cm},
->,shorten >= 3pt,shorten <= 3pt
]

\begin{scope}[xshift=1cm, start chain=going right,node distance=7mm]
\foreach \i in {1,2,3}
\node[ssnode,on chain] (s\i)  {};
\end{scope}


\begin{scope}[xshift=0.5cm,yshift=-4cm,start chain=going right,node distance=7mm]
\foreach \i in {9,10,...,12}
\node[isnode,on chain] (i\i)  {};
\end{scope}

\node [myblue, fit=(s1) (s3), label={[text=myblue]right:$S$}] {};
\node [orange,fit=(i9) (i12),label={[text=orange]right:$\i$}] {};


\draw (i9)  -- (s1);
\draw (i10) -- (s1);
\draw (i11) -- (s1);
\draw (i12) -- (s1);
\draw (i9)  -- (s2);
\draw (i10) -- (s2);
\draw (i11) -- (s2);
\draw (i12) -- (s2);
\draw (i9)  -- (s3);
\draw (i10) -- (s3);
\draw (i11) -- (s3);
\draw (i12) -- (s3);
\end{tikzpicture}}
	\end{subfigure}
	\begin{subfigure}[t]{0.49\textwidth}
		\centering
		\caption{Our Model}
		\resizebox{3.5cm}{\height}{\definecolor{myblue}{RGB}{80,80,160}
\definecolor{mygreen}{RGB}{80,160,80}

\begin{tikzpicture}[thick,
every node/.style={draw,circle},
isnode/.style={fill=orange},
gsnode/.style={fill=mygreen},
ssnode/.style={fill=myblue},
every fit/.style={rectangle,draw,inner sep=1pt, text height=2cm},
->,shorten >= 3pt,shorten <= 3pt
]

\begin{scope}[xshift=1cm, start chain=going right,node distance=7mm]
\foreach \i in {1,2,3}
\node[ssnode,on chain] (s\i)  {};
\end{scope}

\begin{scope}[xshift=0cm,yshift=-2cm,start chain=going right,node distance=7mm]
\foreach \i in {4,5,...,8}
\node[gsnode,on chain] (g\i)  {};
\end{scope}

\begin{scope}[xshift=0.5cm,yshift=-4cm,start chain=going right,node distance=7mm]
\foreach \i in {9,10,...,12}
\node[isnode,on chain] (i\i)  {};
\end{scope}

\node [myblue, fit=(s1) (s3), label={[text=myblue]right:$S$}] {};
\node [mygreen,fit=(g4) (g8),label={[text=mygreen]right:$\g$}] {};
\node [orange,fit=(i9) (i12),label={[text=orange]right:$\i$}] {};


\draw (g4) -- (s1);
\draw (g5) -- (s1);
\draw (g6) -- (s1);
\draw (g7) -- (s1);
\draw (g8) -- (s1);
\draw (g4) -- (s2);
\draw (g5) -- (s2);
\draw (g6) -- (s2);
\draw (g7) -- (s2);
\draw (g8) -- (s2);
\draw (g4) -- (s3);
\draw (g5) -- (s3);
\draw (g6) -- (s3);
\draw (g7) -- (s3);
\draw (g8) -- (s3);
\draw (i9)  -- (g4);
\draw (i10) -- (g4);
\draw (i11) -- (g4);
\draw (i12) -- (g4);
\draw (i9)  -- (g5);
\draw (i10) -- (g5);
\draw (i11) -- (g5);
\draw (i12) -- (g5);
\draw (i9)  -- (g6);
\draw (i10) -- (g6);
\draw (i11) -- (g6);
\draw (i12) -- (g6);
\draw (i9)  -- (g7);
\draw (i10) -- (g7);
\draw (i11) -- (g7);
\draw (i12) -- (g7);
\draw (i9)  -- (g8);
\draw (i10) -- (g8);
\draw (i11) -- (g8);
\draw (i12) -- (g8);
\end{tikzpicture}}
	\end{subfigure}
	\label{fig:us_vs_ADH_tikz}
\end{figure}


%
%


\clearpage

\section{Conclusion}
\label{sec:conclusion}

In this paper we develop a new method for clustering workers and jobs into discrete types that relies on workers' and jobs' choices, rather than observable variables or expert judgments. Our key insight is that linked employer-employee data contain a previously underutilized source of information: millions of worker--job matches, each of which reflects workers' and jobs' perceptions of the workers' skills and the jobs' tasks. We do so by microfounding a classification tool from the network theory literature with a Roy model of workers matching with jobs according to comparative advantage. The link between economic theory and network theory provides the worker types and markets we identify with a rigorous theoretical underpinning and clear interpretability. 

We demonstrate that our network-based worker and job classifications outperform standard worker and job classifications in a number of ways. First, we show that an equilibrium model does a better job of predicting the effects of the Rio de Janeiro Olympics on workers' earnings when workers and jobs are classified using our network-based classifications than when they are classified using standard classifications. Second, we show that reduced form Bartik-style regressions yield larger and more precise estimates of the effects of shocks on workers when workers and jobs are classified using our network-based classifications as opposed to standard classifications. 

A key feature of our classifications is that they simultaneously aggregate and disaggregate workers across occupations. They aggregate workers in different occupations who are revealed to have similar skills (for example, retail and food service workers), while disaggregating workers in the same occupation revealed to have distinct skills (for example course instructors focused on physical versus academic education). Our classifications, therefore, provide value beyond simply choosing the right granularity in, or aggregation of, occupation codes. They identify cohesive groups of workers and jobs that are not too granular to be useful in practical applications.


Although we apply our network-based clustering method to understanding the effects of labor market shocks on workers, this is only the beginning of our research agenda. We are currently working to apply different versions of the method to three different questions. First, we use our method to improve controls for worker skills in wage decompositions. Second, we use our worker and job classifications to improve measures of market power, based on the intuition that if retail and food services jobs compete for the same workers, they belong to the same market, even if they belong to different industries and occupations. Third, we are using closely related techniques to impute occupation and other worker characteristics in the LEHD.

Finally, although our current model abstracts from the role of physical space in the labor market and our empirics therefore focus on a single metropolitan area, we are working to expand our analysis to include geography and apply it to the entire country of Brazil. This will allow us to study the interaction of skills/tasks and geography in determining the scope of labor markets. For example, it will allow us to distinguish between different types of workers, likely with different types of skills, who search for jobs more nationally or more locally. 

Our method is broadly applicable to important questions in labor economics and other fields. In addition to the applications to Bartik-style regressions we discuss in detail, our method may be useful any time researchers need to classify workers and/or jobs. For example, researchers studying how heterogeneous workers match with heterogeneous jobs might classify worker and job heterogeneity using our network-based classifications. The same is true for researchers studing the effects of shocks on workers using structural methods. More broadly, the method we develop may be used to classify agents using revealed preference any time agents' choices lead to a network structure of matches. For example, our method could be adapted to classify products and consumers based on detailed purchasing data, or to cluster financial institutions or countries based on networks of financial or trade flows. This paper provides a blueprint for doing so in a theoretically principled and data-driven way.

\clearpage
\bibliographystyle{aer}	
\bceta

\clearpage
\appendix
\appendixpage

\section{Adding geography}

\label{sec:geography}

If we assume the commuting costs are measured in units of our numeraire good, we can add the cost of worker $i$ commuting to job $j$ to the worker's job choice as follows:
\begin{align*}
\g_{it} = \argmax_{\g \in \{0,1,\dots,\Gamma\}} \psi_{\ig} w_{\g} + \xi_{\g} + CommutingCost_{ij}+ \ve_{i\g t}
\end{align*}

Although we have written the commuting cost for a worker $i$ job $j$ pair, we do not observe commuting costs for individual pairs. However, in the market clearing conditions we are integrating over individual workers and jobs of the same type, so really we would only need an integral of commuting costs (basically, average commuting costs).

%

\section{Network theory details}

\label{app:network_theory}

\subsection{A primer on networks}

\begin{figure}[!htbp]
	\caption{Simple bipartite network}
	\centering
	\includegraphics[width=.69\textwidth]{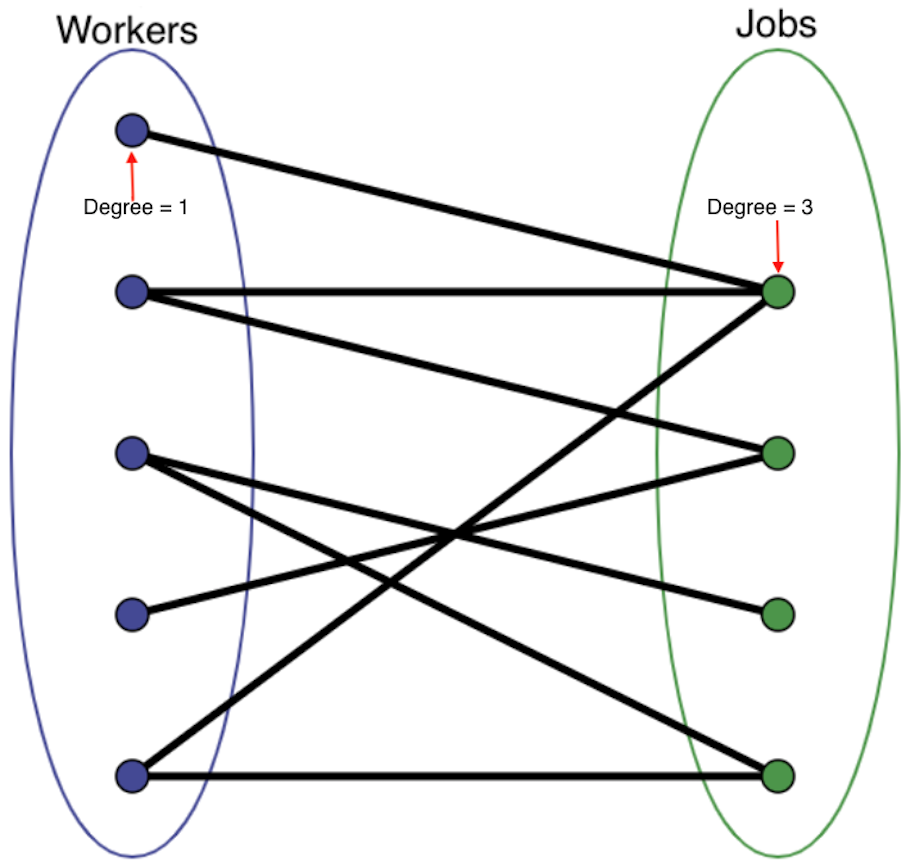}
	\label{fig:simple_bipartite_network}
\end{figure}

``A network is, in its simplest form, a collection of points joined together in pairs by lines'' \citep{Newman2018}. The points are referred to as ``nodes'', and the lines as ``edges.''  In Figure \ref{fig:simple_bipartite_network}, the dots represent nodes and the lines represent edges. Networks can represent a wide variety of phenomena. For example, in an air travel network, airports are nodes and flight paths are edges. Similarly, in a social network, people are nodes and edges represent social relationships like friendship. The labor market, as viewed in LEED, can also be represented as a network. Each node represents an individual worker or job, and each edge represents an employment spell between a worker and a job. 

In a network of worker--job connections like ours, edges connect workers to jobs. This means that there can be no edges between two worker nodes or between two job nodes; only between one worker node and one job node. Networks like this, in which nodes belong to one of two categories and all edges connect nodes in different categories, are known as ``bipartite'' networks. This is reflected in Figure \ref{fig:simple_bipartite_network} by the fact that all worker nodes are in blue on the left, all job nodes are in green on the right, and all edges (black lines) connect a worker to a job. 

There is one more concept we need to introduce before returning our focus to estimation: the ``degree'' of a node. The degree of a node is the number of edges connected to that node. In figure  \ref{fig:simple_bipartite_network}, the first (from the top) worker node has a degree of 1 because it is connected to exactly one edge (black line) while the first job node has a degree of 3. We index workers with $i$ and jobs with $j$. We denote the degree of the node representing worker $i$ $d_i$ and the degree of the job representing job $j$ $d_j$. In Figure \ref{fig:simple_bipartite_network}, $d_{i=1}=1$ and $d_{j=1}=3$. As we discuss below, a worker who changes jobs more frequently will have a higher degree, while a job which hires more  workers at a given time and/or has higher worker turnover will have a higher degree. 

In the next subsection, we show how our model generates a network of worker--job links similar to that in Figure \ref{fig:simple_bipartite_network}, which can be observed using linked employer-employee data.  Then, in the context of our model, we show how to back out latent worker types and markets from this observed network.

\subsection{Bipartite Network Details}

\label{sec:bipartite_network_details}

A network is a collection of nodes (also called ``vertices''), connected to each other by edges. A \emph{bipartite} network is a network in which there are two categories of nodes, and all edges connect a node of one category to a node of the other category. In our application, the two categories of nodes are workers and jobs, and all edges connect an individual worker to an individual job. Alternatively, we could have defined a coworker network in which all of the nodes represent individual workers, and an edge connects pairs of workers who are coworkers. The coworker network is not a bipartite network because any node can be connected via an edge to any other node. 

One way to represent a network is an adjacency matrix, typically denoted $\mathbf{A}$. The typical element of the adjacency matrix, $A_{ij}$, is the number of edges connecting nodes $i$ and $j$. If there are $n$ nodes in the network, then the adjacency matrix will have dimensions $n \times n$. In equation (\ref{eq:bipartite_A}) below, we present an adjacency matrix for a bipartite network. Notice that there are two large blocks of zeros. This reflects the fact that edges only connect edges of different categories. In our case, edges only connect workers to jobs, not jobs to jobs or workers to workers. Suppose there are $n_J$ jobs andd $n_W$ workers, where $n_J+n_W=n$. Jobs are indexed by $(1,\dots,n_J)$ and workers by $(n_J+1,\dots, n)$.

\begin{align*}
\mathbf{A} = 
\begin{blockarray}{ccccccc}
\multicolumn{3}{c}{Jobs} & \multicolumn{3}{c}{Workers} & \\
\multicolumn{3}{c}{$\overbrace{\rule{3cm}{0pt}}$} & \multicolumn{3}{c}{$\overbrace{\rule{3cm}{0pt}}$} & \\
\begin{block}{(cccccc)l}
0 		& \cdots	& 0 		&  A_{1,n_J+1} 	& \dots 	& A_{1,n} 	& \rdelim\} {3}{10pt} \multirow{3}{*}{Jobs } \\
\vdots	& \ddots	& \vdots	& \vdots 		& \ddots 	& \vdots 	& \\
0		& \hdots	& 0	 		& A_{n_J,n_J+1} & \dots 	& A_{n_J,n}	& \\
A_{n_J+1,1} & \hdots	& A_{n_J+1,n_J} & 0 	 	& \dots 	& 0 		&   \rdelim\} {3}{10pt}\multirow{3}{*}{Workers }	\\
\vdots		& \ddots	& \vdots		& \vdots 	& \ddots 	& \vdots	& \\
A_{n,1}		& \hdots	& A_{n,n_J}		& 0 	 	& \dots 	& 0 		& \\
\end{block}
\end{blockarray}
\label{eq:bipartite_A}
\end{align*}

We can also write the adjacency matrix as 
\begin{align*}
\mathbf{A} = 
\begin{pmatrix}
O^{n^J\times n^J} 	& A^{n^J \times n^W} \\ 
A^{n^W \times n^J}  & O^{n^W\times n^W}
\end{pmatrix}
\end{align*}
where $0^{n\times k}$ is an $n \times k$ matrix of zeros, $ A^{n^J \times n^W} = (A^{n^J \times n^W})^T$ and 
\begin{align*}
A^{n^J \times n^W} \equiv 
\begin{pmatrix}
A_{1,n_J+1}   & \dots         & A_{1,n}       \\
\vdots        & \ddots        & \vdots        \\
A_{n_J,n_J+1} & \dots         & A_{n_J,n}
\end{pmatrix}
\end{align*}

\subsection{Stochastic block model details}

\label{sec:SBM_details}
The \emph{stochastic} in stochastic block model indicates that edges in the network are drawn stochastically from a data generating process (DGP). The \emph{block} refers to the block structure of the DGP. Specifically, the SBM assumes that each node in the network belongs to a group $g \in 1,\dots,G$. The probability of an edge between two nodes depends solely on group memberships of the two nodes.\footnote{We have described that standard SBM, as opposed to the degree-corrected version. All of our analysis uses the degree-corrected version, however we ignore that here for simplicity of exposition.} Therefore, we can write a matrix of edge probabilities that has a block structure:

\begin{align*}
EdgeProbability &= 
\begin{blockarray}{ccccc}
g(i)=1 & g(i)=1 & g(i)=2 & g(i)=2 & \\
\begin{block}{(cccc)c}
p_{11} & p_{12} & p_{13} & p_{14} & g(i)=1  \\
p_{21} & p_{22} & p_{23} & p_{24} & g(i)=1  \\
p_{31} & p_{32} & p_{33} & p_{34} & g(i)=2  \\
p_{41} & p_{42} & p_{43} & p_{44} & g(i)=2  \\
\end{block}
\end{blockarray} \\
&= 
\begin{pmatrix}
p_{g_1,g_1} & p_{g_1,g_1} & p_{g_1,g_2} & p_{g_1,g_2}  \\
p_{g_1,g_1} & p_{g_1,g_1} & p_{g_1,g_2} & p_{g_1,g_2}  \\
p_{g_2,g_1} & p_{g_2,g_1} & p_{g_2,g_2} & p_{g_2,g_2}  \\
p_{g_2,g_1} & p_{g_2,g_1} & p_{g_2,g_2} & p_{g_2,g_2}  \\ 
\end{pmatrix}
\end{align*}

In this example, there are four nodes and two groups. Nodes 1 and 2 belong to group 1, as denoted by $g(1) = g(2) =1$. Similarly, nodes 3 and 4 belong to group 2: $g(3)=g(4)=2$. Instead of the edge probability matrix above, which can get quite large as the number of nodes grows, we can describe the matrix with two smaller objects: a vector indicating the group assignment of each node and a $G\times G$ matrix of group-specific edge propensities,\footnote{These are not technically probabilities but they can be normalized to be probabilities. \iffalse{\color{magenta} XX Clarify exactly what these are based on Bernardo's write-up.}\fi} where $G$ is the number of groups. We denote the vector of group assignments $\vec{g}$ and the matrix of group-specific edge propensities $\mathbf{\Omega}$. then
\begin{align*}
\vec{g} = \begin{bmatrix}	1 \\ 1 \\ 2\\ 2	\end{bmatrix}
\end{align*}
and 
\begin{align}
\mathbf{\Omega} = \begin{pmatrix}
p_{g_1,g_1} & p_{g_1,g_2} \\
p_{g_2,g_1} & p_{g_2,g_2}
\end{pmatrix}
\end{align}

Now we describe how to generate a network using the stochastic block model, given parameters. Let $\mathbf{A}$ be the adjacency matrix of a network with $n=4$ nodes and $\vec{g}$ and $\mathbf{\Omega}$ described above, with $\omega_{rs}$ representing an element of $\mathbf{\Omega}$. We assume that edges are placed between each pair of nodes, $i$ and $j$, following a Poisson distribution with mean equal to the edge probability corresponding to the nodes' respective groups: $\omega_{g_i,g_j}$. Therefore, the probability of drawing $A_{ij}$ edges between nodes $i$ and $j$ is
\begin{align*}
P(A_{ij}|\omega_{g_ig_j}, g_i, g_j) = \frac{(\omega_{g_ig_j})^{A_{ij}}}{A_{ij}!} \exp \left( -\omega_{g_ig_j}\right). 
\end{align*}
The probability is slightly different for self-edges (edges connecting a node to itself):\footnote{For more details, see section II of \citet{KarrerNewman2011}.}
\begin{align*}
P(A_{ii}|\omega_{g_ig_i}, g_i) = \frac{(\frac{1}{2}\omega_{g_ig_i})^{A_{ii}/2}}{(A_{ii}/2)!} \exp \left( -\frac{1}{2}\omega_{g_ig_i}\right). 
\end{align*}

The probability of observing the entire network, represented by $\mathbf{A}$, is the product of the probabilities of each element in the adjacency matrix:
\begin{align}
P(\mathbf{A}|\mathbf{\Omega}, \vec{g}) = \prod_{i < j} \frac{(\omega_{g_ig_j})^{A_{ij}}}{A_{ij}!} \exp \left( -\omega_{g_ig_j}\right) \times \prod_i \frac{(\frac{1}{2}\omega_{g_ig_i})^{A_{ii}/2}}{(A_{ii}/2)!} \exp \left( -\frac{1}{2}\omega_{g_ig_i}\right) \label{eq:SBM}
\end{align}
Although equation (\ref{eq:SBM}) presents the standard SBM, this formulation is rarely used in practice. For empirical applications, researchers typically use an extension called the \emph{degree-corrected} stochastic block model (DCSBM). The difference between the SBM and the DCSBM is that the DCSBM allows the expected degree of each node (the number of edges connected to that node) to vary. This more-closely matches real world data and the DCSBM has been shown to have far superior performance in empirical applications than the SBM \citep{KarrerNewman2011}. Let $\vec{d}$ be vector containing the degree of each node, with typical element $d_i$ representing the degree of node $i$. We can write the DCSBM as 
\begin{align}
P(\mathbf{A}|\vec{d}, \mathbf{\Omega}, \vec{g}) = \prod_{i < j} \frac{(d_id_j\omega_{g_ig_j})^{A_{ij}}}{A_{ij}!} \exp \left( -d_id_j\omega_{g_ig_j}\right) \times \prod_i \frac{(\frac{1}{2}d_i^2\omega_{g_ig_i})^{A_{ii}/2}}{(A_{ii}/2)!} \exp \left( -\frac{1}{2}d_i^2\omega_{g_ig_i}\right). \label{eq:DCSBM}
\end{align}

\subsection{Community detection using the stochastic block model}

In Section \ref{sec:SBM_details} we assumed that we know all of the parameters of the model: $\vec{d}$,  $\mathbf{\Omega}$, and $\vec{g}$. However, in actual applications, we typically observe the network $\mathbf{A}$ and the degree distribution $\vec{d}$ and want to recover the group memberships of the nodes $\vec{g}$. (Conditional on knowing $\vec{g}$, we can also compute the empirical edge probabilities matrix $ \mathbf{\hat\Omega}$.)  Therefore, we recover the group memberships of the nodes, $\vec{g}$, by treating equation (\ref{eq:DCSBM}) as a maximum likelihood problem and choosing the group memberships in order to maximize the probability of the observed adjacency matrix $\mathbf{A}$, given the data. We write the likelihood
\begin{align}
\mathcal{L}(A|\vec{g}) = \prod_{i < j} \frac{(d_id_j\omega_{g_ig_j})^{A_{ij}}}{A_{ij}!} \exp \left( -d_id_j\omega_{g_ig_j}\right) \times \prod_i \frac{(\frac{1}{2}d_i^2\omega_{g_ig_i})^{A_{ii}/2}}{(A_{ii}/2)!} \exp \left( -\frac{1}{2}d_i^2\omega_{g_ig_i}\right)
\end{align}
and our task is to choose
\begin{align*}
\hat{\vec{g}} = \arg\max_{\vec{g}} \mathcal{L}(A|\vec{g}) 
\end{align*}

\subsection{Bipartite stochastic block model details}

The bipartite stochastic block model (BiSBM) is an extension of the SBM (Section \ref{sec:SBM_details}) applied to bipartite networks (Section \ref{sec:bipartite_network_details}). The edge probability matrix has the same block structure as in the SBM, however since it is a bipartite network, there are two categories of nodes --- in our case workers and jobs --- and all edges connect a node from one category (a worker) to a node from the other (job).

Suppose there are two types of workers, indexed by $\i \in 1,2$, and two types of jobs, indexed by $\g\in 1,2$. Suppose further that there are 4 individual workers and 4 individual jobs, indexed by $i=1,\dots,4$ and $j=1,\dots,4$, respectively. There are two individual workers and two individual jobs of each type. Denote the probability of an edge between a type $\i$ worker and a job in market $\g$ as $\omega_{\ig}$. Then we have the following edge probability matrix
\begin{align*}
\begin{blockarray}{cccccccccc}
& \multicolumn{4}{c}{Jobs} & \multicolumn{4}{c}{Workers} & \\
& \multicolumn{4}{c}{$\overbrace{\rule{5cm}{0pt}}$} & \multicolumn{4}{c}{$\overbrace{\rule{5cm}{0pt}}$} & \\
& j=1     & j=2	& j=3	  & j=4     & i=1     & i=2     & i=3	  & i=4     & \} \text{Worker/Job Index}\\
& \g=1	& \g=1	& \g=2	& \g=2		& \i=1		& \i=1	& \i=2		& \i=2	& \}\text{Worker/market}\\
\begin{block}{c(cccccccc)l}
j=1,\g=1 & 0       & 0       & 0	   & 0	     & \o_{11} & \o_{11} & \o_{21} & \o_{21} & \rdelim\} {4}{10pt} \multirow{4}{*}{Jobs} \\
j=2,\g=1 & 0       & 0       & 0       & 0       & \o_{11} & \o_{11} & \o_{21} & \o_{21} & \\					  
j=3,\g=2 & 0       & 0       & 0       & 0       & \o_{12} & \o_{12} & \o_{22} & \o_{22} & \\     					  
j=4,\g=2 & 0       & 0       & 0       & 0       & \o_{12} & \o_{12} & \o_{22} & \o_{22} & \\ 					    
i=1,\i=1 & \o_{11} & \o_{11} & \o_{12} & \o_{12} & 0       & 0       & 0       & 0       & \rdelim\} {4}{10pt} \multirow{4}{*}{Workers} \\
i=2,\i=1 & \o_{11} & \o_{11} & \o_{12} & \o_{12} & 0       & 0       & 0       & 0       & \\					  
i=3,\i=2 & \o_{21} & \o_{21} & \o_{22} & \o_{22} & 0       & 0       & 0       & 0       & \\					  
i=4,\i=2 & \o_{21} & \o_{21} & \o_{22} & \o_{22} & 0       & 0       & 0       & 0       & \\					  				  
\end{block}
\end{blockarray}.
\end{align*}   
The primary takeaway from this matrix is that the probability of a connection between a pair of nodes is determined by their group memberships. If worker $i$ belongs to type $\i$ and job $j$ belongs to type $\g$, then the probability of worker $i$ matching with job $j$ is governed by $\omega_{\ig}$. The two blocks of zeros in this matrix reflect the fact that the probability of an edge between two workers or two jobs is zero in a bipartite network. 

We can write the DGP for the BiSBM as we did above for the standard or degree-corrected SBM. Here we will use the degree-corrected version, since that is what we use for estimation The probability of $A_{ij}$ edges between worker $i$ and job $j$ is given by
\begin{align*}
P(A_{ij}|\omega_{g_ig_j}, g_i, g_j, d_i, d_j) =\frac{(d_id_j\omega_{g_ig_j})^{A_{ij}}}{A_{ij}!} \exp \left( -d_id_j\omega_{g_ig_j}\right)
\end{align*}
From this, we can compute the likelihood of the full observed network, represented by the adjacency matrix $\mathbf{A}$. However, it is inmportant to note that the product below is only over pairs of nodes that \emph{belong to opposite categories}. That is, if $i$ indexes workers and $j$ indexes jobs, we are only taking the product over $i,j$ pairs, not $i,i'$ or $j,j'$ pairs. Again, this is because in a bipartite network, edges can only connect nodes that belong to different categories. 
\begin{align}
P(\mathbf{A}|\vec{d}, \mathbf{\Omega}, \vec{g}) = \prod_{i < j} \frac{(d_id_j\omega_{g_ig_j})^{A_{ij}}}{A_{ij}!} \exp \left( -d_id_j\omega_{g_ig_j}\right). 
\end{align}
Notice that this expression lacks the second term found in equation (\ref{eq:DCSBM}), which captures self-edges in which an edge runs connects a node to itself. This is because self-edges are impossible in a bipartite network, since self-edges would connect nodes belonging to the same category (e.g. workers to workers).



\clearpage

\subsection{Visual representation of linked employer-employee data as a network}

\label{sec:visual_representation}

\begin{figure}
	\caption{Representing the data as a network}
	\begin{minipage}{.5\textwidth}
		\centering
		\begin{tabular}{cc}
			\toprule
			WorkerID	& JobID	\\
			\midrule
			1	& 1\_1	\\
			2	& 1\_1	\\
			2	& 2\_1	\\
			3	& 1\_1	\\
			3	& 1\_2	\\
			4	& 1\_2	\\
			5	& 1\_1	\\
			5	& 2\_1	\\
			6	& 1\_2	\\
			6	& 3\_3	\\
			6	& 4\_3	\\
			7	& 4\_3	\\
			8	& 3\_3	\\
			8	& 4\_3	\\
			9	& 4\_3	\\
			9	& 2\_5	\\
			10	& 3\_3	\\
			10	& 4\_3	\\
			10	& 5\_4	\\
			10	& 2\_5	\\
			\bottomrule
		\end{tabular}
	\end{minipage}
	\begin{minipage}{.5\textwidth}
		\centering
		\includegraphics[width = \textwidth]{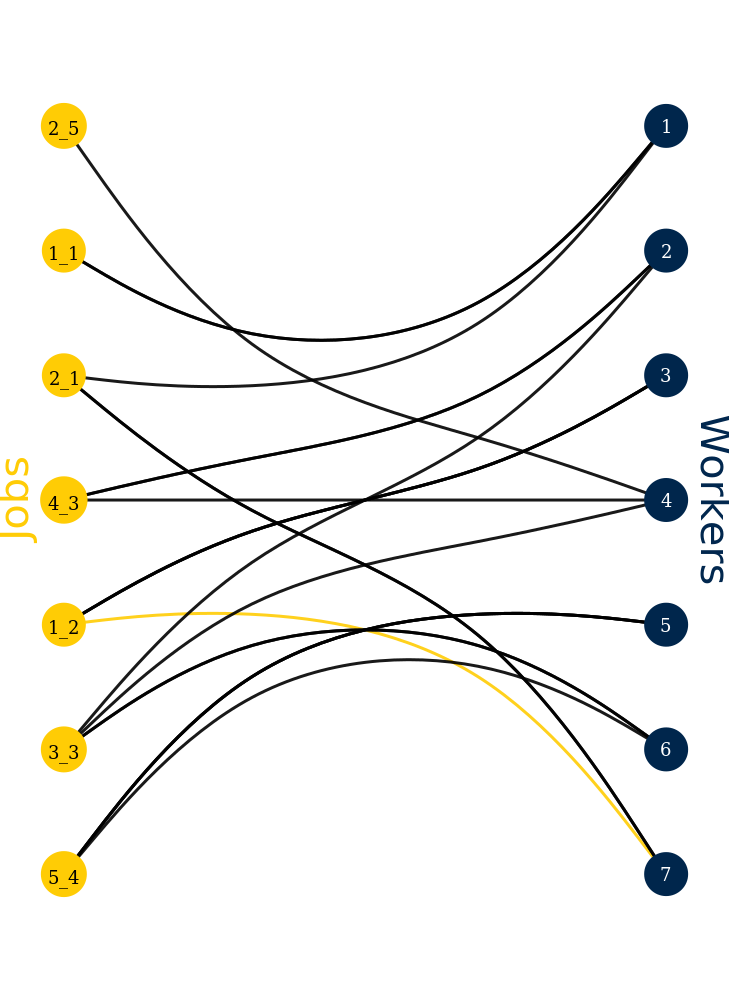} 
	\end{minipage}%
	\label{fig:network_representation}
\end{figure}

Our raw data looks like what is presented in Table \ref{table:sample_leed}, with the exception that we generate the ``JobID'' column ourselves by concatenating the establishment code (`Estab Code') and occupation code (`Occ Code'). However, we only use the two variables `WorkerID' and `JobID' in estimation. Therefore, in Figure \ref{fig:network_representation}, we show the worker and job IDs from the data alongside a network representation of the same data. In the network representation, workers are blue dots on the right, jobs are yellow dots on the left, and black lines represent edges connecting workers to jobs at which they were employed. Finally, in Table \ref{table:adjacency_appendix}, we present an adjacency matrix representation of the same network. 

\begin{table}[h!]
	\caption{Sample linked-employer-employee data}
	\centering
	\begin{tabular}{cccccc}
		\toprule
		WorkerID	& Establishment	& Occupation	& Estab Code	& Occ Code	& JobID	\\
		\midrule
		1	& Walmart	& Cashier	& 1	& 1	& 1\_1	\\
		2	& Walmart	& Cashier	& 1	& 1	& 1\_1	\\
		2	& Kroger	& Cashier	& 2	& 1	& 2\_1	\\
		3	& Walmart	& Cashier	& 1	& 1	& 1\_1	\\
		3	& Walmart	& Greeter	& 1	& 2	& 1\_2	\\
		4	& Walmart	& Greeter	& 1	& 2	& 1\_2	\\
		5	& Walmart	& Cashier	& 1	& 1	& 1\_1	\\
		5	& Kroger	& Cashier	& 2	& 1	& 2\_1	\\
		6	& Walmart	& Greeter	& 1	& 2	& 1\_2	\\
		6	& CVS		& Manager	& 3	& 3	& 3\_3	\\
		6	& Chipotle	& Manager	& 4	& 3	& 4\_3	\\
		7	& Chipotle	& Manager	& 4	& 3	& 4\_3	\\
		8	& CVS		& Manager	& 3	& 3	& 3\_3	\\
		8	& Chipotle	& Manager	& 4	& 3	& 4\_3	\\
		9	& Chipotle	& Manager	& 4	& 3	& 4\_3	\\
		9	& Kroger	& Asst. Mgr	& 2	& 5	& 2\_5	\\
		10	& CVS		& Manager	& 3	& 3	& 3\_3	\\
		10	& Chipotle	& Manager	& 4	& 3	& 4\_3	\\
		10	& Chili's	& Waiter	& 5	& 4	& 5\_4	\\
		10	& Kroger	& Asst. Mgr	& 2	& 5	& 2\_5	\\
		\bottomrule
	\end{tabular} 
	\label{table:sample_leed}
\end{table} 

\begin{table}[h!]
	\centering
	\caption{Adjacency matrix: $\mathbf{A}$}
	\begin{tabular}{c|cccccccc} 
		\toprule
		\textbf{Worker \textbackslash Job} & \textbf{1\_1} & \textbf{1\_2} & \textbf{2\_1} & \textbf{2\_5} & \textbf{3\_3} & \textbf{4\_3} & \textbf{5\_4}  \\
		\midrule
		\textbf{1}                         & 1             & 0             & 0             & 0             & 0             & 0             & 0   \\
		\textbf{2}                         & 1             & 0             & 1             & 0             & 0             & 0             & 0   \\
		\textbf{3}                         & 1             & 1             & 0             & 0             & 0             & 0             & 0   \\
		\textbf{4}                         & 0             & 1             & 0             & 0             & 0             & 0             & 0   \\
		\textbf{5}                         & 1             & 0             & 1             & 0             & 0             & 0             & 0   \\
		\textbf{6}                         & 0             & 1             & 0             & 0             & 1             & 1             & 0   \\
		\textbf{7}                         & 0             & 0             & 0             & 0             & 0             & 1             & 0   \\
		\textbf{8}                         & 0             & 0             & 0             & 0             & 1             & 1             & 0   \\
		\textbf{9}                         & 0             & 0             & 0             & 1             & 0             & 1             & 0   \\
		\textbf{10}                        & 0             & 0             & 0             & 1             & 1             & 1             & 1   \\
	\bottomrule
	\end{tabular}
	\label{table:adjacency_appendix}
\end{table}

\clearpage

\section{Estimating labor supply parameters}
\label{sec:MLE}

This section describes the procedure we use to estimate the labor supply parameters of the model, conditional on the assignments of workers to worker types, $\i(i)$, and jobs to markets, $\g(j)$, described in Section \ref{sec:bisbm}.

\subsection{Estimating $\Psi$ from observed matches}

\label{sec:estimating_psi}

Identification and estimation of the labor supply parameters builds upon  \citet{BonhommeLamadonManresa2019_distributional} and \citet{Grigsby2019}, with the key difference being that we assign both workers to worker types and jobs to markets prior to estimating labor supply parameters and do so in a way that more fully exploits the information revealed by worker--job matches, allowing us to identify a significantly greater degree of worker and job heterogeneity.\footnote{More precisely, \citet{BonhommeLamadonManresa2019_distributional} model workers matching with firms and therefore use k-means clustering to cluster firms on the basis of the firms' earnings distributions, while \citet{Grigsby2019} models workers matching with clusters of occupations identified by combining occupational education requirements with k-means clustering on the basis of occupations' O*NET skills scores. Additionally, neither \citet{BonhommeLamadonManresa2019_distributional} nor \citet{Grigsby2019} actually assign workers to types. Instead,  they employ random effects estimators, in which they identify the distribution of types, rather than assigning any individual worker to a type. As a result, both papers require that flows of worker types between firm/occupation groups form a strongly connected graph (they use the term ``connecting cycle''). This is a strong data requirement and requires them to define worker and firm/occupation groups at a relatively aggregated level, ignoring considerable heterogeneity. By using the network structure of the data to assign workers and jobs to types in a previous step before estimating labor supply parameters, we are able to identify an order of magnitude more worker types and markets, and therefore to allow for much greater heterogeneity.} 

We estimate parameters using a maximum likelihood approach. We assume that individual workers' earnings in period $t$ are observed with multiplicative measurement error $e_{it}$, which has a worker type--market-specific parametric distribution $f_e(e_{it} | \i(i), \g_{it}, \theta_e)$ with unit mean, summarized by parameter vector $\theta_e$. Observed earnings $\omega_{it}$ are therefore
\begin{align}
	\omega_{it} = \psi_{\i(i)\g_{it}} w_{\g_{it}} e_{it}. \label{eq:observed_wages}
\end{align}
Finally, we assume that the earnings measurement errors are serially independent:
\begin{assumption}[Serial independence of earnings measurement error]
	\label{ass:serial_independence}
	The realization of period $t$'s measurement error for worker $i$, $e_{it}$ is independent of the history of errors $\{e_{it'}\}_{t'=1}^{t-1}$, market choices $\{\g_{it'}\}_{t'=1}^{t-1}$, and separations $\{c_{it'}\}_{t'=1}^{t-1}$, conditional on the worker's type, $\i_i$, and current market choice $\g_{it}$.
\end{assumption}

Our model is identified by combining assumption \ref{ass:serial_independence} with assumptions \ref{ass:taste_shocks} and \ref{ass:mobility}, which stated that the market preference parameters $\ve_{i\g t}$ and exogenous separation shocks $c_{it}$ are each serially uncorrelated and independent of all other variables in the model.

Conditional on clustering workers and jobs into types, our data consist of three elements per worker per period: the worker's market choice, $\gamma_{it}$, the worker's earnings, $\omega_{it}$, and the indicator for whether or not the worker changed jobs, $c_{it}$. Observed data are denoted by $\mathbb{X} := \left\{ \gamma_{it},\omega_{it},c_{i t} | t = 1, \ldots, T; i = 1, \ldots, N \right\}$. The parameters are denoted by $\Theta := \left\{ \psi_{\ig}w_\g, \xi_{\g}, \nu, \theta_e | \i = 1,\ldots, I; \g = 1, \ldots, \Gamma  \right\}$. Recall that $\P[\g_{it}|\Theta]$ is the probability of worker $i$ choosing a job in market $\g$ and comes from the Roy model (equation \ref{eq:emp_probs}). Meanwhile, let $f_{\omega}(\omega| \i(i), \g_{it}, \Theta)$ denote the density of observed earnings in period $t$. We construct our likelihood as follows. 

In periods in which workers experience a separation, three pieces of data are generated: a separation indicator $c_{it}$, the worker's new market choice $\g_{it}$, and the worker's earnings $\omega_{it}$. We assume that all workers separate and rematch in the first period for which we have data: $c_{i1}=1$ for all $i$. In periods in which the worker does not separate from their job, we observe only $c_{it}$ and  $\omega_{it}$.\footnote{By only including the worker's market choice in the likelihood in periods in which a separation has occurred, but assuming that all workers separated in period $t=1$, we are ensuring that each match enters the likelihood exactly once. This gives all matches equal weight in the likelihood, regardless of match duration. Alternatively, we could have omitted exogenous separations from the model and assumed that workers make a new choice every period. Under this assumption, persistent matches would indicate that the worker has made the same choice repeatedly and we would put greater weight on persistent matches in estimation.}  Assumptions \ref{ass:mobility} and \ref{ass:meas_error} tell us that realizations of $\omega_{it}$ and $c_{it}$ are independent, and $\g_{it}$ is independent of $\omega_{it}$ conditional on $c_{it}$. Therefore, we write the likelihood of observing $\{\g_{it},\omega_{it}, c_{it}\}$ for an individual worker in period $t$ as
\begin{align*}
	l(\g_{it},\omega_{it}, c_{it}|\mathbb{X}) &= \underset{\text{Separation}}{\underbrace{\left[ f_{\omega}(\omega_{it}|\Theta) \P(\g_{it}|\Theta) \right] }}^{c_{it}}\underset{\text{No separation}}{\underbrace{\left[ f_{\omega}(\omega_{it}|\i(i), \g_{it},\Theta)\right]}}^{1-c_{it}}
\end{align*}

Our assumptions that $\{\g_{it},\omega_{it}, c_{it}\}$ are serially uncorrelated and independent across workers, conditional on the parameters of the data, allow us to write the full likelihood of the data as the product of the individual worker-time likelihoods:

\begin{align}
	\mathcal{L}(\Theta | \mathbb{X}) 	&=  \prod_{i = 1}^N  \prod_{t=1}^{T} l(\g_{it},\omega_{it}, c_{it}|\mathbb{X}) \nonumber \\
	&= \prod_{i = 1}^N  \prod_{t=1}^{T} \underset{\text{Separation}}{\underbrace{\left[ \P(\g_{it}|\Theta)f_{\omega}(\omega_{it}|\i(i), \g_{it},\Theta) \right] }}^{c_{it}}\underset{\text{No separation}}{\underbrace{\left[ f_{\omega}(\omega_{it}|\i(i), \g_{it},\Theta)\right]}}^{1-c_{it}}  
\end{align}
Finally, the log-likelihood is
\begin{align}
	\ell(\Theta | \mathbb{X}) 	&=  \sum_{i=1}^{N}\sum_{t = 1}^T c_{it} \log \P(\g_{it}|\Theta) + \sum_{i=1}^{N}\sum_{t = 1}^T \log f_{\omega}(\omega_{it}|\i(i), \g_{it}, \Theta) \label{eq:log_likelihood}
\end{align}

In order to maximize this likelihood function, we impose a distributional assumption and a normalization:

\begin{assumption}[Distribution of measurement error in wages]
	\label{ass:meas_error}
	$e_{it}$ has a log-normal distribution: $\ln e_{it} \sim \N(0,\sigma_{\ig})$.
\end{assumption}

\begin{assumption}[$\Psi$ normalization] 
	\label{ass:grigsby_normalization}
	The mean productivity level in each market $\g$ is normalized to a constant, $k$:
	\[\sum_{\i} m_{\i} \psi_{\ig} = k \quad \forall \g\]
	where $m_{\i}$ is the mass of type $\i$ workers.
\end{assumption}

Assumption \ref{ass:meas_error} assumes that wages follow a log-normal distribution which is worker type-market specific, following \citet{BonhommeLamadonManresa2019_distributional} and \citet{Grigsby2019}. Assumption \ref{ass:grigsby_normalization} normalizes the $\psi_{\ig}$ to have a mean equal to some constant $k$ within market.

Identification of $\Psi$ comes from two sources: earnings for all employed workers, and market choices for all workers in period $t=1$ and workers who receive exogenous separation shocks in periods $t>1$. Intuitively, ($\i,\g$) matches that pay more and occur more frequently are revealed to be more productive. The relative weight of earnings and market choices is determined by the inverse of the variances of measurement error in wages and idiosyncratic shocks --- if the earnings measurement error $\sigma_{\ig}$ for a worker type--market pair has a relatively high variance, then estimation puts more weight on choices; if the idiosyncratic preference shocks have a relatively high variance (large $\nu$), estimation puts more weight on earnings. The normalization that the mean skill level in each market equals $k$ (Assumption \ref{ass:grigsby_normalization}) converts the distribution of relative skills into a distribution of skill levels. We choose $k$ to maximize the model's ability to match the observed employment rate.\footnote{This normalization is mostly without loss of generality. If one were to double the number of efficiency units of labor each worker supplied to a market, the equilibrium price of labor would halve. However, increasing the number of efficiency units of labor in the economy will impact the fraction of the labor force in employment versus non-employment. This is why we choose $k$ to maximize the model's ability to match the observed employment rate.}

The parameter governing the variance of non-pecuniary benefits, $\nu$, is identified by workers' choices of markets, $\g$. Workers will choose a market that offers their worker type low expected utility (low $\psi_{\ig} w_{\g} + \xi_{\g}$) when they receive a large preference shock draw for that market. Therefore, if workers frequently choose low expected utility markets, it must be because they frequently draw large preference shocks, indicating that the preference shock distribution has a large dispersion parameter, $\nu$. The market amenities parameter $\xi_{\g}$ is a market fixed effect and is identified by the component of the frequency with which workers choose market $\g$ that is common across all worker types $\i$. The relative value of $\xi_{\g}$ to $\xi_{\g'}$ allows the model to match the fact that some high-earning markets, such as doctors, account for a small share of total employment. This is because  $\xi_{\g}$ reflects not just the immediate utility benefits of working in a job in market $\g$, but also reflects broader compensating differentials. In this way, $\xi_{doctor}$ may be low, not because doctor jobs are unpleasant, but because the annualized cost of becoming a doctor --- including medical school --- and maintaining the requisite skills is high. We provide greater detail on identification in appendix \ref{sec:identification_details}.

\subsection{Additional parameters to be estimated or calibrated}

\label{sec:calibration}

We also have the following parameters to estimate or calibrate:
\begin{itemize}
	\item $\beta_{\g s}$ (output elasticity of labor in market $\g$) ---  We calibrate these parameters as the share of the sector $S$ wage bill paid to workers employed in market $\g$ jobs. 
	\item $\eta$ (CES consumption substitution elasticity)--- We calibrate this parameter to 2.\footnote{\citet{BrodaWeinstein2006} estimate this parameter to be 4, however their estimate comes from significantly more disaggregated product categories, so we choose a smaller value. This parameter affects our structural results in Section \ref{sec:model_fit}, but does not affect the reduced form estimates in Section \ref{sec:reduced_form}.}
	\item $a_s$ (demand shifters) --- We calibrate demand shifters to match actual sector output shares, given sector-level prices, for the state of Rio de Janeiro as measured by the Brazilian Institute of Geography and Statistics (IBGE).
\end{itemize}

\clearpage

\section{Model Solution Appendix}

\label{sec:model_solution}

\textbf{Firm's problem}

\emph{This section describes a slightly different version of the firm's problem than we presented in the body of the paper. In the body of the paper we had a set of competitive firms in each sector, whereas in what follows here we have a single representative firm in each sector. }

\begin{align}
\max_{\ell_{\g s}} \quad p_s \prod_{\g} \ell_{\g s}^{\beta_{\g s}} - \sum_{\g} w_{\g} \ell_{\g s}
\end{align}

There are $S$ optimizations with $\Gamma$ choice variables each, giving us $S \times \Gamma$ FOCs.

FOC:
\begin{align}
\ell_{\g s}^D = \frac{p_s \beta_{\g s}\left( \prod_{\g'} {\ell_{\g' s}^D}^{\beta_{\g' s}}\right)}{w_{\g}} \label{eq:firm_FOC}
\end{align}

Combining the $\Gamma$ FOCs for a given sector $S$:
\begin{align}
\ell_{\g s}^D = \frac{\beta_{\g s}}{\beta_{\g' s}}\frac{w_{\g'}}{w_{\g}} \ell_{\g's}^D \label{eq:firm_FOC_ratio}
\end{align}

Plugging in \ref{eq:firm_FOC_ratio} for $\ell_{\g s}^D$ in equation \ref{eq:firm_FOC}, we have
\begin{align}
\ell_{\g s}^D = \left[ p_s \left( \frac{\beta_{\g s}}{w_{\g }} \right)^{1-\sum_{\g'} \beta_{\g' s}} \prod_{\g'} \left(\frac{\beta_{\g' s}}{w_{\g'}} \right)^{\beta_{\g' s}} \right]^{\frac{1}{1-\sum_{\g'} \beta_{\g' s}}} = \ell_{\g s}^D \left(\vec{p}, \vec{w} \right) \label{eq:labor_demand}
\end{align}
which represents labor demand for firm $s$, using only FOCs for firm $s$.\footnote{We could alternatively write this expression as 
	\begin{align*}
	\ell_{\g s}^D =   \left( \frac{\beta_{\g s}}{w_{\g }} \right) \left[ p_s \prod_{\g'}  \left(\frac{\beta_{\g' s}}{w_{\g'}} \right)^{\beta_{\g' s}} \right]^{\frac{1}{1-\sum_{\g'} \beta_{\g' s}}}
	\end{align*}
	}

Since labor is the only factor of production, we can write firm $s$'s product market supply as 
\begin{align}
y_s^S = y_s^S\left( \{\ell_{\g s}^D (\vec{p}, \vec{w}) \}_{\g=1}^{\Gamma} \right) = \prod_{\g} {\ell_{\g s}^D}^{\beta_{\g s}} \label{eq:product_supply}
\end{align}

\textbf{Household's problem}

\begin{align*}
\max_{\{y_s^D\}_{s=1}^S} \quad \underset{U(\{y_s^D\}_{s=1}^S)}{\underbrace{ \left(\sum_s a_s^{\frac{1}{\eta}} {y_s^D}^{\frac{\eta-1}{\eta}}\right)^{\frac{\eta}{\eta-1}} }} \quad \text{s.t.} \quad \sum_s p_s y_s \leq Y
\end{align*}

Lagrangean:
\begin{align*}
\underset{U(\vec{y}^D)}{\underbrace{ \left(\sum_s a_s^{\frac{1}{\eta}} {y_s^D}^{\frac{\eta-1}{\eta}}\right)^{\frac{\eta}{\eta-1}} }} - \lambda \left(  \sum_s p_s y_s - Y \right)
\end{align*} 

FOC:
\begin{align*}
\frac{\eta}{\eta-1} U^{\frac{1}{\eta}} \frac{\eta-1}{\eta} a_s^{\frac{1}{\eta}} {y_s^D}^{-\frac{1}{\eta}} - \lambda p_s = 0
\end{align*}

Simplifying, 
\begin{align*}
U^{\frac{1}{\eta}} a_s^{\frac{1}{\eta}}  {y_s^D}^{-\frac{1}{\eta}} - \lambda p_s = 0
\end{align*}
Rearranging, 
\begin{align}
y_s^D = \frac{U}{\lambda^{\eta}} \frac{a_s}{p_s^{\eta}} \label{eq:hh_FOC}
\end{align}
Next, we plug this into the constraint satisfied with equality ($\sum_s p_s y_s^D = Y $):
\begin{align*}
&\frac{U}{\lambda^{\eta}} \sum_s \left( a_s p_s^{1-\eta} \right) = Y \\
\Rightarrow & \lambda^{\eta} = \frac{U}{Y} \sum_{s'} \left( a_s' p_s'^{1-\eta} \right)
\end{align*}

Plugging this into \ref{eq:hh_FOC}, we have our expression for product demand:
\begin{align}
y_s^D = \frac{ a_s Y}{p_s^{\eta} \sum_{s'} \left(a_{s'} p_{s'}^{1-\eta}\right)} = y_s^D(\vec{p},Y) \label{eq:product_demand}
\end{align}

\textbf{Worker's problem}

\begin{align*}
\max_{\g} \quad w_{\g} \psi_{\ig} + \xi_{\g} + \ve_{i\g}, \quad \ve_{i\g} \sim T1EV(\theta)
\end{align*}

Solving the worker's problem gives labor supply:
\begin{align}
\ell_{\g}^S(\vec{w})  = \sum_{\i} m_{\i} \left( \frac{\exp \left( \frac{\psi_{\ig} w_{\g} + \xi_{\g}}{\nu} \right) }{ \sum\limits_{\g'=0}^{\Gamma} \exp \left( \frac{\psi_{\ig'} w_{\g'} + \xi_{\g'}}{\nu} \right) } \right) \psi_{\ig} \label{eq:labor_supply}
\end{align}

\textbf{Equilibrium}

Equilibrium wages $\vec{w}_{\Gamma \times 1}$ and prices $\vec{p}_{S \times 1}$ must satisfy three market clearing conditions:
\begin{enumerate}
	\item Labor market:
	\begin{align*}
	\sum_s \ell_{\g s}^D = \ell_{g}^S \quad \forall \g\in\{1,\dots,\Gamma\}
	\end{align*}
	\item Product market:
	\begin{align*}
	y_s^D = y_s^S  \quad \forall s \in \{1,\dots,S \}
	\end{align*}
	\item Spending = Income = Wages + Profits
	\begin{align*}
	Y \equiv  \sum_s p_s y_s^D = W + \Pi \equiv \sum_s p_s y_s^S
	\end{align*}
\end{enumerate}
where
\begin{enumerate}
	\item Product demand:
	\[ y_s^D = \frac{ a_s Y}{p_s^{\eta} \sum_{s'} \left(a_{s'} p_{s'}^{1-\eta}\right)} \]
	\item Product supply:
	\[ y_s^S = \prod_{\g} {\ell_{\g s}^D}^{\beta_{\g s}} \]
	\item Labor supply:
	\[ \ell_{\g}^S(\vec{w})  = \sum_{\i} m_{\i} \left( \frac{\exp \left( \frac{\psi_{\ig} w_{\g} + \xi_{\g}}{\nu} \right) }{ \sum\limits_{\g'=0}^{\Gamma} \exp \left( \frac{\psi_{\ig'} w_{\g'} + \xi_{\g'}}{\nu} \right) } \right) \psi_{\ig}  \]
	\item Labor demand:
	\[ \ell_{\g s}^D = \left[ p_s \left( \frac{\beta_{\g s}}{w_{\g}} \right)^{1-\sum_{\g'} \beta_{\g' s}} \prod_{\g'} \left(\frac{\beta_{\g' s}}{w_{\g'}} \right)^{\beta_{\g' s}} \right]^{\frac{1}{1-\sum_{\g'} \beta_{\g' s}}} \]
	\item Budget (which can be plugged in for $Y$ in the product demand equation)
	\[ Y = \sum_s p_s y_{\g s}^S \]
\end{enumerate}

This is enough for equilibrium, which we find numerically using fixed point iteration. The algorithm proceeds as follows:
\begin{enumerate}
	\item Choose vectors of start values for wages $\vec{w}$ and prices $\vec{p}$
	\item Compute labor supply $\ell_{\g}^S(\vec{w})$ given wages $\vec{w}$ following equation \ref{eq:labor_supply}
	\item Compute labor demand $\ell_{\g s}^D \left(\vec{p}, \vec{w} \right)$ given these start values following equation \ref{eq:labor_demand}
	\item Compute the product supply $y_s^S\left( \{\ell_{\g s}^D (\vec{p}, \vec{w}) \}_{\g=1}^{\Gamma} \right)$ implied by the labor demand choice in the previous step following equation \ref{eq:product_supply}
	\item Compute household income $Y = \sum_s p_s y_{\g s}^S$ implied by product supply in the previous step
	\item Compute product demand $y_s^D(\vec{p},Y)$ following equation \ref{eq:product_demand}
	\item Update prices using the update rule $p_s^{t+1} = p_s^{t} \left(\frac{y_s^D}{y_s^S}\right)^{\rho}$, where $\rho$ is a dampening factor that controls the size of the update and $t$ indexes iterations. Intuitively, we increase prices if demand exceeds supply, and decrease them if supply exceeds demand. The size of the update depends on the size of the mismatch between supply and demand. 
	\item Update wages using the update rule $w_{\g}^{t+1} = w_{\g}^{t} \left(\frac{\ell_{\g}^D}{\ell_{\g}^S}\right)^{\rho}$
	\item Repeat steps 2-8 until convergence
\end{enumerate}


\section{Choosing number of worker types and markets}

\label{sec:MDL_details}

Equation \ref{eq:BiSBM} defined the probability of observing our network of worker--job matches, denoted by the adjacency matrix $\mathbf{A}$:
\begin{align} 
P \bigg(\mathbf{A} \bigg|\vec{\i}, \vec{\g}, \vec{d_i}, \vec{d_j} , \mathbf{\mathcal{P}} \bigg)  
&= \prod_{ i,j } \frac{\left(d_i d_j \mathcal{P}_{\i(i)\g(j)}\right)^{A_{ij}}}{A_{ij}!} \exp \left(d_i d_i^J \mathcal{P}_{\i(i)\g(j)} \right) . 
\end{align} 
As \citet{Peixoto2017} shows, we can think of this in Bayesian terms and write the full joint distribution of the data, $\mathbf{A}$, and the parameters,  $\vec{\i}$, $\vec{\g}$, $\vec{d_i}$, and $\vec{d_j}$ as 
\begin{align} 
P \bigg(\mathbf{A}, \vec{\i}, \vec{\g}, \vec{d_i}, \vec{d_j} , \mathbf{\mathcal{P}} \bigg) = 
P \bigg(\mathbf{A} \bigg|\vec{\i}, \vec{\g}, \vec{d_i}, \vec{d_j} , \mathbf{\mathcal{P}} \bigg)   
P \bigg(\vec{d_i}, \vec{d_j}  \bigg|\vec{\i}, \vec{\g}, \mathbf{\mathcal{P}} \bigg) 
P \bigg(\mathbf{\mathcal{P}} \bigg|\vec{\i}, \vec{\g} \bigg)  
P \bigg(\vec{\i}, \vec{\g} \bigg)    \label{eq:bisbm_bayesian}
\end{align}
where $P \bigg(\vec{d_i}, \vec{d_j}  \bigg|\vec{\i}, \vec{\g}, \mathbf{\mathcal{P}} \bigg) $, $P \bigg(\mathbf{\mathcal{P}} \bigg|\vec{\i}, \vec{\g} \bigg)$, and $P \bigg(\vec{\i}, \vec{\g} \bigg) $ are prior probabilities. 

It turns out that this Bayesian formulation has an equivalent information-theoretic interpretation. We can rewrite the joint probability defined in equation (\ref{eq:bisbm_bayesian}) as 
\begin{align*}
	P \bigg(\mathbf{A}, \vec{\i}, \vec{\g}, \vec{d_i}, \vec{d_j} , \mathbf{\mathcal{P}} \bigg) = 2^{-\Sigma}
\end{align*}
where
\begin{align*}
	\Sigma =-\log_2 P \bigg(\mathbf{A}, \vec{\i}, \vec{\g}, \vec{d_i}, \vec{d_j} , \mathbf{\mathcal{P}} \bigg) = \mathcal{S} + \mathcal{L}
\end{align*}
is called the description length of the data and represents the number of bits necessary to encode the data. 
\begin{align*}
	\mathcal{S} = - \log_2 P \bigg(\mathbf{A} \bigg|\vec{\i}, \vec{\g}, \vec{d_i}, \vec{d_j} , \mathbf{\mathcal{P}} \bigg)  
\end{align*}
represents the number of bits necessary to encode the model, conditional on knowing the model parameters, and
\begin{align*}
\mathcal{L} = - \log_2 P \bigg(\vec{\i}, \vec{\g}, \vec{d_i}, \vec{d_j} , \mathbf{\mathcal{P}} \bigg)  
\end{align*}
is the number of bits necessary to encode the model parameters.  $\mathcal{S}$ will be small if the model fits the data well, and $\mathcal{L}$ will be small if the complexity of the model (in our case, the number of worker types and markets) is small.  This implicitly defines a trade-off. As we add more worker types and markets, the model fits the data better, reducing  $\mathcal{S}$; however, we are increasing the complexity of the model and thereby increasing $\mathcal{L}$. MDL resolves this trade-off by minimizing  $\mathcal{S}+\mathcal{L}$.

We choose the assignment of workers to worker types and jobs to markets that maximizes the posterior of the distribution, equation (\ref{eq:bisbm_bayesian}). This is equivalent to choosing the set of parameters that yields the smallest description length, and therefore compresses the data the most. Intuitively, we can think of $\mathcal{L}$ as a penalty term that increases with the number of parameters, and thereby prevents overly complex models. If the number of worker types and markets becomes large, $\mathcal{S}$ will increase, indicating a better model fit, but the penalty term $\mathcal{L}$ will increase as well. The chosen model will therefore be the one that maximizes the quality of the model fit relative to the cost imposed by the penalty term.

For more detail, see \citet{Peixoto2014} and \citet{GerlachPeixotoAltmann2018}.


\section{Identification of Labor Supply Parameters}

\label{sec:identification_details}

Taking  the first order conditions of equation \ref{eq:log_likelihood} with respect to each of the parameters provides intuition for how the parameters are identified. 

\subsection{$\nu$}

\begin{flalign*}
\ell_\nu = 0 &\Rightarrow \sum_{i=1}^{N}\sum_{t = 1}^T  c_{it} \left[ \sum_{\g'} \P(\g'|\Theta)  (\phi_{\i_i\g'} +\xi_{\g'})  - (\phi_{\ig_{it}} +\xi_{\g_{it}}) \right] = 0 \\
\end{flalign*}

Intuitively, $\nu$ will be larger if more workers' actual market choices deviate from the choice those workers would have made in the absence of the preference shock $\ve$. The first term in the bracket, $\sum_{\g'} \P(\g'|\Theta)  (\phi_{\i_i\g'} +\xi_{\g'})$ is the expected systematic (excluding the idiosyncratic component, $\ve$) utility of the optimal market choice for worker $i$ and, and the second term, $\phi_{\ig_{it}} +\xi_{\g_{it}}$ is the systematic utility for worker $i$ in the market they actually chose in period $t$. Intuitively, if this difference is large, it must be because some workers received large idiosyncratic preference shocks, $\ve_{i\g t}$, which caused them to accept otherwise suboptimal jobs and is indicative of a large $\nu$. We can also see this by taking limits. If $\nu$ goes to zero, the $\P(\g|\Theta)$ degenerates to a single point and therefore the difference inside the brackets would be zero. On the other hand, as $\nu$ goes to infinity, the market choice probabilities converge to a uniform distribution and the differences between expected and realized systematic utility will be large.

\subsection{$\xi_{\g}$}

\begin{flalign*}
\ell_{\xi_\g} 	= 0 &\Rightarrow \sum_{i=1}^{N}\sum_{t = 1}^T  c_{it}  \mathbbm{1} \{ \g_{it} = \g \} - \sum_{i=1}^{N}\sum_{t = 1}^T  c_{it} \P(\g|\i_i ;\Theta) = 0 
\end{flalign*}

The above expression chooses $\xi$, which enters the expression through $\P(\g|\i_i;\Theta)$, in order to equate the fraction of job switchers observed to choose market $\g$ with the probability that a given job-switcher would choose $\g$. In otherwords, $\xi$ is identified by market choices.

\subsection{$\phi_{\ig}$}

\begin{flalign*}
\ell_{\phi_{\ig}} = 0 
&\Rightarrow \frac{1}{\sigma^2}\sum_{i=1}^{N}\sum_{t = 1}^T \frac{\log \omega_{it}-\log \phi_{\ig_{it}}}{\phi_{\ig_{it}}}\mathbbm{1} \{ \g_{it} = \g, \i_i = \i \} + \\
&\qquad \qquad + \frac{1}{\nu}\sum_{i=1}^{N}\sum_{t = 1}^T  c_{it} \mathbbm{1}\{ \i_i = \i \} \left[ \mathbbm{1}\{ \g_{it} = \g \} - \P(\g_{it}|\i_i; \Theta) \right] = 0 
\end{flalign*}

The above expression is highly intuitive. It tells us that identification of $\phi_{\ig}$ comes from two sources: earnings for all workers (first term), and market choices for job-switchers (second term). The first term is minimized when $\log \phi_{\ig}$ is close to actual log-earnings $\log \omega_{\it}$. The second term is minimized when the theoretical probability of a type $\i$ job-switcher choosing a job in market $\g$ equals the fraction of type $\i$ job-switchers who actually choose market $\g$ jobs. The relative weight of these terms in calculating the likelihood is determined by the variances of measurement error in wages and idiosyncratic shocks, $\sigma^2$ and $\nu$, respectively. Specifically, if wages are observed with considerable error (large $\sigma^2$) then we put more weight on the second term, which is identified by job changes. On the other hand, if the idiosyncratic preferences have high variance (large $\nu$), then wages are more informative than job changes.

Another thing to notice is that in cases where we observe no matches for a particular ($\i,\g$) pair, identification comes purely from the second term (because $\mathbbm{1}\{\g_{it} =\g,\i_i=\i\}=0$ in the first term). This makes sense, because we do not observe wages for matches that do not occur. Identification based on job choices in the second term relies on the assumption of a T1EV-distributed preference parameter. This is because, in order to achieve a choice probability of zero to match the count of observed matches, $\phi_{\ig} + \xi_{\g}$ will be forced towards $-\infty$. In practice, we will do something to handle zeros because we do not want to set $\phi_{\ig}+ \xi_{\g} =-\infty$. This allows us to achieve identification of the entire $\Phi$ matrix despite sparsity in observed ($\i,\g$) matches, although identification for sparse parts of $\Phi$ relies strongly on functional form assumptions. While identification based on functional form assumptions is suboptimal, we are doing so primarily for ($\i,\g$) pairs that rarely match, so imprecise estimation of these parameters will have minimal effect on our actual results. On the other hand, moving away from non-parametric identification allows us to identify a much higher degree of productivity heterogeneity.

More technically, if an ($\i,\g$) cell has zero matches, i.e. if $\mathbbm{1}\{ \g_{it} = \g, \i_i = \i \} = 0$ for all $i,t$, then the FOC above will be reduced to $ \sum_{i=1}^{N}\sum_{t = 1}^T  c_{it} \mathbbm{1}\{ \i_i = \i \} \P(\g_{it}|\i_i; \Theta) = 0$. This implies that there is no solution to the MLE problem, as $\phi_{\ig} + \xi_{\g}$ would have to go to minus infinity to make the FOC equation zero. A potential way to handle this is to add a small positive constant inside the last FOC brackets multiplied by the indicator $ \mathbbm{1} \left\{\sum_{i=1}^{N}\sum_{t = 1}^T  c_{it} \mathbbm{1}\{ \g_{it} = \g, \i_i = \i \}  = 0\right\}$.

\subsection{$\lambda$}

Note that we have dropped $\ig$ subscripts here, but the estimation would be approximately the same with the subscripts.

\begin{flalign*}
\ell_\lambda 	= 0 &\Rightarrow \frac{1}{\lambda} \left(\sum_{i=1}^{N}\sum_{t = 2}^T c_{it}  \right)  - \frac{1}{1-\lambda} \left((T-1)N-\sum_{i=1}^{N}\sum_{t = 2}^T c_{it} \right) = 0 \\
&\Rightarrow (1-\lambda) \left(\sum_{i=1}^{N}\sum_{t = 2}^T c_{it}  \right) = \lambda \left((T-1)N-\sum_{i=1}^{N}\sum_{t = 2}^T c_{it} \right) \\
&\Rightarrow \left(\sum_{i=1}^{N}\sum_{t = 2}^T c_{it}  \right) = \lambda (T-1)N \\
&\Rightarrow \hat{\lambda} = \frac{\sum_{i=1}^{N}\sum_{t = 2}^T c_{it}}{(T-1)N}
\end{flalign*}

\subsection{$\sigma$}

Again, we have dropped $\ig$ subscripts here, but the estimation would be approximately the same with the subscripts.

We proceed taking derivatives w.r.t. $\sigma$, knowing that $f_{\omega}(\omega|\Theta) = \frac{1}{\omega\sigma\sqrt{2\pi}}e^{-\frac{1}{2} \left(\frac{\log \omega - \log \phi_{\ig}}{\sigma}\right)^2} = \frac{1}{\omega\sigma} \phi\left(\frac{\log \omega - \log \phi_{\ig}}{\sigma}\right)$ and that $\log f_{\omega}(\omega|\Theta) = -\log(\omega \sqrt{2\pi})-\log\sigma -\sigma^{-2}\frac{1}{2} \left(\log \omega - \log \phi_{\ig}\right)^2$

\begin{flalign*}
\ell_\sigma 	= 0 &\Rightarrow \sum_{i=1}^{N}\sum_{t = 1}^T \frac{\partial \log f_{\omega}(\omega_{it}|\Theta)}{\partial \sigma} = 0 \\
&= -\frac{NT}{\sigma} + \sigma^{-3} \sum_{i=1}^{N}\sum_{t = 1}^T \left(\log \omega_{it} - \log \phi_{\ig_{it}}\right)^2 = 0\\
&\Rightarrow \hat\sigma^2 = \sum_{i=1}^{N}\sum_{t = 1}^T \frac{\left(\log \omega_{it} - \log \hat\phi_{\ig_{it}}\right)^2}{NT}
\end{flalign*}

\section{Measurement error}

\label{app:misclassification}

\def\tx{\tilde X}
\def\ty{\tilde Y}

The Bartik regressions in equation \ref{eq:bartik_reg} can be written
\begin{align*}
	\Delta Earnings_g = \beta_0 + \beta_1 Bartik_g + \ve_g
\end{align*}
where
\begin{align*}
	Bartik_{g} = \sum_m \left( Exposure_{g m} \times Shock_m \right)
\end{align*}
The earnings variable depends only on worker classifications, $g$, however the Bartik instrument depends on both worker and job classifications, $g$ and $m$. This means that worker classification error will affect both the LHS and the RHS, while job classification error will affect only the RHS. 

For simplicity, Let $Y=\{\Delta Earnings_g\}_{g=1}^G$, $X=\{Bartik_g\}_{g=1}^G$, and $U=\{\ve_g\}_{g=1}^G$. Then our regression model is 
\[ Y = X \beta + U\]
However we measure both $X$ and $Y$ with additive measurement error, $V_X$ and $V_Y$. Denote our measures of $X$ and $Y$, $\tilde X$ and $\tilde Y$, respectively, where
\begin{align*}
	\tx &= X + V_X \\
	\ty &= Y + V_Y 
\end{align*}

If we estimate the regression using the noisy measures $\tilde X$ and $\tilde Y$ we obtain
\begin{align*}
	\tilde \beta &= (\tx^T\tx)^{-1}(\tx^T\ty) 
\end{align*}
For simplicity, let's assume that $X$, $V_X$, and $V_Y$ are orthogonal to the regression error term $\ve$. Asymptotically,
\begin{align*}
	\tilde \beta &\overset{p}{\rightarrow} \frac{Cov(X+V_X, Y+V_Y)}{Var(X+V_X)} \\
	&= \frac{Cov(X+V_X, X\beta+U + V_Y)}{Var(X+V_X)} \\
	&= \frac{\beta Var(X) + Cov(X,U) + Cov(X,V_Y) + \beta Cov(X,V_X) + Cov(V_X,U) + Cov(V_X,V_Y)}{Var(X) + Var(V_X) + 2Cov(X,V_X)}
\end{align*}
For simplicity, and because we are focusing on the problem of measurement error rather than endogenous regressors, we assume that the regression error $U$ is independent of both X and $V_X$: $U\indep X,V_X$. This implies that $Cov(X,U)=Cov(V_X,U)=0$ and allows us to simplify the above expression to
\begin{align*}
	\tilde \beta &\overset{p}{\rightarrow} \frac{\beta Var(X) + \beta Cov(X,V_X) + Cov(X,V_Y) + Cov(V_X,V_Y)}{Var(X) + Var(V_X) + 2Cov(X,V_X)}
\end{align*}

The true coefficient $\beta$ can be written
\[ \beta = \frac{Cov(X,Y)}{Var(X)}\]	
and in our application we can reasonably assume $\beta>0 \Leftrightarrow Cov(X,Y)>0$.

To ascertain the direction of the bias created by measurement error we compare $\tilde \beta$ to $\hat\beta$. Theoretically, the direction of the bias is ambiguous. However, we can determine the sign of the bias under different assumptions about the covariances. 

The simplest assumption would be that all of the covariances involving measurement error terms are 0: $Cov(X,V_Y) = Cov(X,V_X) = Cov(V_X,V_Y) = 0$. This is equivalent to classical measurement error, giving us the familiar attenuation bias result:
\[ \tilde \beta \overset{p}{\rightarrow}  \frac{Cov(X,Y)}{Var(X) + Var(V_X)} < \hat \beta \overset{p}{\rightarrow}  \frac{Cov(X,Y)}{Var(X) }.\]

However, we almost certainly have non-classical measurement error, so let's consider what the bias would be under more reasonable assumptions. Suppose we randomly assigned workers and jobs to groups. Then both $\tx$ and $\ty$ would simply be equal to the overall means: $\tilde X_g = \bar X \hspace{4pt} \forall g$ and  $\tilde Y_g = \bar Y \hspace{4pt} \forall g$. This means that for large values of $Y$, $\tilde Y<Y$ and similarly for $X$. This implies that $Cov(X,V_X)<0$ and $Cov(Y,V_Y)<0$. Combining this with the fact that $Cov(X,Y)>0$ implies that $Cov(X,V_Y)<0$, $Cov(Y,V_X)<0$, and $Cov(V_X,V_Y)>0$. 

Therefore,
\begin{align*}
	\tilde\beta \overset{p}{\rightarrow} \frac{\beta Var(X) + \beta \overset{<0}{\overbrace{Cov(X,V_X)}} + \overset{<0}{\overbrace{Cov(X,V_Y)}} + \overset{>0}{\overbrace{Cov(V_X,V_Y)}}}{Var(X) + \underset{>0}{\underbrace{Var(V_X)}} + \underset{<0}{\underbrace{2Cov(X,V_X)}}}
\end{align*}
In this case it is theoretically ambiguous whether $\tilde\beta>\hat\beta$ or $\tilde\beta<\hat\beta$. Empirically, we consistently find that $\tilde\beta<\hat\beta$. This means that it must be the case that the terms that tend to reduce $\tilde\beta$ --- $Var(V_X)$, $\beta Cov(X,V_X)$, and $Cov(X,V_Y)$ --- must dominate the terms that increase $\tilde\beta$ --- $Cov(V_X,V_Y)$ and $2 Cov(X,V_X)$.

\begin{figure}
	\centering
	\caption{Coefficient estimates with worker and job misclassification}
	\includegraphics[width=\textwidth]{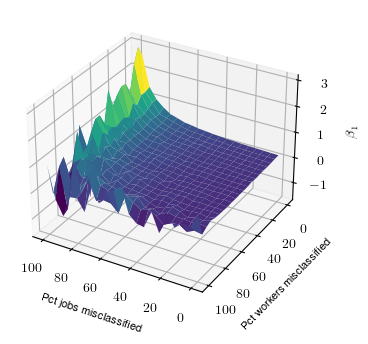}\\
	\label{fig:misclassification_demo_coef}
\end{figure}

\begin{figure}
	\centering
	\caption{$R^2$ values with worker and job misclassification}
	\includegraphics[width=\textwidth]{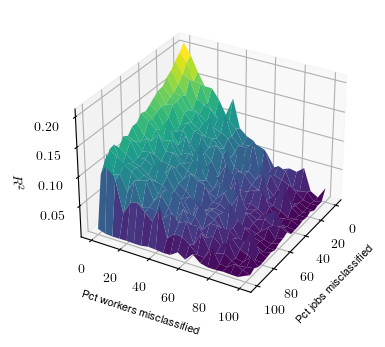}\\
	\label{fig:misclassification_demo_r2}
\end{figure}

We demonstrate this point through a simulation. We simulate a shock as described in Section \ref{sec:reduced_form_fake_olympics}. We estimate a series of regressions on changes in earnings by worker type on the Bartik instrument with jobs classified by market, however in each regression we randomly misclassify some percentage of workers and jobs. We loop from 0 to 100 percent of workers misclassified in intervals of five percent, and within each loop perform the same loop from 0 to 100 percent of jobs misclassified. We present the coefficients on the Bartik instrument in Figure \ref{fig:misclassification_demo_coef} and the $R^2$ values in Figure \ref{fig:misclassification_demo_r2}. $R^2$ values decline approximately monotonically with the degree of misclassification in both the worker and job dimensions, as expected. By contrast, there is much less of a coherent story with the regression coefficients. Again, this is consistent with the theoretical prediction that the effect of misclassification on regression coefficients is indeterminate.

\clearpage

\section{Proof that $A_{ij}$ follows a Poisson distribution}

\label{app:poisson_proof}

If an individual worker $i$ only searched for a job once, then the probability of worker $i$ matching with job $j$ would be equal to $\P_{ij} = \mathcal{P}_{\ig} d_j $ and $A_{ij}$ would follow a Bernoulli distribution: 
\[ A_{ij} \sim Bernoulli(\mathcal{P}_{\ig} d_j ). \]
However, since worker $i$ searches for jobs $c_i\equiv \sum_{t=1}^T c_{it}$ times, $A_{ij}$ is actually the sum of $c_i$ Bernoulli random variables, and is therefore a Binomial random variable. Conditional on knowing $c_i$, 
\[ A_{ij}|c_i \sim Binomial(c_i, \mathcal{P}_{\ig} d_j ).\]
However, we still need to take into account the fact that $c_i$ is a Poisson-distributed random variable with arrival rate $d_i$. Consequently, the unconditional distribution of $A_{ij}$ is Poisson as well: 
\[ A_{ij} \sim Poisson( d_i d_j \mathcal{P}_{\ig}  ).\]

We prove this fact by multiplying the conditional density of $A_{ij}|c_i$ by the marginal density of $c_i$ to get the joint density of $A_{ij}$ and $c_i$, and then integrating out $c_i$.

\begin{flalign*}
	P(A_{ij},c_i) = \underset{Bin(c_i, d_j P_{\iota\gamma}) }{\underbrace{P(A_{ij}|c_i)}} \quad \times \quad \underset{Poisson(d_i)}{\underbrace{P(c_i)}} \\
\end{flalign*}

Deriving the joint distribution:
\begin{flalign*}
	P(A_{ij},c_i) =& \binom{c_i}{A_{ij}} (d_j P_{\iota\gamma})^{A{ij}}(1-d_j P_{\iota\gamma})^{c_i - A{ij}} \times \frac{d_i^{c_i} \exp{(-d_i})}{c_i!} \\
\end{flalign*}

We want to find out the marginal distribution of $A_{ij}$:
\begin{flalign*}
	P(A_{ij}) &= \sum_{c_i=0}^\infty P(A_{ij},c_i) \\
	&= \sum_{c_i=0}^\infty \binom{c_i}{A_{ij}} (d_j P_{\iota\gamma})^{A{ij}}(1-d_j P_{\iota\gamma})^{c_i - A{ij}} \times \frac{d_i^{c_i} \exp{(-d_i})}{c_i!} \\
	&= \sum_{c_i=0}^\infty \frac{c_i!}{A_{ij}!(di-A_{ij})!} (d_j P_{\iota\gamma})^{A{ij}}(1-d_j P_{\iota\gamma})^{c_i - A{ij}} \times \frac{d_i^{c_i} \exp{(-d_i})}{c_i!} \\
	&= \frac{(d_j P_{\iota\gamma})^{A{ij}}\exp{(-d_i})}{A_{ij}!}  \sum_{c_i=0}^\infty \frac{1}{(di-A_{ij})!} (1-d_j P_{\iota\gamma})^{c_i - A{ij}} d_i^{c_i} \\
\end{flalign*}

If the summation term is equal to 
\begin{equation}\label{eq_bracket}
	\sum_{c_i=0}^\infty \frac{1}{(di-A_{ij})!} (1-d_j P_{\iota\gamma})^{c_i - A{ij}} d_i^{c_i} = d_i^{A_{ij}} \exp{(d_i (1 - d_j P_{\iota\gamma}))}
\end{equation}
then $P(A_{ij}) = \frac{(d_i d_j P_{\iota\gamma})^{A{ij}}\exp{(-d_i d_j P_{\iota\gamma}})}{A_{ij}!}$, i.e. $A_{ij}$ would be Poisson distributed: \\
\[ A_{ij} \sim Poisson(d_i d_j P_{\iota\gamma}) \] 

\vspace{2em}
Proving (\ref{eq_bracket}) is equivalent to proving the following equality:
\begin{flalign*}
	1 =& \frac{1}{d_i^{A_{ij}} \exp{(d_i (1 - d_j P_{\iota\gamma}))}}  \sum_{c_i=0}^\infty \frac{1}{(di-A_{ij})!} (1-d_j P_{\iota\gamma})^{c_i - A{ij}} d_i^{c_i} \\
\end{flalign*}

Proof:
\begin{flalign*}
	& d_i^{-A_{ij}} \exp{(-d_i (1 - d_j P_{\iota\gamma}))}  \sum_{c_i=0}^\infty \frac{1}{(di-A_{ij})!} (1-d_j P_{\iota\gamma})^{c_i - A{ij}} d_i^{c_i} = \\
	&= \sum_{c_i=0}^\infty \frac{\exp{(-d_i (1 - d_j P_{\iota\gamma}))}}{(di-A_{ij})!} (1-d_j P_{\iota\gamma})^{c_i - A{ij}} d_i^{c_i-A_{ij}} \\
	&= \sum_{c_i=0}^\infty \frac{\exp{(-d_i (1 - d_j P_{\iota\gamma}))}}{(di-A_{ij})!} (d_i(1-d_j P_{\iota\gamma}))^{c_i - A{ij}} \\
	&\text{We assume $\lambda = d_i (1 - d_j P_{\iota\gamma})$ for simplicity and we apply a change of variables $z = c_i - A_{ij}$} \\
	&= \sum_{z=0}^\infty \frac{\exp{(-\lambda)}}{z!} \lambda^z \text{, knowing that in our problem $c_i \geq A_{ij}$, i.e. $z \geq 0$}. \\
	&= 1 \\
	&\text{Since we have the p.d.f. of a Poisson r.v. inside the summation, i.e. $z \sim Poisson(\lambda)$  } \square\\ 
\end{flalign*}

Therefore, we have 

\[ A_{ij} \sim Poisson(d_i d_j P_{\iota\gamma}) \qed \]

\section{Worker and firm fixed effects}

Following \citet{BonhommeHolzheuLamadonManresaMogstadSetzler2020} and others, we decompose the variance in workers' log earnings into a component explained by worker fixed effects, a component explained by firm fixed effects, and a component explained by the covariance between worker and firm fixed effects. We find that firm effects explain 16\% of the variance in log earnings in our data and the covariance  between worker and firm effects explains 11\%. However, \citet{BonhommeHolzheuLamadonManresaMogstadSetzler2020} show that estimates of the firm effects component are subject to considerable upward bias due to limited mobility of workers between firms. Therefore, building upon the approach of \citet{BonhommeLamadonManresa2019_distributional,BonhommeLamadonManresa2019_discretizing}, we re-estimate the model at the group level, replacing firm effects with market ($\g$) effects. Using this grouped-data approach, we find that the share of the variance explained by market effects, as opposed to firm effects, falls to 1.2\% and the share of variance explained by worker--market covariance is 2.6\%.

\end{document}